\documentclass{aa}  

\usepackage{graphicx}
\usepackage{txfonts}
\usepackage{lipsum}
\usepackage{subcaption}         
\usepackage{lscape}             
\usepackage{placeins}           
                                
\usepackage[colorlinks=true,allcolors=blue]{hyperref}
\usepackage{orcidlink}
\providecommand{\orcid}[1]{\orcidlink{#1}}
\usepackage{comment}
\usepackage{booktabs}
\usepackage{lipsum}
\newcommand{\micmap}[1]{\textbf{\textcolor{red}{Michela: {#1}}}}

\begin{document}

   \title{N/O-enhanced chemical enrichment by massive stars:\\ the interplay of optically-thick winds and rotation}


%
%
%

   \author{Boyuan Liu
          \inst{1}\orcid{0000-0002-4966-7450}\thanks{\href{mailto:boyuan.liu@uni-heidelberg.de}{boyuan.liu@uni-heidelberg.de}}
          \and Michela Mapelli\inst{1,2,3,4}\orcid{0000-0001-8799-2548}
          \and Lumen Boco\inst{1}\orcid{0000-0003-3127-922X}
          \and Xihan Ji\inst{5,6}\orcid{https://orcid.org/0000-0002-1660-9502}
          }

   \institute{Universit\"at Heidelberg,  Zentrum f\"ur Astronomie, Institut f\"ur Theoretische Astrophysik,  Albert Ueberle Str. 2, D-69120 Heidelberg, Germany
       \and
             Universit\"at Heidelberg, Interdiszipli\"ares Zentrum f\"ur Wissenschaftliches Rechnen, D-69120 Heidelberg, Germany
        \and
        INFN-Padova, Via Marzolo 8, I–35131 Padova, Italy
        \and
        Dipartimento di Fisica e Astronomia Galileo Galilei, Università di Padova, Vicolo dell’Osservatorio 3, I–35122 Padova, Italy
        \and
        Kavli Institute for Cosmology, University of Cambridge, Madingley Road, Cambridge CB3 0HA, UK
        \and
        Cavendish Laboratory, University of Cambridge, 19 JJ Thomson Avenue, Cambridge CB3 0HE, UK
             }

   \date{Received XXXX, 2026}

 
  \abstract
    {Recent JWST observations have revealed a sample of high-redshift galaxies with highly enhanced N/O abundance ratios, $\log\rm (N/O)\gtrsim -1.1$. Strong optically-thick winds of rotating 
    massive stars going through the Wolf-Rayet (WR) phase provide a viable explanation for such N/O enhancements, supported by the direct spectroscopic WR signatures discovered in a few galaxies. Such strong winds are also needed to reproduce the single WR stars in the Small Magellanic Cloud (SMC). Here, we compute the metal yields of massive ($\sim 20-100\ \rm M_\odot$), metal-poor stars using two sets of \textsc{mesa} single stellar evolution models with and without optically-thick winds in a dense grid of metallicities ($Z_\star\sim 0.001-0.0045$) and initial spins ($\omega\equiv\Omega/\Omega_{\rm crit}\sim 0-0.7$). Considering an initial spin distribution based on observations of O-type stars in the SMC, we find that both sets produce strong N/O enhancements up to $\rm \log(N/O)\sim 0.1$ within a few Myr after a starburst. The models with optically-thick winds undergo a rapid transition from high N/O to the `normal' state of $\rm \log(N/O)\sim -1.5$ at $t\sim 3-4$~Myr driven by C/O-rich winds, while the transition is slower and weaker without optically-thick winds. 
    The optically-thick-wind scenario can reproduce the nebular C, N, O, and He abundance patterns of three representative high-$z$ galaxies showing direct WR signatures: RXCJ2248-ID, Sunburst Arc, and MARTA 4327, with stellar metallicities and ages similar to those inferred from SED-fitting. This can be interpreted as a coherent evolution sequence captured before ($t\lesssim 3$~Myr), during ($t\sim 3-4$~Myr), and after ($t\gtrsim 5$~Myr) the transition. In contrast, the optically-thin-wind-only scenario ejects much less C and O via winds and cannot reproduce Sunburst Arc and MARTA 4327 without fine-tuning of initial conditions. 
    This highlights the importance of optically-thick winds and rotation in massive stellar evolution and feedback. }


   \maketitle
   \nolinenumbers

\section{Introduction}\label{sec:intro}

Recent JWST observations have revolutionised our understanding of chemical enrichment in the first billion years of cosmic history, revealing a growing population of galaxies at $z\gtrsim 4$ that exhibit extreme nitrogen-to-oxygen abundance ratios \citep[e.g.,][]{Schaerer2024,Isobe2025,Ji2026,Morel2026,Rusakov2026,Watanabe2026}. These N/O-enhanced galaxies (NOEGs) show $\log(\rm N/O)>-1.1$ at sub-solar metallicities, deviating significantly from the local scaling relations that give $\log(\rm N/O)\sim -1.4$ \citep[e.g.,][]{Nicholls2017,Cataldi2026}. The most extreme objects, including GHZ9 at $z=10.15$ \citep{Napolitano2025}, CEERS-1019 at $z=8.68$ \citep{Isobe2023,Larson2023,Marques-Chaves2024}, GS\_3073 at $z=5.55$ \citep{Ubler2023,Ji2024,Isobe2025}, and UNCOVER-45924 at $z=4.46$ \citep{Labbe2024,Ji2026} 
reach highly supersolar N/O ratios $\log(\rm N/O)\gtrsim 0$, pointing to efficient CNO-cycle processing. These objects are typically extremely compact ($R_{\rm e}\lesssim 100\ \rm pc$), exhibit high star formation rate surface densities ($\Sigma_{\rm SFR}\gtrsim 10^2\ \rm M_\odot\ yr^{-1}\ kpc^{-2}$), and show high-ionisation lines (e.g., He II and C IV) indicative of hard radiation fields \citep[e.g.,][]{Topping2024,Schaerer2026,Tang2026,Umeda2026}.

The origin of such N/O enhancements is currently under intense investigation, and several mechanisms have been proposed, including massive stars (with initial masses of $m_\star\sim 20-100\ \rm M_\odot$) entering the Wolf-Rayet (WR) phase, very/extremely massive or supermassive stars ($m_\star\sim 100-10^5\ \rm M_\odot$), intermediate-mass stars ($m_\star\sim 1-8\ \rm M_\odot$) in the asymptotic giant branch (AGB), and tidal disruption events \citep[see the summary by][]{Watanabe2026}. The WR scenario is particularly relevant as direct spectroscopic signatures of WR stars-- i.e., broad He II]$\lambda$4686 emission and the blue (N III]$\lambda$4640) and red (C IV]$\lambda$5808) bumps-- have been detected in several high-$z$ galaxies, many of which also show highly enhanced N/O ratios \citep[e.g.,][]{Rivera-Thorsen2024,Berg2026,Curti2026}. 
Moreover, the spectrum of GN-z11 at $z=10.6$ \citep{Cameron2023,Tacchella2023,Maiolino2024,Alvarez-Marquez2025,Ji2025}, one of the most famous NOEGs \citep[with $\log(\rm N/O)\sim -0.35$,][]{Senchyna2024}, shows a broad N IV]$\lambda$1486 component that can possibly arise from WR stars \citep{Chen2026,Nakane2026}. Strong He~II]$\lambda$4686 emission (with a He~II]$\lambda4686$/H$\beta$ ratio of $\sim 0.03$) is also detected in the stacked spectrum of star-forming galaxies at $z\sim 4-10$, which shows a strong N/O enhancement $\log (\rm N/O)\sim -0.2$ as well \citep{Rusakov2026,Umeda2026}. These observations provide evidence that WR stars are actively enriching the interstellar medium (ISM) in the early Universe with CNO-cycle-processed material. 


Indeed, previous chemical evolution models \citep[e.g.,][]{Kobayashi2024,Fukushima2024,Marques-Chaves2024,Berg2026,Watanabe2024,Watanabe2026} have successfully reproduced the abundance patterns of individual NOEGs using the metal yields of WR stars from \citet{Limongi2018}. 
However, they typically rely on specific assumptions about star formation and galactic inflow/outflow histories, initial stellar rotation velocities, and, most importantly, massive stellar evolution physics that essentially determine the metal yields. 
It is timely to revisit the metal yields from massive stars, taking into account recent progress in understanding WR star formation in metal-poor environments based on state-of-the-art simulations and local observations. In particular, it has been shown by \citet{Boco2025} that optically-thick winds, combined with quasi-chemically homogeneous evolution driven by rotational mixing, allow single metal-poor stars with initial masses down to $m_\star=20\ \rm M_\odot$ to self-strip their envelopes and become WR stars even at the metallicity of the Small Magellanic Cloud (SMC), $Z\sim 0.002-0.004$, without requiring binary companions. 
This 
provides a possible explanation for the 
observation of single WR stars 
in the SMC, and highlights the importance of taking into account both rotation and optically-thick winds in massive stellar evolution. Such optically-thick winds also strongly affect the fates of massive stars in supernova explosions and massive black hole formation \citep{Simonato2025,Torniamenti2026}.


Here, we aim to clarify the roles played by rotation and optically-thick winds in shaping the metal yields from metal-poor massive stars ($m_\star\sim20-100\ \rm M_\odot$) and to evaluate whether the observed NOEGs can be explained by their metal enrichment under reasonable assumptions on initial (stellar) spin distribution and metal mixing with the ISM. We compute the metal yields of massive stars using two sets of \textsc{mesa} single stellar evolution models with and without optically-thick winds, spanning a dense grid of metallicities ($Z\sim 0.001-0.0045$) and initial spins ($\omega\equiv\Omega/\Omega_{\rm crit}\sim 0-0.7$), incorporating metal yields from both stellar winds and supernovae (SNe). 
By integrating these yields over the initial mass function (IMF) and initial spin distribution, and by modelling the mixing of stellar ejecta with the ISM through a phenomenological parameter, we predict the chemical evolution of the enriched medium following a starburst, focusing particularly on the N/O ratio. We compare our predictions to the detailed CNO and He abundances of three representative high-$z$ galaxies with direct WR signatures: RXCJ2248-ID at $z=6.1$ \citep{Berg2026}, Sunburst Arc at $z=2.37$ \citep{Rivera-Thorsen2024}, and MARTA 4327 at $z=2.2$ \citep{Curti2026}. These systems show a considerable range of metallicities ($12+\log(\rm O/H)\sim 7.7-8.5$) and N/O ratios from highly enhanced ($\log(\rm N/O)\sim -0.4$) to normal ($\log(\rm N/O)\sim -1.4$), providing an ideal laboratory to test our models. 
We also compare the predicted total N yields with the N mass budgets of NOEGs inferred from observations to check if enough N is available from our massive stars.

The paper is organised as follows. In Section~\ref{sec:method}, we describe our stellar evolution models, the methods for computing wind metal yields, SN metal yields, and the models for stellar population synthesis and metal mixing. In Section~\ref{sec:comp_obs}, we discuss the observational context, focusing on the abundance patterns of the three benchmark galaxies with direct WR signatures. In Section~\ref{sec:res}, we present the chemical composition of the metal yields from individual stars and stellar populations, and evaluate the strength and duration of N/O enhancements. In Section~\ref{sec:diss}, we discuss the implications of our results for understanding high-$z$ observations and massive stellar evolution, covering the models that can reproduce the three benchmark galaxies, N mass budgets of NOEGs, and alternative/supplementary enrichment mechanisms. We summarise our conclusions in Section~\ref{sec:conclusion}.

\section{Methods}\label{sec:method}
In this Section, we first outline the stellar evolution models developed by \citet{Boco2025}, which are used to derive wind metal yields and predict pre-SN states (Sec.~\ref{sec:stellar}). Then we explain our scheme for computing SN metal yields (Sec.~\ref{sec:sn}) based on the SN models of \citet{Woosley2017} and \citet{Liu2025mzsfr}. We further introduce a scheme to modify the CNO yields (from secondary production) to account for non-solar initial stellar abundance patterns through post-processing (Appendix~\ref{sec:non_solar}). Next, we develop a simple stellar population synthesis model 
and a phenomenological model for the mixing of stellar yields with the ISM to predict the abundance pattern of the enriched medium (Sec.~\ref{sec:sps_mix}). 

Throughout the paper, we mostly express the abundance ratio of two elements, $\rm X_1$ and $\rm X_2$, with respect to solar as $\rm [X_1/X_2]\equiv\log(X_1/X_2)-\log(X_1/X_2)_{\odot}$, given the (proto)solar abundance ratio $\rm \log(X_1/X_2)_{\odot}$ from \citet{Asplund2009} with a bulk metal mass fraction of $\rm Z_\odot=0.0142$. For He, we use the mass fraction $Y$, which is related to the He number fraction $y$ or ${\rm \log(He/H)}\equiv\log y$ by $Y\simeq 4y/(1+4y)$. We only consider stable isotopes, i.e., C means $\rm ^{12}C$, etc. 
We use the iron abundance $\rm [Fe/H]$ as the indicator of stellar metallicity, assuming that the iron opacity plays the dominant role in shaping overall stellar evolution. We use $\rm [Fe/H]\equiv \log(Z_\star/\rm Z_\odot)$ to convert SED-derived stellar metallicity $Z_\star$ (i.e., metal mass fraction) from observational studies to $\rm [Fe/H]$ unless specially clarified, although the relation does not hold for $\alpha$-enhanced, metal-poor stars. This inconsistency should not affect the interpretation of our results, as most stellar population synthesis models used in SED-fitting also adopt the solar abundance pattern for metal-poor stars, such as early versions of BPASS before the introduction of $\alpha$-enhanced models in v2.2.3 and thereafter \citep{Byrne2023,Byrne2025}. 

\subsection{Massive stellar evolution and wind metal yields}\label{sec:stellar}

We derive the wind metal yields from the single stellar evolution models presented in \citet{Boco2025}, which have been extended to span a slightly denser and broader grid of initial stellar masses, metallicities, and spins. The models were computed using the Modules for Experiments in Stellar Astrophysics (\textsc{mesa}; version r12115, \citealp{Paxton2011,Paxton2013,Paxton2015,Paxton2018,Paxton2019}), consideirng stars with 11 zero-age main sequence (ZAMS) masses $m_\star=20$, 25, 30, 35, 40, 50, 60, 70, 80, 90, and 100~$\rm M_\odot$ at 8 metallicities $Z_{\star} = 0.001$, 0.0015, 0.002, 0.0025, 0.003, 0.0035, 0.004, and 0.0045, and 9 initial spins $\omega\equiv \Omega /\Omega_{\rm crit} = 0$, 0.3, 0.45, 0.48, 0.5, 0.55, 0.6, 0.65, and 0.7, with the critical rotation rate defined by $\Omega_{\rm crit}^{2} = (1 - \Gamma_{\rm e})Gm_{\star}/R_{\star}^{3}$, where $\Gamma_{\rm e} =\chi_{\rm e} L/(4\pi G c m_\star)$ is the electron scattering Eddington parameter (with $\chi_{\rm e}$ being the electron scattering opacity, $G$ the gravity constant, and $c$ the speed of light), $R_\star$ is the stellar radius, and $L$ is the stellar luminosity. We run two sets of models with and without optically-thick winds, implemented following the approach of \citet{Sabhahit2023}.  
Below, we summarise the key ingredients of the \textsc{mesa} runs, and refer the reader to \citet[]{Boco2025} for details.

Our simulations employ the initial composition (for the solar abundance pattern) and opacity tables of \citet{Grevesse1998}, the default equation of state tables in \textsc{mesa} \citep{Rogers1996}, and the Ledoux criterion for convective regions with mixing length parameter $\alpha_{\rm MLT} = 1.5$. Following \citet{Sabhahit2023}, overshooting is modeled using the exponential criterion with coefficient $f_{\rm ov} = 0.03$ \citep[which is favoured by asteroseismology of massive stars of $m_\star\sim 12-40\ \rm M_\odot$ in the MW,][]{Lechien2026} during the main sequence (MS, when the central H fraction is above 0.01) and $f_{\rm ov} = 0.01$ during subsequent evolution. The MLT++ prescription is turned on for post-MS evolution, which artificially reduces the superadiabatic gradient in radiation-dominated, convective envelopes to avoid numerical instabilities. The rotational mixing efficiency follows the \citet{Heger2000} calibration with $f_{c} = 1/30$ (ratio of the turbulent viscosity to the diffusion coefficient) and $f_\mu = 0.05$ (ratio between the actual molecular weight gradient and the value used for computing the mixing coefficients). The models are evolved until the central temperature reaches $T_c= 10^{9.55}\ \rm K$ during the O-burning phase. 

The key innovation of these models is the implementation of a self-consistent switch between optically-thin and optically-thick wind regimes following the formalism of \citet{Sabhahit2023}. In the optically-thin regime, the standard \citet{Vink2001} wind prescription is used. When the wind efficiency parameter $\eta\equiv \dot{m}_{\star}v_\infty/(L/c)$ exceeds the threshold $\eta_{\rm switch}$, the steep optically-thick wind scaling from \citet{Vink2011} is activated. The switch point is determined by $\eta_{\rm switch} = 0.75/[1 + (v_{\rm esc}^{2}/v_\infty^2)]$, based on dynamically consistent stellar atmosphere models computed by the Potsdam Wolf–Rayet (PoWR) code \citep{Sander2017,Sander2020a}, where $v_{\rm esc} \equiv \sqrt{2Gm_\star/R_\star}$ and $v_\infty\equiv a\,{}Z^{0.2}\sqrt{m_\star(1-\Gamma_{\rm e})/R_\star}$ \citep{Lamers1995} are the escape and wind terminal velocities, with $a=2.6$ (1.3) above (below) the bi-stability jump at $T_{\rm eff}= 2.5 \times 10^{4}\ \rm K$. In practice, if optically-thick winds are included, they are turned on when the Eddington parameter $\Gamma_{\rm e}$ exceeds a critical value $\Gamma_{\rm e, switch}$, defined by $\eta(\Gamma_{\rm e,switch}) = \eta_{\rm switch}(\Gamma_{\rm e,switch})$, for stars with effective temperatures $T_{\rm eff}\sim 4000-10^{5}$~K. Here, $\eta$ is computed using the optically-thin mass-loss rate of \citet{Vink2001}, which follows $\eta\propto Z^{0.85}(v_\infty/v_{\rm esc})^{-1.23}\propto Z^{0.6}$ and $\eta\propto Z^{0.85}(v_\infty/v_{\rm esc})^{-1.60}\propto Z^{0.53}$ above and below the bi-stability jump. 
For red supergiants (RSGs) with $T_{\rm eff}<4000$~K, we always use the wind prescription of \citet{deJager1988}, while for $T_{\rm eff} > 10^{5}$~K, we use the maximum between the WR wind mass loss rate from \citet{Vink2017} and \citet{Sander2020}. 
We also implement the enhancement of mass loss by rotation following the prescription of \citet{Maeder2000}: $\dot{m}_\star(\omega)/\dot{m}_\star(\omega=0) = (1 - \Gamma)^{1/\alpha-1} / [1 - \Gamma - (4/9)(v_{\rm rot}/v_{\rm cri})^{2})]^{1/\alpha-1}$ with $\alpha = 0.52$, $v_{\rm crit}=\sqrt{2 G m_\star/(3 R_{\rm pb} )}$, and $\Gamma=\chi L/(4\pi G c m_\star)$ being the total Eddington parameter, given the polar radius $R_{\rm pb}$ and rotation velocity $v_{\rm rot}$.

Once the wind mass loss rate $\dot{m}_\star$ is known, the cumulative yield of an element X can be computed as $m_{\rm X}(t)=-\int_0 ^{t} \dot{m}_{\star}(t')Z_{\rm X}(t')dt'$, where $Z_{\rm X}$ is the surface mass fraction of X.



\subsection{Supernova metal yields}\label{sec:sn}

In addition to the metal yields from winds predicted by the stellar evolution models, we derive the SN metal yields from the stellar structure at the last snapshot\footnote{The models end in the O-burning phase when the central temperature reaches $T_{c}=10^{9.55}\ \rm K$ with well-defined He and CO cores, based on which we derive the SN yields.}. For pulsational pair-instability supernovae (PPISNe), we derive the remnant mass $m_{\rm rem}$ from the simulated CO core mass $m_{\rm CO, core}$ and He core mass $m_{\rm He,core}$ using linear interpolation of the $m_{\rm rem,W17}(m_{\rm CO, core})$ and $m_{\rm rem,W17}(m_{\rm He,core})$ relations from \citet{Woosley2017}, which hold for $m_{\rm CO, core}\in [26.3,54.14]\ \rm M_\odot$ and $m_{\rm He,core}\in [32,64]\ \rm M_\odot$. Here, the upper core mass limits are never reached by our models of $m_\star\le 100\ \rm M_\odot$ stars, and we only trigger PPISNe when $m_{\rm He,core}> 32\ \rm M_\odot$. In practice, we have $m_{\rm rem}=0.5[m_{\rm rem,W17}(m_{\rm CO, core})+m_{\rm rem,W17}(m_{\rm He,core})]$ when both relations are applicable, and $m_{\rm rem}=m_{\rm rem,W17}(m_{\rm He,core})$ when the CO core-remnant mass relation is not applicable (i.e., $m_{\rm CO, core}<26.3\ \rm M_\odot$). Once $m_{\rm rem}$ is known, we model PPISNe as super winds by ejecting all layers above $m_{\rm rem}$ without any synthesis of new elements during the explosions/pulsations. 

For core-collapse supernovae (CCSNe) and failed supernovae, we interpolate and rescale the yields adopted by \citet{Liu2025mzsfr} which combine the \textsc{parsec} stellar evolution tracks of \citet{Costa2025} with the yield tables of \citet{Limongi2003} and \citet{Chieffi2004} using the delayed formalism for explodability and remnant mass of \citet{Fryer2012}, following the method of \citet{Goswami2021}. First, we identify the SN model of \citet{Liu2025mzsfr} that best matches the initial stellar metallicity (in terms of $\log Z_{\star}$) and final CO core mass of each stellar evolution model in our grid. 
Next, we calculate the fallback fraction of the layers above the CO core as
\begin{align}
    f_{\rm fb}=\frac{\tilde{m}_{{\rm rem}}-\tilde{m}_{{\rm CO, core}}}{\tilde{m}_{{\rm preSN}}-\tilde{m}_{{\rm CO, core}}}\ ,
\end{align}
where $\tilde{m}_{{\rm rem}}$, $\tilde{m}_{{\rm CO, core}}$, and $\tilde{m}_{{\rm preSN}}$ are the remnant mass, pre-SNe CO core and total masses of the best-match model. The same fraction is applied to our pre-SN structure (whose properties are denoted by symbols without tilde), such that the total mass ejected from the SN event and the remnant mass can be estimated as $m_{\rm ej,SN}=(1-f_{\rm fb})(m_{\rm preSN}-m_{\rm CO,core})$ and $m_{\rm rem}=m_{\rm CO,core}+f_{\rm fb}(m_{\rm preSN}-m_{\rm CO,core})$. The He/H yield $m_{\rm He/H,SN}$ is then estimated as the total He/H mass in layers above $m_{\rm rem}$. 

We further assume that the \textit{newly-synthesized} $\rm X=C$, N, O mass is proportional to the mass of layers in between the He and CO core boundaries, such that the relevant yield from \citet{Liu2025mzsfr} should be scaled by a factor
\begin{align}
    f_{\rm scale}=\frac{m_{\rm He,core}-m_{\rm CO,core}}{\tilde{m}_{\rm He,core}-\tilde{m}_{\rm CO,core}}\ .
\end{align}
The final yield $m_{\rm X,SN}$ is the summation of the newly-synthesized mass $\Delta m_{\rm X,SN}$ and that initially in the ejecta $m_{\rm X,SN,0}$, i.e.,
\begin{align}
    &m_{\rm X,SN}=\Delta m_{\rm X,SN} + m_{\rm X,SN,0}\ ,\\
    &\Delta m_{\rm X,SN}=f_{\rm scale}(\tilde{m}_{\rm X,SN}-\tilde{m}_{\rm X,SN,0})\ ,\\
    &m_{\rm X,SN,0}=m_{\rm H,SN}A_{\rm X}(Z_\star/\rm Z_{ \odot})\times 10^{\rm \log(X/H)_{\odot}}\ ,\label{eq:mxsn0}
\end{align}
Here $\tilde{m}_{\rm X,SN}$ is the total yield of X from \citet{Liu2025mzsfr}, and $\tilde{m}_{\rm X,SN,0}$ is the corresponding mass initially in the ejecta, which is calculated with Eq.~\ref{eq:mxsn0} by replacing $m_{\rm H,SN}$ and $Z_\star$ with $\tilde{m}_{\rm H,SN}$ and $\tilde{Z}_\star$ of the best-match SN model. 

\subsection{Stellar population synthesis and mixing of stellar yields} 
\label{sec:sps_mix}

Since our stellar evolution models only involve massive single stars with $m_\star\ge 20\ \rm M_\odot$, we focus on the metal enrichment of young stellar populations with ages $t_{\rm age}\lesssim 10$~Myr. In practice, we stop the calculation at the first moment when a $m_\star=20\ \rm M_\odot$ star dies. 
For such young populations, the exact form of star formation history is not expected to have a significant impact. Therefore, we consider an instantaneous starburst by default. We also explore the case of continuous star formation at a constant rate to evaluate the potential effects in Appendix~\ref{apdx:csfr}, which indeed do not change our conclusions. Less massive ($m_\star<20\ \rm M_\odot$) stars formed in the same starburst do not contribute much, because they have not evolved enough to produce SNe and strong winds (at least for single stellar evolution). The contributions of older stars can be absorbed into the (pre-starburst) ISM composition. The population-average cumulative yield of element X from massive stars $m_\star\ge 20\ \rm M_\odot$ per unit stellar mass formed is
\begin{align}
    \frac{\hat{M}_{\rm X}(Z_\star)}{M_\star}=\frac{\int_0^1 p(\omega)\left[\int_{20\ \rm M_\odot}^{m_{\max}} m_{\rm X}(m_\star,\omega,Z_\star)\Phi(m_\star)dm_\star\right] d\omega}{M_\star}\ ,\label{eq:mz}
\end{align}
where $m_{\rm X}(m_\star,\omega,Z_\star)$ is the yield of a single star with initial mass $m_\star$, spin $\omega$, and metallicity $Z_\star$, $\Phi(m_\star)$ is the IMF, $p(\omega)$ is the initial spin distribution (assumed to be independent of $m_\star$ and $Z$), and $M_\star$ is the corresponding total stellar mass $M_\star =\int_{0.01\ \rm M_\odot}^{m_{\max}} m_{\star}\,{}\Phi(m_\star)\,{}dm_\star$. The time dependence of $m_{\rm X}$ and $\hat{M}_{\rm X}$ is not written out for conciseness. By default, we adopt the canonical \citet{Kroupa2001} IMF with $m_{\max}=\rm 100\ \rm M_\odot$. We focus on two models of $p(\omega)$:
\begin{itemize}
    \item The spin distribution inferred from the observed rotation velocities of O stars in the SMC derived by \citet[see their fig.~2]{Boco2025}, which we call the SMC spin distribution\footnote{The cumulative distribution function can be well fitted by $F(\omega)=\sum_{i=1}^{4}b_{i}\omega^{i}$ with $b_1=1.451$, $b_2=2.336$, $b_3=-5.084$, and $b_4=2.294$. The bins of our spin grid, $\omega=0$, 0.3, 0.45, 0.48, 0.5, 0.55, 0.6, 0.65, and 0.7, correspond to the $\sim0$, 53, 75, 79, 82, 87, 91, 95, and 97 percentiles of the spin distribution. Their weights are set following the trapezoidal rule as $W_{0}=0.228$, $W_{0.3}=0.411$, $W_{0.45}=0.129$, $W_{0.48}=0.031$, $W_{\rm 0.5}=0.040$, $W_{\rm 0.55}=0.050$, $W_{\rm 0.6}=0.039$, $W_{\rm 0.65}=0.029$, and $W_{\rm 0.7}=0.042$. }. 
    \item $p(\omega)=\delta(\omega-0.7)$, an extreme fast-rotating model in which all stars have $\omega=0.7$.
\end{itemize}

If the yield $\hat{M}_{\rm X}$ is mixed with the ISM of mass $M_{\rm ISM}$, the mass of element X with nucleon number $A_{\rm X}$ in the enriched medium (normalised by the total stellar mass $M_\star$) can be written as:
\begin{align}
    M_{\rm X}/M_\star=\hat{M}_{\rm X}/M_\star + f_{\rm mix}f_{\rm dilution}\times X_{\rm H}A_{\rm X}10^{\rm [Fe/H]+[X/Fe]_{\rm pre}}\ ,\label{eq:mx}
\end{align}
where $f_{\rm mix}\equiv M_{\rm ISM}/M_{\star}$ is the mixing factor, $f_{\rm dilution}=0.5$ is a dilution factor, assuming that the pre-starburst ISM is mixed with the same amount of inflowing pristine gas\footnote{The inflowing gas from the intergalactic and circumgalactic medium can already be pre-enriched. This can have non-negligible effects on chemical evolution \citep{Citro2026}, which we have ignored for simplicity.} during the starburst, $X_{\rm H}=0.76$ is the primordial H mass fraction, ${\rm [Fe/H]}$ and $\rm [X/Fe]_{\rm pre}$ are the stellar, i.e., pre-starburst ISM, metallicity and X-Fe abundance ratio. We treat $f_{\rm mix}$ as a phenomenological free parameter governing the strength of mixing, which can be adjusted to reproduce the abundance pattern of a certain observed system (as done so in Sec.~\ref{sec:obs}). The exact physical meaning and values of $f_{\rm mix}$ can only be acquired with detailed simulations of star formation and stellar feedback \citep[e.g.,][]{Saitoh2026,Shi2026vms}, beyond the scope of this study. Nevertheless, the values of $f_{\rm mix}$ inferred from observations can still provide some hints. Given the masses of individual elements from Eq.~\ref{eq:mx}, the abundance ratio of two elements, $\rm X_1$ and $\rm X_2$, is simply
\begin{align}
    {\rm [X_1/X_2]}=\log[(M_{\rm X_1}/M_{\rm X_2})(A_{\rm X_2}/A_{\rm X_1})]-\rm \log(X_1/X_2)_\odot\ .\label{eq:abund_ratio}
\end{align}


%


\section{Comparison with observations}\label{sec:comp_obs}

\begin{table*}[htbp]
    \caption{Properties of three representative high-$z$ systems with spectroscopic signatures of WR stars and the chemical abundances measured from the stacked spectrum of star-forming galaxies in lensing-cluster fields at $z\sim 4.5-10.1$ (explained in detail in Sec.~\ref{sec:comp_obs}). The range of $t_{\rm age}$ given in the brackets show the results of multiple references, broader than that of the primary reference in front of the brackets. The CNO abundance ratios with respect to solar can be converted to absolute abundance ratios with $\rm [N/O]\equiv \log(N/O)+0.872$, $\rm [C/O]\equiv \log(C/O)+0.264$, and $\rm [O/H]\equiv \log(O/H)+3.312$.}
    {\centering
    \begin{tabular}{cc|cccc|ccc}
     \toprule
     Object & Redshift & $\rm [O/H]$ & $\rm [N/O]$ & $\rm [C/O]$ & $Y$ & $t_{\rm age}\ [\rm Myr]$ & [Fe/H] \\ 
     \midrule
     RXCJ2248-ID$^a$ & $6.1$ & $-0.935_{-0.023}^{+0.023}$ & $0.482_{-0.035}^{+0.035}$ & $-0.531_{- 0.052}^{+0.052}$ & $0.33_{-0.09}^{+0.05}$ & $1.8_{-0.4}^{+0.7}$ ($0.7-13.5$) & $-0.935_{-1.065}^{+0.023}$ \\
     Sunburst Arc$^b$ & $ 2.37$ & $-0.498_{-0.06}^{+0.11}$ & $0.412_{-0.07}^{+0.17}$ & $-0.302_{-0.003}^{+0.128}$ & $0.32_{-0.03}^{+0.01}$ & $4.5\pm 0.5$ ($1.4-5$) & $-0.85_{+0.15}^{-0.15}$ \\
     MARTA 4327$^c$ & $2.2$ & $-0.218_{-0.15}^{+0.11}$ & $-0.528_{-0.07}^{+0.07}$ & - & $0.253_{\rm -0.008}^{+0.008}$ & $6.3_{-1.3}^{+1.6}$ ($2.5-7.9$) & $-0.598_{-0.266}^{+0.246}$ \\
     \midrule
     GAll stacked$^d$ & $4.5-10.1$ & $-1.028_{-0.04}^{+0.06}$ & $0.66_{-0.24}^{+0.24}$ & $-0.47_{-0.06}^{+0.07}$ & - & - & - \\
    \bottomrule
    \end{tabular}}\\
    {\small $(a)$ \citet{Topping2024,Berg2026,Claeyssens2026,Moreschini2026}, $(b)$ \citet{Chisholm2019,Pascale2023,Rivera-Thorsen2024,Moreschini2026}, $(c)$ \citet{Curti2026,Moreschini2026}, $(d)$ \citet{Umeda2026}}
    \label{tab:obs}
\end{table*}

To put our theoretical predictions in the context of observations, we follow the conventional definition of N/O enhancement as $\rm \log(N/O)>-1.1$ \citep[e.g.,][]{Bhattacharya2026,Rusakov2026}, i.e., $\rm [N/O]>-0.228$. We further combine this threshold with additional criteria, $\rm [C/O]<0.2$, $\rm [O/H]<-0.1$, $\rm [N/C]>0$, and $Y<0.44$, to delineate the region occupied by most observed NOEGs \citep[]{Ji2026,Rusakov2026} in the chemical abundance space. We want to understand how the enriched medium enters and leaves this region, which we call the N/O-enhanced (NOE) phase, and how strong N/O enhancement can be achieved.  

We further compare our predictions with the detailed chemical abundances of three benchmark high-$z$ systems that exhibit direct spectroscopic signatures of WR stars: RXCJ2248-ID, Sunburst Arc, and MARTA 4327 \citep{Pascale2023,Topping2024,Rivera-Thorsen2024,Yanagisawa2024,Berg2026,Curti2026,Moreschini2026}. Thanks to the unprecedented sensitivity and wavelength coverage of JWST, and strong magnifications by lensing (for the former two), the CNO (and He) abundances of these galaxies are well characterised (with measurement uncertainties $\lesssim 0.15$~dex), spanning a considerable range of metallicities ($\rm 12+\log(O/H)\sim 7.7 - 8.5$) and N/O ratios $-1.4\lesssim \rm \log(N/O)\lesssim -0.4$. The detections of WR signatures further provide independent constraints on the stellar populations powering the nebula lines. Therefore, these systems serve as an exceptional laboratory to test whether massive stellar evolution and chemical enrichment models can reproduce the observed diversity of chemical abundance patterns and WR signatures in NOEGs. We summarise the nebula abundances and stellar population properties adopted for comparison in Table~\ref{tab:obs} and explain the references below. 
\begin{itemize}
    \item \textbf{RXCJ2248-ID} is a galaxy at $z= 6.1$, strongly lensed by the Abell S1063 cluster. We consider the CNO abundances measured by \citet[see their table 3]{Berg2026} using a four-zone ionisation model for the brightest image (RXCJ2248-ID3): $\rm 12+\log(O/H)=7.753\pm 0.023$, $\rm \log(N/O)=-0.390\pm 0.035$, and $\rm\log(C/O)=-0.795\pm 0.052$. Note that the O/H and C/O measurements here adopt high electron densities $n_{\rm e}\sim 1-3\times 10^5\ \rm cm^{-3}$ derived from UV diagnostics for high-ionisation regions, which are expected to better trace the metal yields from young, massive (WR) stars than low-density-sensitive optical lines\footnote{Realistic star-forming galaxies/regions contain gas across a wide range of densities traced by different nebular lines \citep{Mendez-Delgado2026}. Assuming a single low density can bias the O (N and He) abundance to smaller (larger) values by up to $\sim 1$~dex \citep{Yanagisawa2024,Martinez2025,Hsiao2026,Moreschini2026}. Massive stars are expected to form in dense central hubs of collapsing clouds \citep[e.g.,][]{Miao2026}, and their feedback can pressurize stellar ejecta and surrounding gas to high densities \citep{Pascale2023,Shi2026vms}. Therefore, we always consider the abundance measurements based on high-density-sensitive UV lines (e.g., N III], N IV], O III], and C III]) even if those from low-density-sensitive optical lines (e.g., N II] and O II]) are also available, noting that the two can be significantly different in chemically stratified ISM \citep[e.g.,][]{Pascale2023,Ji2024,Moreschini2026,SinghRai2026cno}.}. The N/O measurement is the weighted mean of the values for different ionisation zones, which are consistent within uncertainties, and is dominated by the contribution of the high-density, high-ionisation zone. These results are in good agreement with those obtained by \citet{Moreschini2026} using a similar multi-zone model. 
    For the He abundance, we adopt $12+\rm \log(He/H)=11.09_{-0.04}^{+0.09}$ from \citet{Moreschini2026}, corresponding to a He mass fraction $Y=0.33_{-0.09}^{+0.05}$, which is consistent with the results of \citet{Yanagisawa2024} when high densities $n_{\rm e}\sim 10^3- 10^5\ \rm cm^{-3}$ are considered in the prior. For the stellar population properties, we take $t_{\rm age}=1.8_{-0.4}^{+0.7}\ \rm Myr$ derived from SED-fitting by \citet{Topping2024}. 
    We also consider a broader age range $t_{\rm age}=1.6_{-0.9}^{+11.9}\ \rm Myr$ \citep{Berg2026,Claeyssens2026}. Here, an relatively old age of $t_{\rm age}\sim 12$~Myr is preferred with the presence of an active galactic nucleus \citep[AGN,][]{Moreschini2026}. The stellar metallicity is not well-constrained, although the metal-poor nature is evident from the low nebular oxygen abundance $\rm -0.958\lesssim [O/H]\lesssim-0.912$. This nebular metallicity (of high-density gas) approximately places an upper limit on the \textsc{stellar} metallicity, considering that dense regions are not expected to undergo significant metal dilution by incoming pristine gas. On the other hand, $Z_\star=10^{-4}\sim 10^{-2}\ \rm Z_\odot$ is found by \citet{Moreschini2026} from SED-fitting on a coarse grid of $\log Z_\star=-1.7,-2.7,-4$, which we take as the lower limit. In the end, we estimate the stellar metallicity of RXCJ2248-ID to be in the range $\rm [Fe/H]\in [-2, -0.912]$. 
    \item \textbf{Sunburst Arc} is the brightest known strongly-lensed galaxy at Cosmic Noon ($z=2.37$), hosting a compact LyC-emitting clump that appears to be a super star cluster with WR signatures \citep[e.g.,][]{Rivera-Thorsen2024}. 
    \citet{Pascale2023} and \citet{Moreschini2026} have shown that a two-component model is needed to reproduce all observed nebular lines, featuring a N/O-enhanced, compact (central) component with $n_{\rm e}\sim 10^{5}\ \rm cm^{-3}$ and a subdominant, N/O-normal, extended component with $n_{\rm e}\sim 10^2\ \rm cm^{-3}$. Here, we use the CNO and He abundances measured by \citet[see their table~2 and D.1]{Moreschini2026} for the compact component: $\rm 12+\log(O/H)=8.19_{-0.06}^{+0.11}$, $\rm \log(N/O)=-0.46_{-0.07}^{+0.17}$, $\rm \log(C/O)=-0.566_{-0.003}^{+0.128}$, and $12+\rm\log(He/H)=11.08_{-0.06}^{+0.02}$, based on stacked spectrum of multiple images from MUSE \citep{Pascale2023} and JWST \citep{Welch2025}. \citet{Rivera-Thorsen2024} find that the blue (]$\lambda\sim 4550-4750$~\AA) and red/orange (]$\lambda\sim 5600-6000$~\AA) WR bumps in the stacked spectrum of Sunburst Arc can be best explained by a stellar age of $t_{\rm age}\sim 4-5$~Myr, and a stellar metallicity of $Z_\star/\rm Z_\odot\sim 0.1-0.2$ (i.e., $\rm [Fe/H]\in [-1,-0.7]$), which we take as the primary reference. Earlier SED-fitting analysis obtained a similar metallicity, and a slightly younger (light-weighted) age, $t_{\rm age}\sim 1.4-4$~Myr \citep{Chisholm2019,Pascale2023}. 
    \item \textbf{MARTA 4327} is a star-forming galaxy at $z=2.223$ with the highest-redshift detections of WR signatures without lensing. According to \citet{Moreschini2026}, unlike RXCJ2248-ID and Sunburst Arc, the N/O ratio of MARTA 4327 is rather normal, $\rm \log(N/O)=-1.40_{-0.07}^{+0.07}$, given its higher metallicity $\rm 12+\log(O/H)=8.47_{-0.15}^{+0.11}$. Here, we consider the models of \citet{Moreschini2026} that include the contributions of high-density regions (up to $n_{\rm e}\sim 10^{7}\ \rm cm^{-3}$), which significantly improves the nebular lines fit quality. The C abundance is not known due to the lack of rest-frame UV coverage of the C III]$\lambda1908$ line by JWST. The He mass fraction is estimated by \citet{Curti2026} to be $Y=0.253\pm 0.008$, also lower than those of RXCJ2248-ID and Sunburst Arc. It is found by \citet{Curti2026} that the WR spectra bumps can be best reproduced by a stellar population at $t_{\rm age}\sim 5-7.9$~Myr with $Z_\star=0.2\ \rm Z_\odot$ and $\rm [\alpha/Fe]=0.4$, matching the observed nebular $\rm [O/H]\sim -0.22$ and $\alpha$ enhancement $\rm [O/Fe]=0.38\pm 0.22$. Therefore, we take the stellar metallicity as $\rm [Fe/H]=[O/H]-[O/Fe]=-0.598_{-0.266}^{+0.246}$, combining the uncertainties in $\rm [O/H]$ and $\rm [O/Fe]$ by geometric mean. As a secondary reference, \citet{Moreschini2026} inferred a similar stellar metallicity $\log Z_\star=-2.7$ and a younger age $t_{\rm age}=2.5$~Myr. 
\end{itemize}
Among these systems, RXCJ2248-ID and Sunburst Arc belong to the highly NOE (HNOE) regime with $\log\rm(N/O)>-0.6$, where the high N/O ratios have only been measured from high-density-sensitive UV (N III], N IV], and O III]) lines, unlike the moderately NOE ($-1.1<\log\rm(N/O)<-0.6$) systems observed mostly by optical (O II] and N II]) lines \citep[][]{Cameron2026,Rusakov2026}. 
We are particularly interested in RXCJ2248-ID, whose chemical pattern is very similar to that measured using UV diagnostics from the stacked JWST/NIRSpec high-resolution ($R\sim 1000$) GAll spectrum of $z\sim 4.5-10.1$ star-forming galaxies in lensing-cluster fields: $12+\rm \log(O/H)=7.66_{-0.04}^{+0.06}$, $\rm [N/O]=0.66_{-0.24}^{+0.24}$, and $\rm [C/O]=-0.47_{-0.06}^{+0.07}$ \citep[see their table 5]{Umeda2026}. The stacked spectrum also shows a high He~II]$\lambda4686$/H$\beta$ ratio of $\sim 0.03$ that could be attributed to hard ionising radiation from WR stars. It is also found by \citet[][]{Rusakov2026} that the stacked spectrum of young ($t_{\rm age}<30$~Myr) and high-N/O ($\log(\rm N/O)>-0.96$) NOEGs shows a broadened He~II line with $\rm FWHM\sim300-600\ \rm km\ s^{-1}$ (i.e., the blue WR bump) similar to that of RXCJ2248-ID. 
Therefore, if our stellar population synthesis and chemical enrichment models can explain RXCJ2248-ID, they should be generally applicable to (the high-density, high-ionisation regions traced by UV lines in) high-$z$ star-forming galaxies. 




\section{Results}\label{sec:res}

\begin{figure*}
    \centering
    \includegraphics[width=1\textwidth]{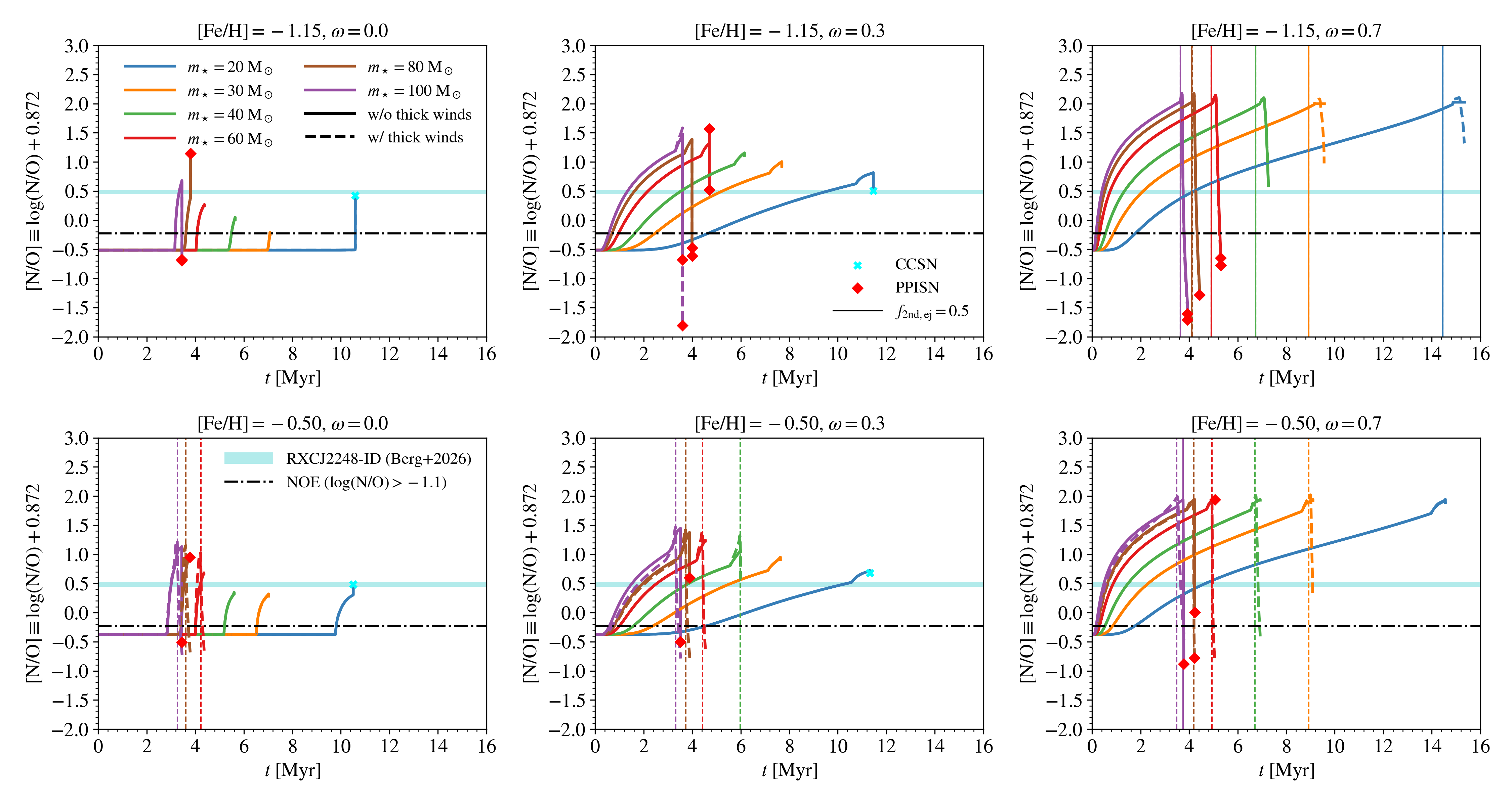}
    \vspace{-19pt}
    \caption{Nitrogen-to-oxygen abundance ratio (with respect to solar) of the cumulative mass loss from individual stars of initial masses $m_\star=20$, 30, 40, 60, 80, 100~$\rm M_\odot$ (curves from right to left in different colors), metallicities $\rm [Fe/H]=-1.15$ (top row) and -0.5 (bottom row) and spins 
    $\omega=0$ (left column), 0.3 (middle column), and 0.7 (right column). 
    The thick solid and dashed curves show the models without and with optically-thick winds. The corresponding thin vertical lines label the timing when primary CNO production starts to win over secondary production (i.e., $f_{\rm 2nd,ej}<0.5$). There are only thin solid vertical lines in the top-right panel because the lines of models with and without optically-thick winds coincide with each other. The CCSN and PPISN events are labelled by the crosses and diamonds. The horizontal dashed-dotted line shows the [N/O] threshold for the N-enhanced (NOE) phase defined by $\rm\log(N/O)>-1.1$. The horizontal shaded band shows the abundance ratio of RXCJ2248-ID \citep{Berg2026}. }
    \label{fig:no_ind}
\end{figure*}

\begin{figure*}
    \centering
    \includegraphics[width=1\textwidth]{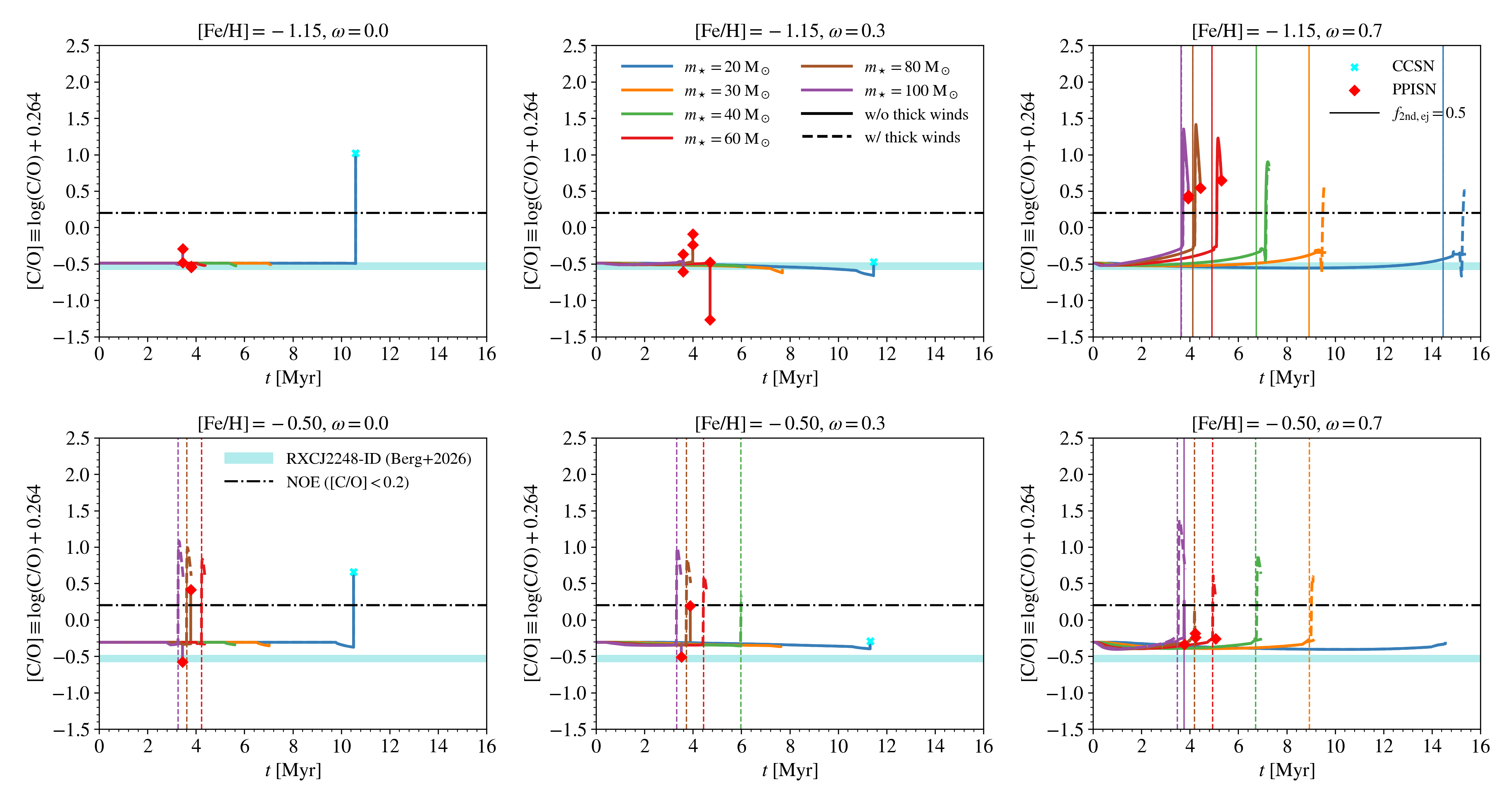}
    \vspace{-19pt}
    \caption{Same as Fig.~\ref{fig:no_ind} but for the carbon-to-oxygen abundance ratio. The horizontal dashed-dotted line shows the maximum [C/O] seen in N/O-enhanced (NOE) galaxies.}
    \label{fig:co_ind}
\end{figure*}

\begin{figure*}
    \centering
    \includegraphics[width=1\textwidth]{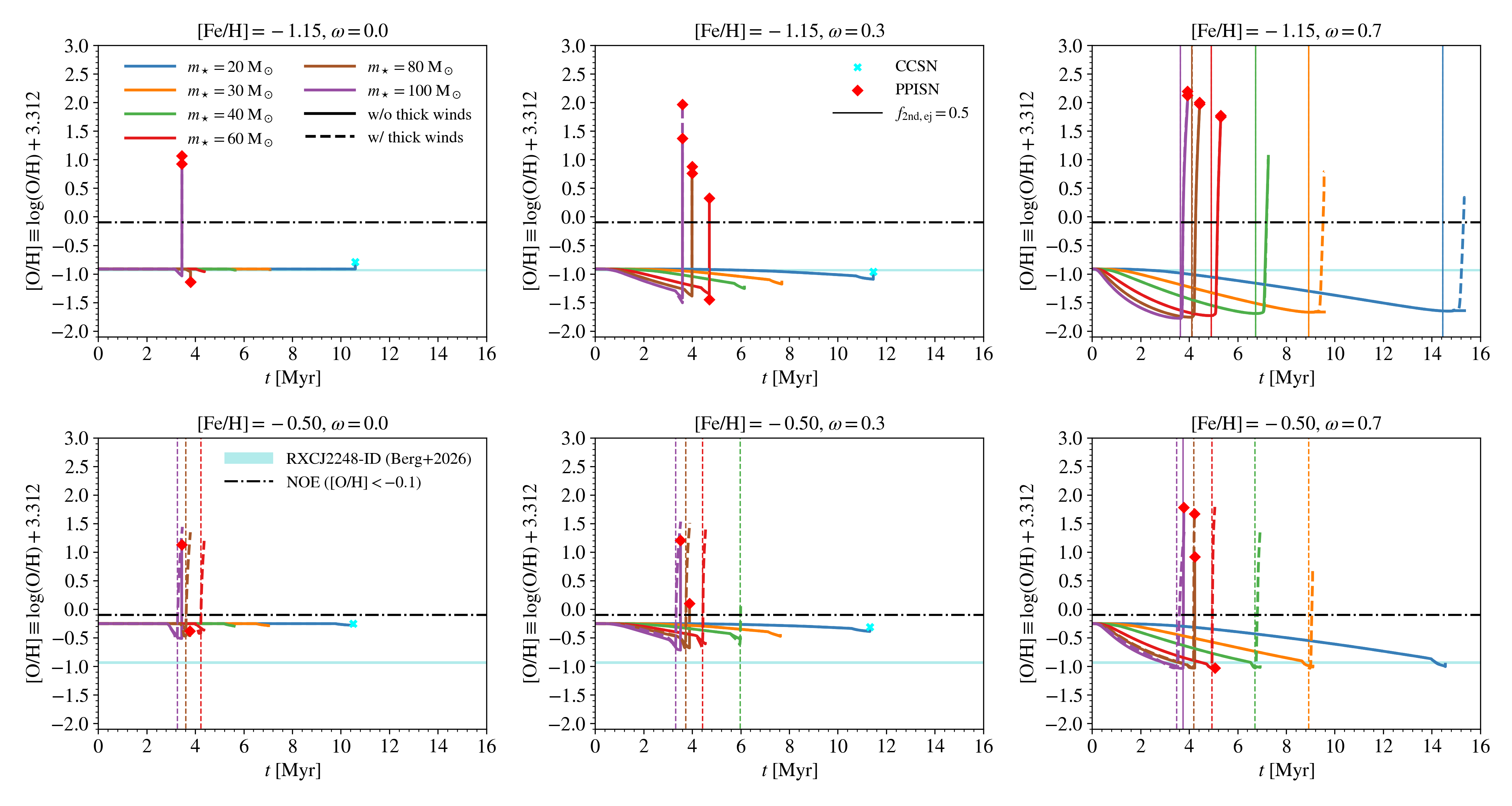}
    \vspace{-19pt}
    \caption{Same as Fig.~\ref{fig:no_ind} but for O abundance. The horizontal dashed-dotted line shows the maximum [O/H] seen in N/O-enhanced (NOE) galaxies.}
    \label{fig:oh_ind}
\end{figure*}

\begin{figure*}
    \centering
    \includegraphics[width=1\textwidth]{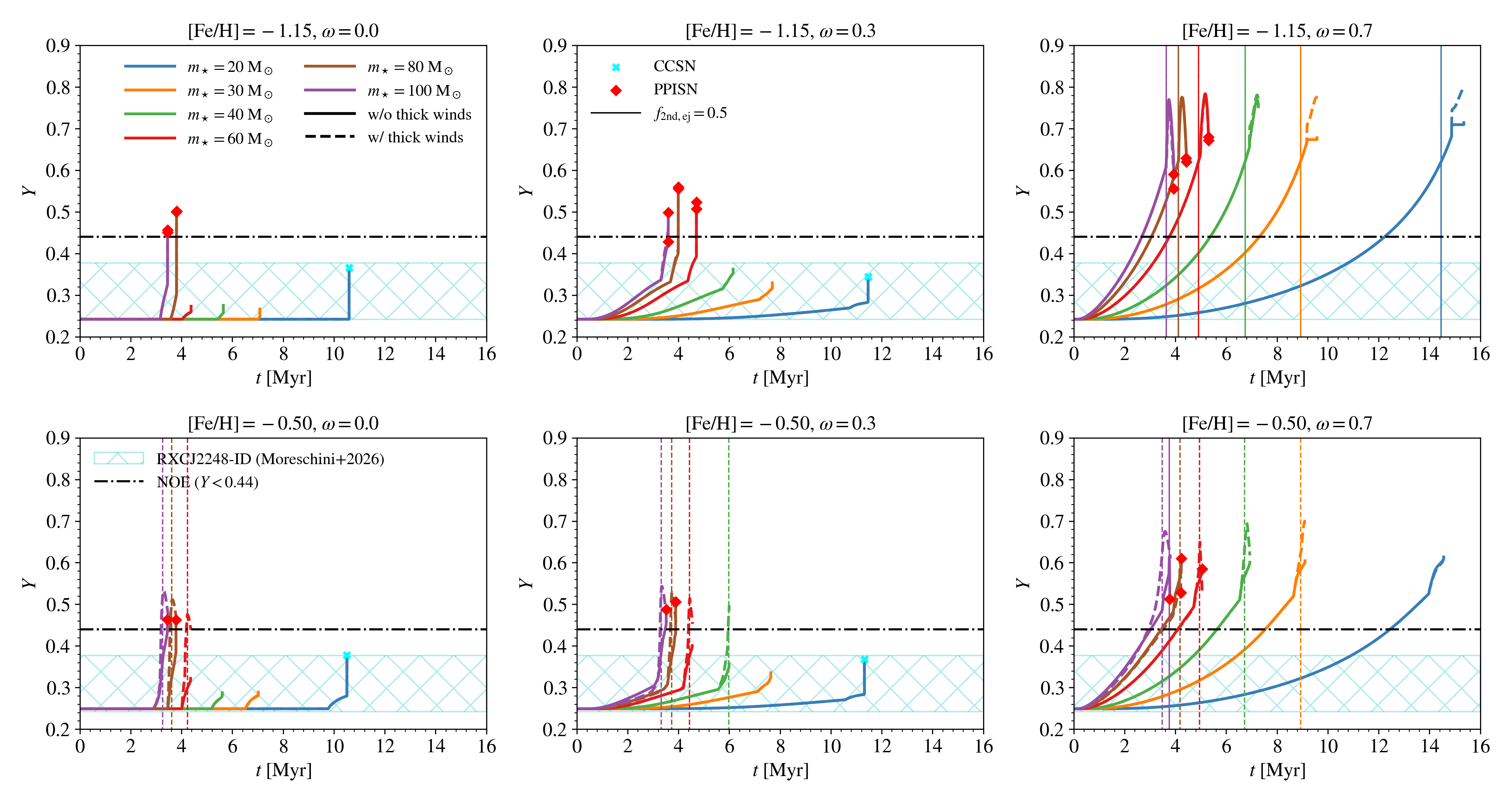}
    \vspace{-19pt}
    \caption{Same as Fig.~\ref{fig:no_ind} but for the He mass fraction. The hatched region shows the range of $Y$ measured by \citet{Moreschini2026} for RXCJ2248-ID with high electron densities $n_{\rm e}\sim 10^3-10^7\ \rm cm^{-3}$ appropriate for regions emitting strong N lines. The horizontal dashed-dotted line shows the maximum $Y$ seen in NOEGs. }
    \label{fig:yhe_ind}
\end{figure*}

Here, we first present the metal yields from individual stars and discuss their dependence on stellar mass, metallicity, spin, and wind prescription (Sec.~\ref{sec:ind}). Then, we show the evolution tracks of pure stellar ejecta from stellar populations of different metallicities, spin distributions and wind models in the chemical abundance space (Sec.~\ref{sec:pop}). Finally, to provide an overview, we evaluate the strength and duration of the (H)NOE phase for each stellar population with varying degree of mixing with the ISM (Sec.~\ref{sec:ne}).

\subsection{Metal yields from individual stars}\label{sec:ind}

Figs.~\ref{fig:no_ind}--\ref{fig:yhe_ind} show the [N/O], [C/O], [O/H], and $Y$ of the cumulative ejecta of individual stars as functions of time, comparing the models with  and without  optically-thick winds for different metallicities and rotations. 
The evolution is insensitive to the wind model for the majority of the star's lifetime, as optically-thick winds can only occur in late stages when the star becomes luminous and hot enough.

\subsubsection{Early evolution}
The early evolution, including most of MS, follows two pathways depending on the initial spin: 
\begin{enumerate}
    \item  For $\omega\lesssim 0.3$, rotational mixing is inefficient, and the surface/wind composition remains as the initial condition during early evolution. 
    \item When rotational mixing is efficient given $\omega\gtrsim 0.3$, [N/O] keeps increasing by up to $\sim 2.5$~dex, reaching $\rm [N/O]\sim 2$, accompanied by the decrease of [O/H] by up to $\sim 0.8$~dex and the rise of He abundance up to $Y\sim 0.6$, while [C/O] shows much weaker variation ($\lesssim0.2$~dex). This strong surface N enrichment is caused by rotational mixing that brings up materials from hot H-burning regions approaching the CNO equilibrium. For a given stellar mass, $\rm [N/O]$ is generally higher as $\omega$ increases and $[\rm Fe/H]$ decreases, since rotational mixing is more efficient in faster-rotating and metal-poorer (thus more compact) stars. Most ($\gtrsim 80\%$) of the N yield is newly synthesized (primary N production), rather than converted from pre-existing C and O (secondary N production), while the overall fraction of primary production of CNO nuclei remains below 50\%, except for the most metal-poor ($\rm [Fe/H]\lesssim-0.75$), fastest-rotating ($\omega\gtrsim 0.6$) models with the strongest rotational mixing and the smallest initial CNO abundances. 
\end{enumerate}

\subsubsection{Late evolution}
In later stages (last $\sim10$\% of the star's lifetime), the initial mass, rotation, metallicity, and wind prescription show a more complex interplay. Below, we identify the key trends in two regimes defined by metallicity (corresponding to the top and bottom rows in Figs.~\ref{fig:no_ind}--\ref{fig:yhe_ind}):

\begin{enumerate}

\item For relatively metal-poor ($\rm [Fe/H]\lesssim -0.75$) stars in our grid, self-stripping down into the CO core can be achieved just by optically-thin winds and/or PPISNe given fast rotation ($\omega\gtrsim 0.45$). This leads to a strong late enhancement of [O/H] up to $\rm [O/H]\sim 3$, often accompanied by the rise of [C/O] up to $\rm [C/O]\sim 1.4$, and a sharp (moderate) drop of [N/O] ($Y$) down to $\rm [N/O]\sim -2$ ($Y\sim 0.8$), as illustrated in the top-right panels of Figs.~\ref{fig:no_ind}--\ref{fig:yhe_ind}. Once the stripping of the CO core happens, primary production dominates the overall CNO yield ($f_{\rm 2nd,ej}\lesssim0.5$). 

Here, the most massive stars with $m_\star\sim 70-100\ \rm M_\odot$ undergo PPISNe, which drive late C/O enrichment regardless of the pre-SN wind mass loss history. In this massive regime, when winds are not strong enough to remove C-rich layers before SNe (see the left/middle-top panels of Figs.~\ref{fig:co_ind} and \ref{fig:oh_ind}), the stars produce large C and O yields via strong pair-instability pulsations, which significantly enhance $\rm [O/H]$ but can only increase $\rm [C/O]$ moderately (up to $\rm [C/O]\sim 0.3$). This PPISN-dominated scenario is more likely to occur with smaller $\omega$ and $\rm [Fe/H]$. It is prevalent when optically-thick winds are not considered. When pre-SN winds can already remove C-rich layers, the resulting CO cores are less massive and undergo weaker pulsations that release less oxygen, such that $\rm [C/O]$ is also significantly enhanced. This wind-dominated scenario applies to most massive stars when optically-thick winds are included, except for a few special cases in which wind-driven stripping is inefficient due to initially slow rotation ($\omega\lesssim 0.45$) or strong spin-down by optically-thin winds. 

Less massive ($m_\star\lesssim 70\ \rm M_\odot$) stars can produce large C/O yields via optically-thin winds only in a mass window, which is $\sim 35-70\rm \ M_\odot$ at $\rm [Fe/H]=-1.15$, becomes narrower with larger $\rm[Fe/H]$, and eventually vanishes at $\rm [Fe/H]>-0.75$. This can be explained by the increase of wind mass loss rate with stellar metallicity, which can spin down massive stars early-on, thus reducing their late-stage wind mass loss. 

\item  For metal-richer ($\rm [Fe/H]\gtrsim -0.75$) stars, optically-thick winds play more important roles, especially for low/moderate initial rotation ($\omega\lesssim 0.45$) that is expected to be more common (corresponding to $\sim75\%$ of O stars observed in the SMC). Once included, they are generally active above a critical initial spin 
that generally decreases with $m_\star$ and [Fe/H], ranging from $\omega_{\rm crit}=0$ for $m_\star\gtrsim 50-70\ \rm M_\odot$ and $\omega_{\rm crit}\sim 0.3-0.7$ for $m_\star\sim 20-40\ \rm M_\odot$. 

For $\omega\gtrsim 0.3$, the enrichment of C and O is mainly driven by optically-thick winds once they are included, which can enhance the O abundance up to $\rm [O/H]\sim 2$ and the C/O ratio up to $\rm [C/O]\sim 1$. The enhanced O yield reduces $\rm [N/O]$ down to $[\rm N/O]=-1.5$. These effects are stronger for more massive stars with lower metallicities. Without optically-thick winds, the stars with $m_\star\sim 20-70\ \rm M_\odot$ that do not undergo powerful PPISNe only produce small C/O yields, so that the N/O enhancement is maintained till the end with $\rm [O/H]\lesssim -0.2$, $\rm [C/O]\lesssim -0.3$, and $\rm [N/O]\sim 0.5-1.9$, set by the CNO cycle and rotational mixing. More massive ($\gtrsim 70\ \rm M_\odot$) stars produce considerable C/O yields via PPISNe, but still smaller than those from optically-thick winds.

\end{enumerate}

In both metallicity regimes, non-rotating stars only produce N/O enhancements in the late stage, reaching up to $[\rm N/O]\sim 1.2$. Optically-thick winds can strip the CO cores of massive stars with $m_\star\sim 50-100\ \rm M_\odot$, reducing the N/O ratio back to $\rm [N/O]\lesssim -0.5$, while less massive stars keep a moderate N/O enhancement ($-0.2\lesssim [\rm N/O]\lesssim 0.6$) till the end. Besides, only in this case ($\omega=0$), CCSNe of $m_\star\sim 20-25\ \rm M_\odot$ stars contribute to the N and C yields, resulting in $\rm [N/O]\sim 0.5$ and $\rm [C/O]\sim 0.7-1$. In contrast, for rotating stars with $\omega\gtrsim0.3$, CCSNe only contribute to He enrichment (increasing $Y$ by up to $\sim 0.1$) due to large fallback fractions of He cores enlarged by rotational-mixing.

\subsection{Metal yields from stellar populations}\label{sec:pop}

\begin{figure*}
    \centering
    \includegraphics[width=0.95\textwidth]{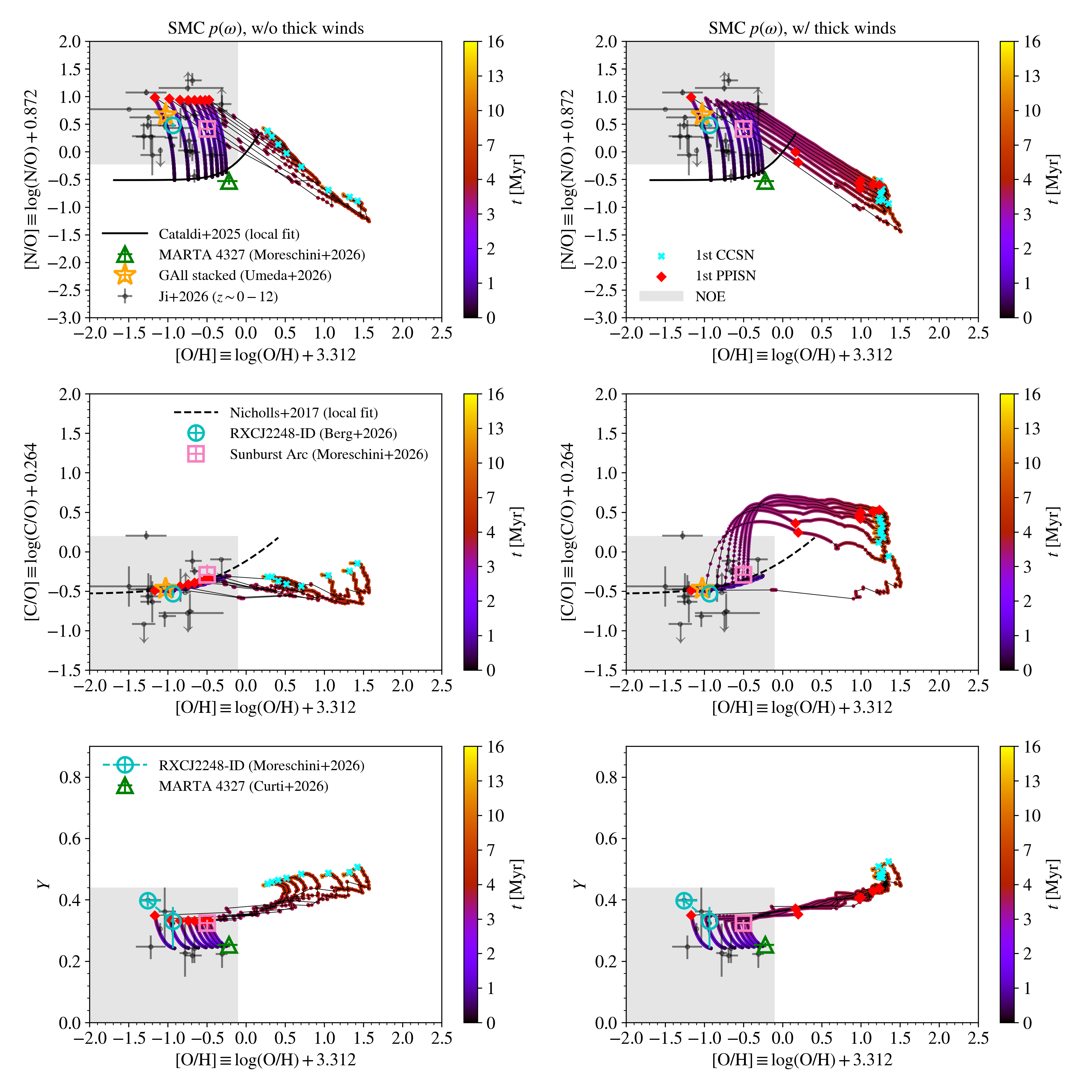}
    \vspace{-10pt}
    \caption{Abundance patterns of population-averaged cumulative stellar ejecta that include the contributions of stars with different initial spins according to the observed spin distribution of O stars in the SMC \citep{Boco2025} for the models without (left) and with (right) optically-thick winds. Different curves show the stellar populations of different metallicities, as reflected by their starting points ($t=0$). The top, middle, and bottom rows show the evolution of [N/O], [C/O], and $Y$ with [O/H], color coded by time after an instantaneous starburst. The first CCSN and PPISN events in a stellar population are labeled by the crosses and diamonds, respectively. The shaded regions denote the abundance-ratio space generally occupied by observed NOE galaxies, for which we show the sample compiled by \citet{Ji2026} with the small squares. The big circle, square, hexagon, and star show the abundance patterns of three representative high-$z$ galaxies with WR signatures, RXCJ2248-ID, Sunburst Arc, and MARTA 4327 \citep{Berg2026,Curti2026,Moreschini2026}, and the GAll stacked spectrum of $z\sim 4.5-10.1$ star-forming galaxies in lensing-cluster fields \citep{Umeda2026}. Here, the O and He abundance measurements are highly sensitive to the assumed electron density. Therefore, we show two measurements for RXCJ2248-ID in the $Y$-[O/H] space (bottom row), with low densities $n_{\rm e}\sim 10^2\ \rm cm^{-3}$ \citep{Yanagisawa2024}, as adopted by \citet{Ji2026}, and high densities $n_{\rm e}\sim 10^3-10^7\ \rm cm^{-3}$ \citep{Moreschini2026}. The latter is preferred because high-ionisation, high-density regions are expected to closely trace the metal yields of young, massive stars. For similar reasons, we only show the [N/O], [C/O], and [O/H] measurements by \citet{Berg2026} using UV-derived high densities $n_{\rm e}\sim 1-3\times 10^5\ \rm cm^{-3}$. For Sunburst Arc, we take the CNO and He abundance measurements from \citet{Moreschini2026}, focusing again on the high-density NOE region. For and MARTA 4327, we use the N and O abundances from \citet{Moreschini2026} and the (stellar) He abundance from \citet{Curti2026}. The solid and dashed curves show the [N/O]-[O/H] and [C/O]-[O/H] scaling relations from local observations \citep{Nicholls2017,Cataldi2026}.}
    \label{fig:abundance_avg}
\end{figure*}

\begin{figure*}
    \centering
    \includegraphics[width=0.95\textwidth]{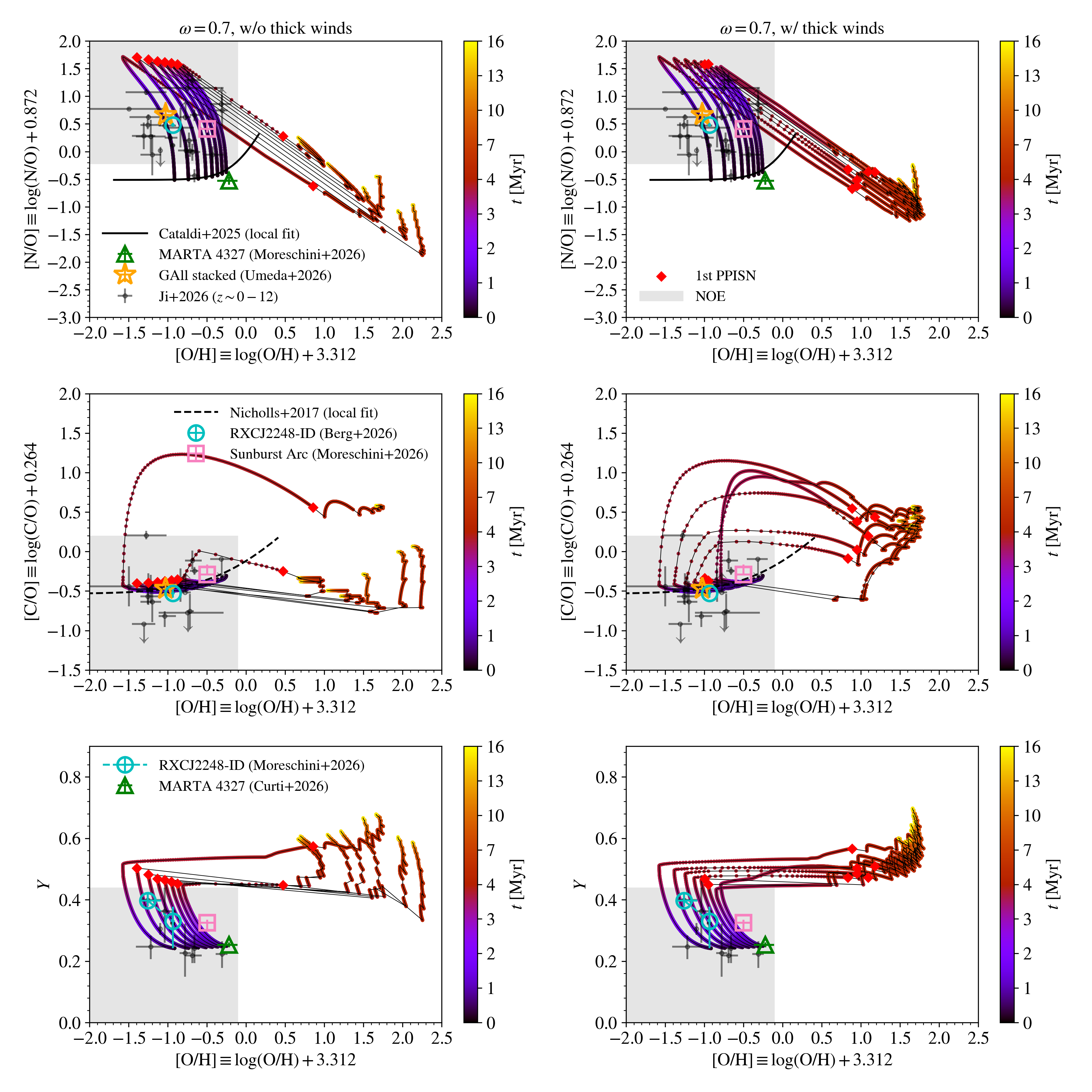}
    \vspace{-10pt}
    \caption{Same as Fig.~\ref{fig:abundance_avg} but assuming all stars are initially fast rotating with $\omega=0.7$.}
    \label{fig:abundance_opt}
\end{figure*}

We compute the metal yields from single-age stellar populations of different metallicties by integrating the yields of individual stars over the IMF and initial spin distribution with Eq.~\ref{eq:mx}. Fig.~\ref{fig:abundance_avg} shows the evolution of [N/O] (top), [C/O] (middle), and $Y$ (bottom) with $\rm [O/H]$ ($x$ axis) and time (color-coding) in the integrated cumulative yields for the SMC spin distribution, comparing the models without (left) and with (right) optically-thick winds. For comparison with observations, we plot the region occupied by most observed NOEGs, together with the sample of NOEGs complied by \citet[small dots]{Ji2026}, the three representative galaxies with WR signatures (big open triangles, circles, and squares), and the results of the stacked spectrum of $z\sim 4.5-10.1$ star-forming galaxies in lensing-cluster fields (big open stars), as detailed in Sec.~\ref{sec:comp_obs}. The evolution generally goes through three stages:
\begin{enumerate}

    \item The early evolution within $t\lesssim 3-4\ \rm Myr$ after the starburst is dominated by the N-rich winds of stars with efficient rotational mixing ($\omega\ge 0.3$), reaching a peak $\rm [N/O]\sim 1$ as $\rm [O/H]$ decreases slightly by $\sim 0.2$~dex. Meanwhile, $\rm [C/O]$ remains almost unchanged, while the He mass fraction increase to $Y\sim 0.36$. The models with and without optically-thick winds show minor differences. 

    \item Following the peak of $[\rm N/O]$, a sharp transition happens as the most massive ($m_\star\sim80-100\ \rm M_\odot$) stars approach their death, enhancing $\rm [O/H]$ while bringing down $\rm [N/O]$ (by up to $\sim 2.5$~dex). Meanwhile, $Y$ does not change much during the transition, as the He-rich layers of massive stars have already been lost. 
    
    In the models without optically-thick winds, the transition is mainly driven by O enrichment from PPISNe. The increase (decrease) of [O/H] ([N/O]) is stronger for metal-poorer stars, which produce larger O yields from more powerful PPISNe with larger He cores due to more efficient rotational mixing. 
    The change of $\rm [C/O]$ is small during the transition and it remains below $-0.3$, as expected for PPISN-dominated enrichment. 
    
    In contrast, once optically-thick winds are included, the transition typically happens before the first PPISN event at a slightly smaller $\rm [N/O]$, and is driven by the C/O-rich winds of $m_\star\sim 50-100\ \rm M_\odot$ stars. In this case, the dependence of $\rm [O/H]$ and [N/O] on stellar metallicity is much weaker, with all models reaching $\rm [O/H]\sim 1-1.5$ and $\rm [N/O]\lesssim-0.5$ by $t=4$~Myr. Besides, $\rm [C/O]$ rises rapidly after $t\sim 3$~Myr, peaks at $\rm [C/O]\sim 0.4-0.7$ with $-0.5\lesssim\rm [O/H]\lesssim 0$, and slightly falls down by $\sim 0.2$~dex as $\rm [O/H]$ keeps increasing towards its maximum around $t\sim 4$~Myr. This reflects the wind-driven stripping process of the CO core. One exception is the most metal-poor ($\rm [Fe/H]=-1.15$) case, where winds cannot strip the CO cores of most massive stars with $\omega\le 0.45$ before PPISNe, which then produce large O yields that ``dilute'' C. 

    \item After the transition, $\rm [N/O]$ and $Y$ increase as $\rm [O/H]$ decreases. This is driven by the N/He-rich, C/O-poor winds of less massive ($m_\star\sim 20-80\ \rm M_\odot$) stars that are still in their early evolution stages. Without optically-thick winds, these stars typically do not produce large C/O yields, such that $\rm [C/O]$ remains small except for the models with $\rm [Fe/H]\lesssim -0.85$. In these cases, initially fastest-rotating ($\omega\gtrsim 0.6$) stars (with $m_\star\gtrsim 35\ \rm M_\odot$) can remove C-rich layers by optically-thin winds, thus boosting $\rm [C/O]$ up to $\sim 0$ by $t\sim 10$~Myr. The post-transition evolution of CNO abundance patterns is weaker when optically-thick winds are included, because they enable less massive stars to produce C/O-rich winds. The CCSN yields of $m_\star\sim 20-25\ \rm M_\odot$ stars cause minor evolution at $t\sim 8-10$~Myr, although there are $\sim 20$ times more such stars than $m_\star\sim 80-100\ \rm M_\odot$ stars according to the \citet{Kroupa2001} IMF. 
    
\end{enumerate}

Fig.~\ref{fig:abundance_opt} shows the chemical evolution of the yields from purely fast-rotating ($\omega=0.7$) stars. In this extreme case, the models with and without optically-thick winds show generally smaller differences and stronger early-stage N/O enhancements up to $\rm [N/O]\sim 1.7$ 
and late-stage C/O enhancements up to $[\rm C/O]\sim 1.2$. Here, even without optically-thick winds, massive ($m_\star\sim 70-100\ \rm M_\odot$) stars produce comparable/larger C/O yields from powerful PPISNe when aided by fast rotation ($\omega\gtrsim 0.6$) at $\rm [Fe/H]\lesssim -0.7$. 
Metal-richer stars still produce less O without optically-thick winds due to their weaker/rarer PPSINe. Another noticeable difference is that the late-stage enhancement of [C/O] does not occur at $-0.7\lesssim\rm [Fe/H]\lesssim-0.55$ in the optically-thin-wind-only models, which is a feature of PPISN-dominated enrichment. This is caused by wind-driven spin-down of massive stars that suppresses their late-stage wind mass loss, which can be revived if switching to optically-thick winds is allowed. 

In general, our model evolution tracks cover a substantial part of the NOE region in the [N/O]--[O/H] and $Y$--[O/H] spaces occupied by NOEGs in observations, reaching peak N/O ratios of $\rm [N/O]\sim 1-1.7$, i.e., $\log(\rm N/O)\sim 0.1-0.8$, comparable to the highest values observed \citep{Ji2026}. However, the metal-poorest NOEGs with $\rm [O/H]\lesssim -1$ need more metal-poor stellar populations than considered here or strong dilution of O yields to explain \citep[e.g.,][]{Limongi2018,Roberti2024, Costa2025,Nandal2026}. 
The model tracks typically go out of the NOE region at $t\sim 3$~Myr after the starburst due to C enrichment by winds in the models with optically-thick winds or in the extreme case of purely fast-rotating ($\omega=0.7$) stars. The transition is instead driven by O enrichment from PPISNe in the optically-thin-wind-only models with the more realistic SMC spin distribution. Mixing with the ISM can further move the tracks around to cover more cases or reproduce a certain system, as discussed below.


\subsection{The N/O-enhanced phase: strength and duration}\label{sec:ne}

\begin{figure*}
    \centering
    \includegraphics[width=1\textwidth]{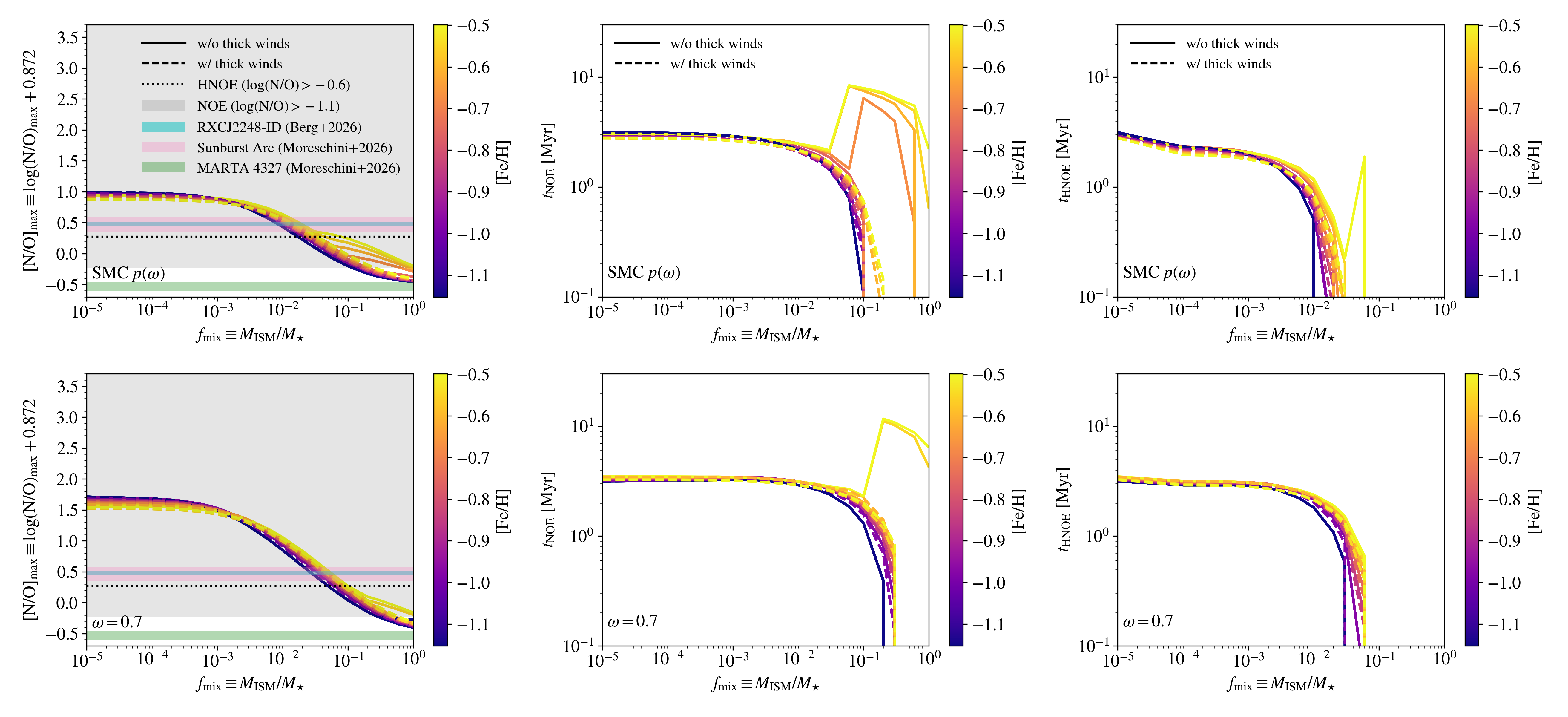}
    \vspace{-20pt}
    \caption{Maximum [N/O] (left) and durations of the NOE (middle) and HNOE (right) phases as functions of the ISM mixing efficiency $f_{\rm mix}$ for single-age stellar populations (colour-coded by stellar metallicity [Fe/H]) without (solid) and with (dashed) optically-thick winds. The top row shows the results assuming the SMC spin distribution \citep{Boco2025}, while the bottom row assumes that all stars are initially fast-rotating with $\omega=0.7$. On the left panels, we plot the observed abundance ratios of RXCJ2248-ID \citep{Berg2026}, Sunburst Arc, and MARTA 4327 \citep{Moreschini2026} with the cyan, pink, and green horizontal bands (reflecting $1\sigma$ uncertainties), respectively. The NOE phase with $\rm\log(N/O)>-1.1$ and HNOE phase with $\rm\log(N/O)>-0.6$ are marked by the shaded region and the horizontal dotted line. }
    \label{fig:NOE_avg}
\end{figure*}

The results shown above describe the stellar yields, without accounting for the role of the ISM. 
In reality, the stellar ejecta will be mixed with the surrounding ISM, and the observed metal lines trace this enriched medium. To better evaluate the ability of our models in reproducing observed NOEGs, we model the effect of mixing with a phenomenological parameter, $f_{\rm mix}\equiv M_{\rm ISM}/M_\star$, the ratio of the ISM mass mixed with the stellar ejecta and the total stellar mass formed (Sec.~\ref{sec:pop}). In this subsection, we explore how $f_{\rm mix}$ affects the general properties of the enriched medium, focusing on the maximum N/O ratio achieved $\rm [N/O]_{\max}$ and the durations spent in the NOE ($t_{\rm NOE}$) and HNOE ($t_{\rm HNOE}$) phases with $\log\rm(N/O)>-1.1$ and $\log\rm(N/O)>-0.6$ (see Sec.~\ref{sec:comp_obs} for full definitions). 

Fig.~\ref{fig:NOE_avg} compares the evolution of $\rm [N/O]_{\max}$ (left), $t_{\rm NOE}$ (middle), and $t_{\rm HNOE}$ (right) with $f_{\rm mix}$ in the models without (solid) and with (dashed) optically-thick winds of different stellar metallicities ($\rm [Fe/H]$, color-coding). 
The top row considers the ``realistic'' SMC spin distribution. The effects of mixing are minor at $f_{\rm mix}\lesssim 10^{-3}$ with $\rm [N/O]_{\max}\sim 1$, $t_{\rm NOE}\sim 3$~Myr, and $t_{\rm HNOE}\sim 2-3$~Myr reflecting the yields themselves for both wind prescriptions. 
As $f_{\rm mix}$ further increases, $\rm [N/O]_{\max}$ decreases and drops out of the HNOE regime at $f_{\rm mix}\sim 0.02$. In this process, $t_{\rm NOE}$ gradually decreases to $\sim 2$~Myr, while $t_{\rm HNOE}$ drops rapidly below $\sim1\ (0.1)$~Myr at $f_{\rm mix}\gtrsim 0.01$~(0.02). 
Thereafter, $t_{\rm NOE}$ goes down to $\lesssim 0.1$~Myr and $\rm [N/O]_{\max}$ drops below the NOE threshold at $f_{\rm mix}\sim 0.1$ in the models with optically-thick winds, and the evolution shows weak dependence on $\rm [Fe/H]$. However, without optically-thick winds, the decrease of $\rm [N/O]_{\max}$ with $f_{\rm mix}$ is slower at higher $\rm [Fe/H]$ in the metal-rich models with $\rm[Fe/H]\gtrsim-0.7$, some of which even show an increase of $t_{\rm NOE}$ ($t_{\rm HNOE}$) up to $\sim 8$~(2)~Myr before dropping down rapidly again at $f_{\rm mix}\gtrsim 0.6$~(0.06). 
This is caused by the small O yields in the lack of powerful PPISNe and strong CO-core-stripping winds. Indeed, in these cases, the peak $\rm [N/O]$ is slightly larger by $\sim 0.1$~dex without optically-thick winds (Fig.~\ref{fig:abundance_avg}), and requires more O from the ISM to suppress. Besides, when the O yield is small, moderate mixing with the ISM can reduce $\rm [O/H]$ back to the NOE regime of $\rm [O/H]<-0.1$ in the post-PPISN era while keeping $\log(\rm N/O)>-1.1$. 

The bottom row of Fig.~\ref{fig:NOE_avg} shows the results of the extreme fast-rotating ($\omega=0.7$) scenario. The difference between the models with and without optically-thick winds is much smaller in the evolution of $\rm [N/O]_{\max}$ with $f_{\rm mix}$, because the massive stars in both cases have strong enough winds to expose the CO core, which produce similar CNO yields around the $\rm [N/O]$ peaks (Fig.~\ref{fig:abundance_opt}). Compared with the case of the SMC spin distribution, the $t_{\rm NOE}$ and $t_{\rm HNOE}$-$f_{\rm mix}$ relations show similar tends, but generally shifted to larger $f_{\rm mix}$ by a factor of $\sim 2$. This simply reflects the larger N yields from the winds of highly fast-rotating ($\omega=0.7$) stars than those of more common, slower-rotating ($\omega\lesssim 0.45$) stars. The increase of $t_{\rm NOE}$ ($t_{\rm HNOE}$) with $f_{\rm mix}$ at $f_{\rm mix}\gtrsim 0.02$ is rarer (absent) because PPISNe provide enough O yields even without optically-thick winds.

In conclusion, we find that the (H)NOE signatures of massive ($m_\star\gtrsim 20\ \rm M_\odot$) stellar winds can be preserved for a considerable period $\sim 1-3$~Myr after an instantaneous starburst if the mixing of stellar ejecta with the ISM is generally inefficient, i.e., when the mixing efficiency $f_{\rm mix}$ remains below a critical value $f_{\rm mix,crit}\sim 0.1$~(0.01). The inefficiency of mixing, at least for the high-density, high-pressure NOE regions probed by UV lines, is consistent with the N budget inferred from observations and predicted by hydrodynamic simulations (see Sec.~\ref{sec:MN}). Excluding optically-thick winds can extend the NOE phase up to $\sim 8$~Myr and enhance its survivability against mixing (with $f_{\rm mix,crit}$ increased by up to a factor of 10) for relatively metal-rich ($\rm [Fe/H]\gtrsim -0.7$) stellar populations. However, such extended NOE phases may not be favourable, considering the existence of young ($t_{\rm age}\lesssim 10$~Myr) stellar populations with WR signatures but no N/O enhancement, as discussed below (Sec.~\ref{sec:obs}). Our conclusion is insensitive to the exact form of star formation history. We see very similar trends for continuous star formation at a constant rate, albeit with smaller $f_{\rm mix,crit}$ by a factor of $\sim 3$, as shown in Appendix~\ref{apdx:csfr}. The predicted characteristic timescale $t_{\rm (H)NOE}\sim 1-3$~Myr implies that a few percent of high-$z$ galaxies will be captured in the (H)NOE phase, considering typical star-formation duty cycles of $\sim 100$~Myr. This occurrence rate is consistent with the finding of \citet{Rusakov2026} that $\sim 3-18\%$ galaxies (spectroscopically detectable by JWST NIRSpec) at $z\sim 4-8.5$ are NOEGs, and among them $\sim 30\%$ belong to the HNOE regime with $\log (\rm N/O)>-0.6$.

\section{Discussion}\label{sec:diss}

\subsection{Reproducing high-$z$ galaxies with Wolf-Rayet signatures}\label{sec:obs}

\begin{figure*}
    \centering
    \includegraphics[width=0.78\textwidth]{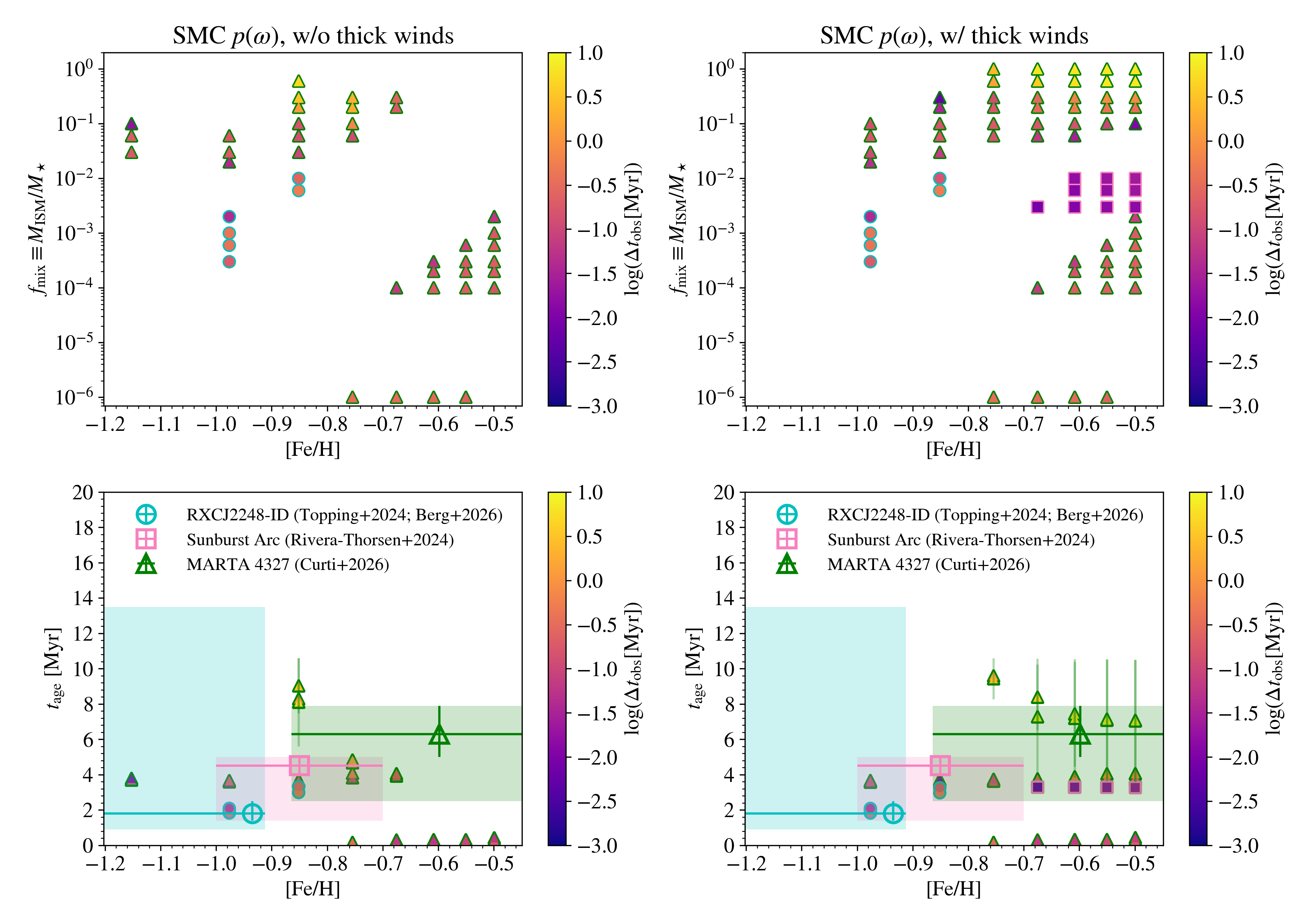}
    \vspace{-10pt}
    \caption{Fitting the abundance patterns of RXCJ2248-ID, Sunburst Arc, and MARTA 4327 with chemical enrichment by single-age stellar populations, considering the yields of stars with different initial spins following the observed spin distribution of O stars in the SMC \citep{Boco2025} without (left column) and with (right column) optically-thick winds. The top (bottom) row plots the successful models in the space of $f_{\rm mix}$ ($t_{\rm age}$)-[Fe/H], as small data points colour-coded by $\Delta t_{\rm obs}$, where $t_{\rm age}$ is the middle point of the period $\Delta t_{\rm obs}$ when the predicted abundances are within 2$\sigma$ of the observed values (see Fig.~\ref{fig:abundance_avg}). We plot the results without any mixing ($f_{\rm mix}=0$) at $f_{\rm mix}=10^{-6}$. In the bottom row, we plot the stellar ages and metallicities of RXCJ2248-ID \citep{Topping2024,Berg2026}, Sunburst Arc \citep{Rivera-Thorsen2024}, and MARTA 4327 \citep{Curti2026} inferred from their spectra as big open circles, squares, and triangles, respectively. The associated shaded regions consider a broader range of $t_{\rm age}$ from multiple references than that from the primary reference denoted by the thick errorbars (see Sec.~\ref{sec:obs} for details). 
    The shapes of model data points follow their fit targets. For the models with $\Delta t_{\rm obs}>1\ \rm Myr$, we show the period $t\in [t_{\rm age}-0.5\Delta t_{\rm obs}, t_{\rm age}+0.5\Delta t_{\rm obs}]$ with the thin errorbars.}
    \label{fig:fit_avg}
\end{figure*}

\begin{figure*}
    \centering
    \includegraphics[width=0.78\textwidth]{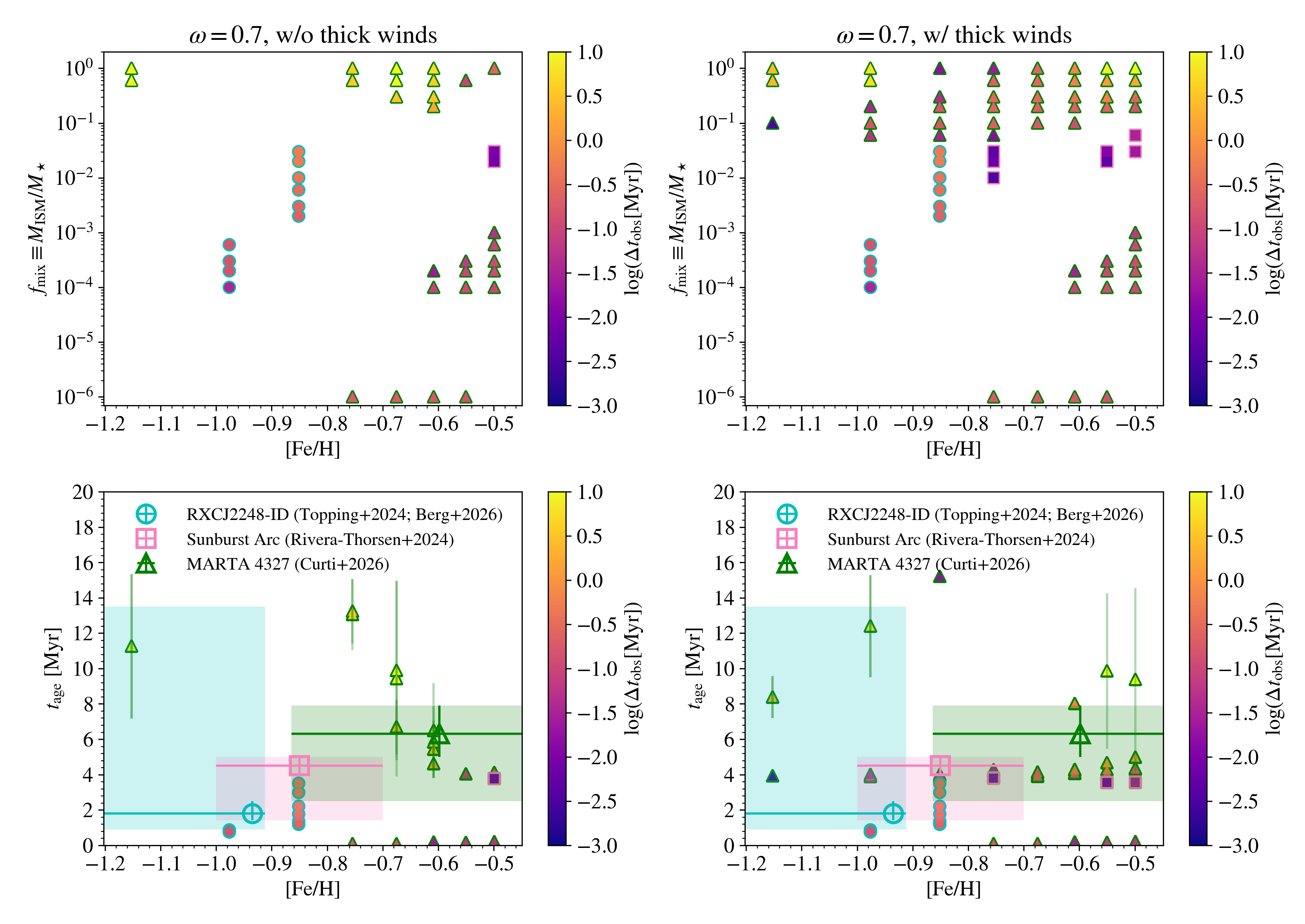}
    \vspace{-10pt}
    \caption{Same as Fig.~\ref{fig:fit_avg} but assuming all stars are initially fast rotating with $\omega=0.7$.}
    \label{fig:fit_opt}
\end{figure*}

To demonstrate the applications of our metal yields, we search for the stellar population and metal mixing parameters (i.e., stellar spin distribution, wind prescription, metallicity $\rm [Fe/H]$, age $t_{\rm age}$, and ISM mixing efficiency $f_{\rm mix}$) required to explain the (nebular) abundance patterns ([N/O], [O/H], [C/O], and $Y$) of three representative high-$z$ galaxies with WR signatures: RXCJ2248-ID, Sunburst Arc, and MARTA 4327, whose properties are summarised in Table~\ref{tab:obs} (see Sec.~\ref{sec:comp_obs} for details). For a given combination of stellar spin distribution and wind prescription, we first identify the combinations of $\rm [Fe/H]$ and $f_{\rm mix}$ that can reproduce the abundance pattern of a target galaxy within $2\sigma$ uncertainties, and note down the stellar age $t_{\rm age}$ and duration $\Delta t_{\rm obs}$ of the period when model predictions agree with observations (within the lifetimes of $m_\star=20\ \rm M_\odot$ stars). Then we compare the $\rm [Fe/H]$ and $t_{\rm age}$ obtained in this process with those independently inferred from SED fitting of stellar spectra features. A model is (more) favourable when consistency is found in $\rm [Fe/H]$ and $t_{\rm age}$ (and $\Delta t_{\rm obs}$ is longer). Note that this consistency check is illustrative rather than conclusive, because the SED fitting results are based on previous stellar evolution models \citep[e.g., BPASS,][]{Stevance2020}, not the ones used in this work to fit the \textit{nebular} abundance patterns. A fair check must perform SED fitting with the synthetic stellar spectra of the same stellar evolution tracks underlying the metal yields. This is a highly non-trivial task that in principle involves entangled modelling of stellar atmospheres and stellar winds \citep[see, e.g.,][]{Toala2024,Lefever2025,Lefever2026,Sander2026,Sander2026wno,Voje2026} beyond the scope of this paper.

Fig.~\ref{fig:fit_avg} shows the results for the SMC spin distribution, comparing the models without (left) and with (right) optically-thick winds. The top panels show the successful models on the grid of $\rm [Fe/H]$ and $f_{\rm mix}$, colour-coded by $\Delta t_{\rm obs}$, while the bottom panels plot the abundance-pattern-based stellar population properties on top of those estimated from SED fitting in the $\rm [Fe/H]$-$t_{\rm age}$ space. The successful models generally occupy larger regions in the $\rm [Fe/H]$-$f_{\rm mix}$ space when optically-thick winds are included. 

For RXCJ2248-ID, both wind prescriptions lead to similar successful models with $f_{\rm mix}\sim 0.001$ and 0.01 at $\rm [\rm Fe/H]=-0.98$ and $-0.85$, respectively. The former are preferred by their better agreement with the SED-fitting estimates of low metallicities $\rm [\rm Fe/H]\lesssim-0.9$ \citep{Topping2024,Moreschini2026}. The observed [N/O], [C/O], [O/H], and $Y$ are simultaneously reproduced within $2\sigma$ for a relatively short period of $\Delta t_{\rm obs}\sim 0.4$~Myr at $t_{\rm age}\sim 2-3$~Myr after the starburst, consistent with the young ages $t_{\rm age}=1.8_{-0.4}^{+0.7}$ from SED-fitting \citep{Topping2024}. The relevant model tracks spend in total $t_{\rm HNOE}\sim 2$~Myr in the HNOE regime ($\log\rm(N/O)>-0.6$) as [N/O] keeps increases over time towards the peak $[\rm N/O]\sim 0.9$ until C/O-rich winds reverse the trend at $t\sim 3$~Myr. Our results support the picture that the chemical enrichment in RXCJ2248-ID is dominated by WN stars with little WC contribution, as evidenced by the non-detection of C IV ]$\lambda\lambda$5803,5814 emission and the relatively narrow/weak He II ]$\lambda$4687 line compared with Sunburst Arc and MARTA 4327 \citep{Berg2026}. The same conclusion has been drawn from the chemical evolution model of \citet{Kobayashi2024} using the yields of \citet{Limongi2018}, which also considers a substantial fraction ($\sim 50\%$) of rotating ($\omega\gtrsim 0.2$) stars and reproduce RXCJ2248-ID at a slightly later stage $t_{\rm age}\sim 4.2$~Myr.

For Sunburst Arc, when optically-thick winds are included, we find successful models with $f_{\rm mix}\sim 0.003-0.01$, $\rm -0.7\lesssim[Fe/H]\lesssim -0.5$, and $t_{\rm age}\sim 3.3$~Myr. These correspond to a rapid ($\Delta t_{\rm obs}\sim 0.01-0.02$~Myr) transition driven by the C-rich optically-thick winds from massive ($m_\star\sim 80-100\ \rm M_\odot$) stars in the WC phase. In this phase, the enrichment of C and O is just enough to reproduce the observed [C/O] and [O/H], yet still unable to remove the N/O enhancement. The stellar age is within the broad range found by SED-fitting, $t_{\rm age}\sim 1.4-5$~Myr \citep{Chisholm2019,Rivera-Thorsen2024}, while the metallicity is marginally consistent with the upper limit from SED-fitting. It will be shown below that better agreement can be achieved when we consider continuous star formation (Appendix~\ref{apdx:csfr}). 
However, there are no successful models in the optically-thin-wind-only case, which cannot produce enough C/O-rich winds during the transition (see Fig.~\ref{fig:abundance_avg}). This alone does not necessarily rule out the optically-thin-wind-only scenario, because the C/O ratio is sensitive to the pre-starburst ISM composition. We have found by numerical experiments that when the pre-starburst ISM is more C/O-rich (e.g., $\rm [C/O]_{\rm pre}=-0.2$) than that adopted by default ($\rm [C/O]_{\rm pre}\sim -0.5$) from the local scaling relation of \citet{Nicholls2017}, both wind prescriptions can provide successful models for Sunburst Arc in longer periods up to $\Delta t_{\rm obs}\sim 1$~Myr. Nevertheless, \citep{Rivera-Thorsen2024} find that a considerable fraction ($11\%$) of WC stars in the WR population is required to explain the C~IV feature in Sunburst Arc's spectrum, which is a natural feature in the picture of optically-thick-wind-driven transition\footnote{The transition phase is rather short ($\sim 0.01-0.02$~Myr) and difficult to observe, as the winds of WC stars tend to cause higher [C/O] and lower [N/O] than observed in Sunburst Arc. For this reason, \citet{Tapia2024} concludes that WR stars should not contribute to the chemical enrichment of Sunburst Arc, seemingly in tension with the detection of WC signatures \citep{Rivera-Thorsen2024}. This tension can be reconciled if the C-rich winds are not retained efficiently in the nebula at $t\lesssim 5$~Myr. In this case, the transition can be prolonged and easier to observe.}. 
Without optically-thick winds, very high initial spins ($\omega\sim 0.7$) are needed to evolve $m_\star\sim 80-100\ \rm M_\odot$ stars into the WC phase at $t\sim 4$~Myr. Such high spins are extremely rare ($\lesssim 3\%$) in observations of O stars in the SMC, needless to say that the PPISNe of slower-rotating (shorter-lived) massive stars can remove the N/O enhancement even before WC stars show up. 


For MARTA 4327, which is not N/O enhanced with $\log(\rm N/O)\sim -1.4$, there are two groups of successful models at $f_{\rm mix}\lesssim 0.002$ and $f_{\rm mix}\gtrsim 0.02$. The former correspond to the very early-phase ($t_{\rm age}\lesssim 0.5$~Myr) when significant N enrichment from winds has not built up, which is inconsistent with the SED-derived ages $t_{\rm age}\sim 2.5-8$~Myr \citep{Curti2026,Moreschini2026}. The latter instead capture the late stage ($t_{\rm age}\sim 4-9$~Myr) when the N/O enhancement has been suppressed by O enrichment, and proper mixing with the ISM leads to the right [O/H]. Here, a few successful models with optically-thick winds have $t_{\rm age}\sim 7$~Myr with $\Delta t_{\rm obs}\sim7$~Myr given $f_{\rm mix}\sim 0.6$ and $\rm -0.7\lesssim[Fe/H]\lesssim -0.5$, in good agreement with the SED fitting results of \citet{Curti2026}. Without optically-thick winds, the successful models are either too metal-poor ($\rm [Fe/H]\lesssim -0.9$) or too old ($t_{\rm age}\gtrsim 8$~Myr) or more difficult to observe with shorter $\Delta t_{\rm obs}\sim 0.1-2$~Myr. 
The reason is that these models struggle at producing enough O to remove the N/O enhancement within $t\sim 8$~Myr after the starburst (before most WR stars die). Besides, without optically-thick winds, the $\rm -0.7\lesssim[Fe/H]\lesssim -0.5$ models do not produce strong C-rich winds (Fig.~\ref{fig:abundance_avg}), which is in tension with the finding from SED-fitting that 9\% of WR stars in MARTA 4327 are WC stars \citep{Curti2026}. 


Fig.~\ref{fig:fit_opt} shows the results for the extreme case in which all stars are initially fast-rotating ($\omega=0.7$). Again, there are more successful models with optically-thick winds included. 
RXCJ2248-ID is reproduced at younger ages $t_{\rm age}\sim 1$~Myr and smaller $f_{\rm mix}$ due to the stronger N/O enhancement by both wind prescriptions. Sunburst Arc can only be reproduced with optically-thick winds in the stellar metallicity range $-1\lesssim \rm [Fe/H]\lesssim -0.7$ from SED fitting. 
Interestingly, MARTA 4327 can also be reproduced in the thin-wind-only case, where powerful PPISNe produce large O yields (Fig.~\ref{fig:abundance_opt}). 

Beyond the two initial spin models based on SMC observations and assuming purely fast-rotating ($\omega=0.7$) stars, we also explored a reference case of purely non-rotating stars, which produces much weaker N/O enhancements ($\rm [N/O]\lesssim 0.5$) and thus, cannot reproduce RXCJ2248-ID and most metal-poor ($\rm [O/H]\lesssim -0.8$) systems observed in the HNOE regime ($\rm [N/O]\gtrsim 0.3$). Therefore, we do not show the results of this case. This highlights the necessity of taking into account rotation in the modelling of massive stellar evolution and feedback \citep[see also][]{Prantzos2018}.

In general, when we consider the ``realistic'' spin distribution based on observations of O stars in the SMC, the models with optically-thick winds can fit the abundance patterns of RXCJ2248-ID, Sunburst Arc, and MARTA 4327 with stellar ages and metallicities in broad agreement with those estimated from SED fitting. Moreover, in this case, RXCJ2248-ID, Sunburst Arc, and MARTA 4327 fits into a coherent evolution sequence captured before ($t\lesssim 3$~Myr), during ($t\sim 3-4$~Myr), and after ($t\gtrsim 5$~Myr) the transition driven by C/O-rich winds from the NOE regime to ``normal'' low [N/O]. Interestingly, the inferred ISM-stellar-ejecta mixing efficiency $f_{\rm mix}\equiv M_{\rm ISM}/M_\star$ increases over time from $f_{\rm mix}\lesssim 10^{-2}$ at $t\lesssim 3$~Myr to $f_{\rm mix}\sim 0.3-1$ at $t\gtrsim 5$~Myr, consistent with the picture that mixing becomes stronger as the momentum/energy injection of stellar feedback builds up and pushes into a larger volume beyond the vicinity of massive stars \citep[e.g.,][]{Saitoh2026,Shi2026vms}. In contrast, without optically-thick winds, the moderate C/O enhancement and spectral signatures of WC stars in Sunburst Arc cannot be explained, and the models also struggle to simultaneously reproduce the abundance pattern, stellar age and metallicity of MARTA 4327. These problems can be marginally mitigated if most stars (at least those in the ZAMS mass range $m_\star\sim 70-100\ \rm M_\odot$) are born as extremely fast rotators with $\omega\gtrsim0.6$ ($v_{\rm rot}\gtrsim 600\ \rm km\ s^{-1}$). It will be shown in Appendix~\ref{apdx:csfr} that we arrive at the same conclusions when assuming continuous star formation at a constant rate (rather than the default case of instantaneous starbursts presented here). Select best-fit models are discussed in Appendix~\ref{apdx:best_fit}.

\subsection{Nitrogen mass budget}\label{sec:MN}

\begin{figure}
    \centering
    \includegraphics[width=1\linewidth]{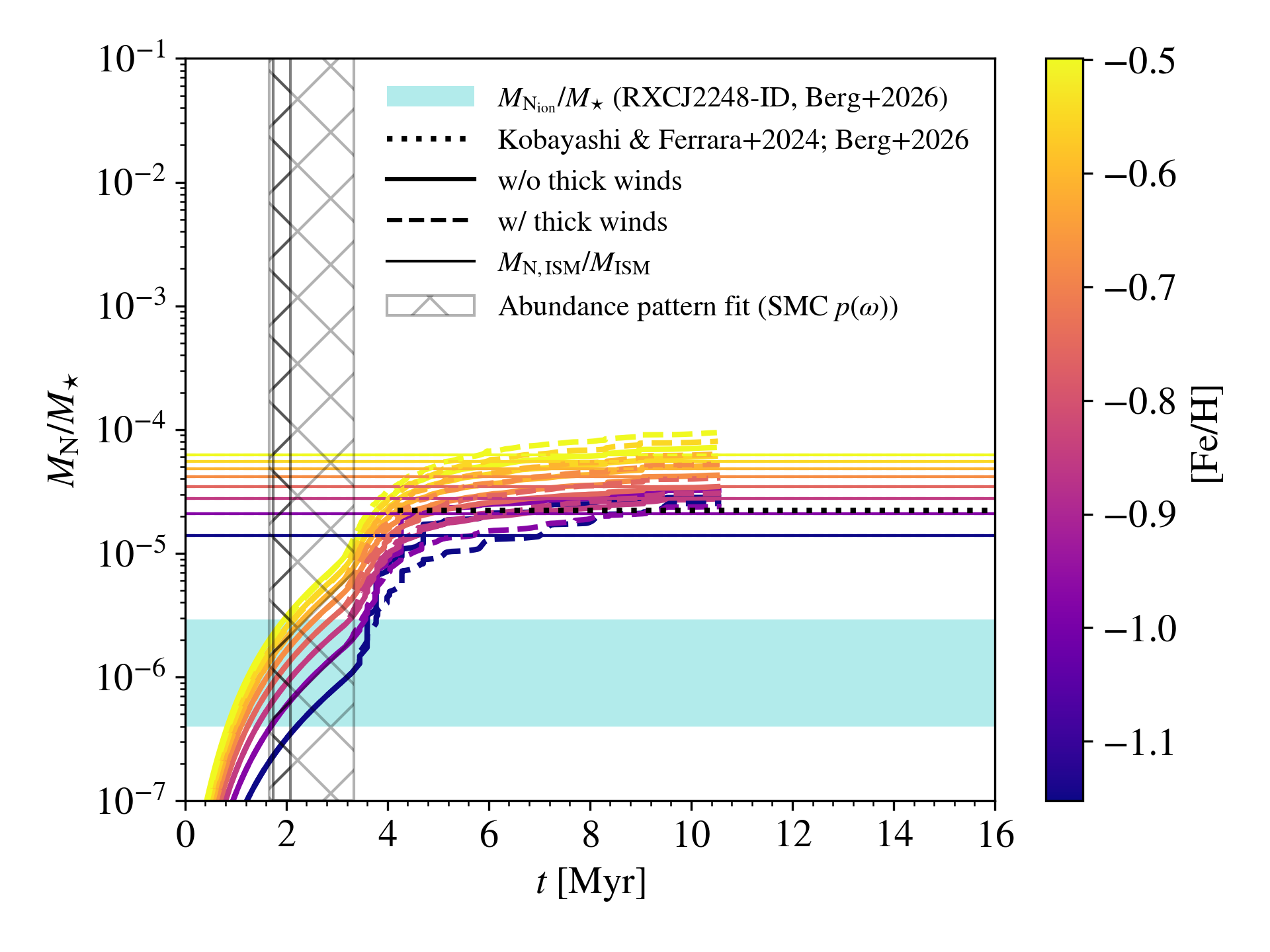}
    \vspace{-25pt}
    \caption{Cumulative N yield normalised by the total stellar mass (thick curves) as a function of time for single-age stellar populations without (solid) and with (dashed) optically-thick winds. Each population, colour-coded by the initial stellar metallicity [Fe/H], includes the contributions of stars with different initial spins using the observed spin distribution of O stars in the SMC \citep{Boco2025}. The horizontal thin lines show the mass fraction of N in the pre-starburst ISM. The vertical wide (narrow) hatched band shows the period in which our models with $\rm [Fe/H]=-0.98$ (and -0.85) can fit the abundance pattern of RXCJ2248-ID within 2 $(1)\ \sigma$. The horizontal shaded band shows the mass of ionised N in RXCJ2248-ID measured by \citet{Berg2026}. The horizontal dotted line shows the N mass produced over the duration of the second starburst ($\sim 4.2$~Myr) in the dual-burst chemical evolution model of \citet{Kobayashi2024} modified by \citet{Berg2026} to reproduce RXCJ2248-ID with reduced contributions from WC stars. } 
    \label{fig:MN_avg}
\end{figure} 

We have shown that the metal yields of our rotating massive star models with optically-thick winds can explain the \textit{relative} abundance ratios between C, N, O, He, and H of representative high-$z$ galaxies. Now, we further look into the absolute N mass budget, focusing on RXCJ2248-ID. We also comment on the cases of Sunburst Arc, the stacked spectrum of $z\sim 4.5-10.1$ star-forming galaxies in lensing-cluster fields, GN-z11, and CEERS-1019. 

Fig.~\ref{fig:MN_avg} shows the cumulative N yield normalised by the total stellar mass for our single-age stellar populations without (solid) and with (dashed) optically-thick winds, assuming the SMC spin distribution. For comparison, we plot the mass of ionised N in RXCJ2248-ID $M_{\rm N_{\rm ion}}\sim 18.2-57.5\ \rm M_\odot$ (shaded band) estimated by \citet{Berg2026} with a simple photoionisation model, i.e., $M_{\rm N_{\rm ion}}/M_{\star}\sim 10^{-6}$, normalised by an SED-derived total stellar mass of $M_\star=1.96\times 10^{7}\ \rm M_\odot$. 
The N yields increase rapidly over time before slowing down at $t\gtrsim 4$~Myr and saturate at $M_{\rm N}/M_\star\sim 3-9\times 10^{-5}$ at $t\sim 10$~Myr, above the observed ionised N budget by $\sim 1.5$~dex. Our results agree well with that of the chemical evolution model of \citet{Kobayashi2024}, $M_{\rm N}/M_{\star}\sim 2\times 10^{-5}$ at $t\sim4$~Myr. This is not surprising, since their model also considers a substantial fraction ($\sim 50\%$) of rotating ($\omega\sim 0.2-0.45$) stars based on the yields of \citet{Limongi2018}. It will be shown in Appendix~\ref{apdx:comp_model} that the models of \citet{Limongi2018} predict comparable/larger N yields than ours. 
Unlike the model of \citet{Kobayashi2024} that reproduces the abundance pattern of RXCJ2248-ID at $t_{\rm age}\simeq4.2$~Myr with a larger N yield than the observed $M_{\rm N_{\rm ion}}$ \citep{Berg2026}, our models with $\rm [Fe/H]\sim -1$ reproduce RXCJ2248-ID at $t\sim 2-3$~Myr after the starburst (see Sec.~\ref{sec:obs} and Appendix~\ref{apdx:csfr}), in slightly better agreement with the SED-derived age $t_{\rm age}=1.8_{-0.4}^{+0.7}$ \citep{Topping2024}. The N yield at this earlier stage is well within the range of $M_{\rm N_{\rm ion}}$ inferred from observations, and the contribution of N from the ISM is negligible ($\lesssim 2\%$) given $f_{\rm mix}\lesssim 10^{-3}$. This implies that it is mostly the N-rich stellar winds that power the strong N emission in a high-density ($n_{\rm e}\gtrsim 10^5\ \rm cm^{-3}$), high-pressure ($\gtrsim 10^9\ \rm K\ cm^{-3}$), ionised state, and that a substantial fraction of the N yield is retained in this phase. 

A similar picture emerges in the recent hydrodynamic simulations of giant molecular clouds by \citet{Shi2026vms}. They find that the stellar ejecta dominate the N mass budget $M_{\rm N}/M_\star\sim 10^{-6}$ in the high-pressure phase at $t\sim 3-4$~Myr after the onset of star formation\footnote{The main sources of N enrichment in the simulations of \citet{Shi2026vms} are non-rotating very massive ($m_\star\sim 100-300$) stars modeled by \citet{Costa2025}, which only produce large N-yields in late evolution stages $t\gtrsim 2$~Myr. Therefore, the NOE phase driven by stellar winds shows up later in their case than predicted by our models of rotating massive stars with N-rich winds during MS evolution. }. This picture also applies to the case of Sunburst Arc \citep[with $M_\star\sim 10^7\ \rm M_\odot$,][]{Pascale2023,Rivera-Thorsen2024}, which \citet{Shi2026vms} aim at reproducing. As shown in Sec.~\ref{sec:obs} and Appendix~\ref{apdx:csfr}, our models can reproduce Sunburst Arc with $t_{\rm age}\sim 3-4$~Myr and $f_{\rm mix}\lesssim 0.01$, showing $M_{\rm N}/M_\star\sim 10^{-6}-10^{-5}$, which is generally consistent with the estimates made by \citet{Pascale2023} if mixing with the ISM is inefficient ($f_{\rm mix}\lesssim 0.02$). Both in our phenomenological models and the simulations of \citet{Shi2026vms}, the stellar N yield only accounts for a small fraction (in the order of $f_{\rm mix}$) of the total N mass budget of the cloud (mostly as low-pressure gas) at this early stage, although they dominate the high-pressure component. The simulations further show that the high-pressure gas is removed later-on ($t\gtrsim 4$~Myr), and the stellar N yield can account for up to $\sim 20$\% of the total N mass. This is consistent with our finding that the stellar N yield cannot dominate the total N mass budget unless the total mass of well-mixed, enriched medium is very small $M_{\rm ISM}/M_{\star}\ll2$, as $M_{\rm N}/M_\star$ can reach at most 2 times of $M_{\rm N,ISM}/M_{\rm ISM}$. This naturally explains the absence of N/O enhancement in systems like MARTA 4327 observed at a later stage ($t\gtrsim 5$~Myr) compared with RXCJ2248-ID and Sunburst Arc, where efficient mixing ($f_{\rm mix}\gtrsim 0.3$) has significantly diluted the stellar N yield with O in the ISM.    

Finally, we compare our predictions with the N mass budgets inferred for the stacked spectrum of $z\sim 4.5-10.1$ star-forming galaxies in lensing-cluster fields, GN-z11, and CEERS-1019, all belonging to the HNOE regime with $\log(\rm N/O)>-0.6$, with the first two even showing tentative signatures of WR stars \citep{Chen2026,Nakane2026,Umeda2026}. 
For the stacked spectrum of $z\sim 4.5-10.1$ star-forming galaxies in lensing-cluster fields, it is estimated by \citet{Umeda2026} that the N mass budget in the HNOE gas is $M_{\rm N}\sim 15\ \rm M_\odot$ given a total stellar mass of $M_\star\sim 10^{7.7}\ \rm M_\odot$, such that $M_{\rm N}/M_\star\sim 3\times 10^{-7}$, well below our N yields at $t\gtrsim 2$~Myr. For GN-z11, \citet{Senchyna2024} estimate the ionised N mass as $M_{\rm N_{\rm ion}}\sim 10^{4}\ \rm M_\odot$ assuming $n_{\rm e}=10^{3}\ \rm cm^{-3}$, while the total stellar mass is $M_\star\sim 10^{9}\ \rm M_\odot$ \citep{Bunker2023,Tacchella2023,Alvarez-Marquez2025}, leading to $M_{\rm N}/M_\star\sim 10^{-5}$, which is comparable to/smaller than our N yields at $t\gtrsim 5$~Myr. This estimate should be regarded as an upper limit because the mass should be smaller given for higher $n_{\rm e}\sim 10^5\ \rm cm^{-3}$ traced by UV diagnostics as $M_{\rm N_{\rm ion}}\propto n_{\rm e}^{-1}$. \citet{Senchyna2024} also derives the N yield per O-type star of GN-z11 as $M_{\rm N}/N_{\rm O-type}\sim 0.01\ (10^{3}\ {\rm cm^{-3}}/n_{\rm e})~\rm M_\odot$, which is again consistent with our results $\bar{m}_{\rm N}\equiv M_{\rm N}/N(m_\star \ge20\ \rm M_\odot)\sim 0.01-0.035\ \rm M_\odot$. For CEERS-1019, \citep{Marques-Chaves2024} estimates the total mass of ionised gas as $M_{\rm ion}\sim 1.2\times 10^{5}\ \rm M_\odot$ with $n_{\rm e}=10^{5}\ \rm cm^{3}$, which corresponds a N mass of $M_{\rm N_{\rm ion}}\sim 25\ \rm M_\odot$ given $\log(\rm N/O)=-0.18$ and $\log\rm (O/H)+12=7.7$. The resulting N-to-stellar mass ratio is rather small, $M_{\rm N_{\rm ion}}/M_\star\sim 10^{-8}$, given $M_\star\sim 10^{9.3}\ \rm M_\odot$. 

In general, the $M_{\rm N}/M_\star$ values of observed NOEGs can be well explained by our N yields of massive stars considering a reasonable distribution of initial stellar spins \citep{Boco2025}. The large scatter of observed $M_{\rm N}/M_\star$ can be attributed to measurement uncertainties and diverse fractions ($\sim 0.01-1$) of stellar N yields retained in the high-density, high-pressure, ionised phase observed at different stages. 



\subsection{Caveats}\label{sec:caveats}
Although the aforementioned phenomenological models can reproduce representative high-$z$ galaxies with WR signatures with a coherent evolution sequence for the first $\sim 10$~Myr after a starburst based on the metal yields from massive ($m_\star\sim 20-100\ \rm M_\odot$) stars with rotation and optically-thick winds, several caveats in our stellar evolution models can affect this interpretation:

First, mass loss from massive stars, especially in the metal-poor regime, is still poorly understood \citep[e.g.,][]{Backs2024,Toala2024,Lefever2025,Sabhahit2026,Sander2026,Trigg2026}. Although our approach based on \citet{Sabhahit2023} can reproduce most properties of single WR stars in the SMC with optically-thick winds, it is definitely not a perfect solution that works in all evolution stages. On the one hand, it has been shown in \citet[see their fig.~9]{Boco2025} that our models overpredict the mass loss rate in the WR phase ($T_{\rm eff}>10^5\ \rm K$) by a factor of $\sim 10-100$. Weaker mass loss of WR stars is also favoured by the direct transition from WN to WO (without any WC phase showing optical C lines) found in observations \citep{Sander2026wno}. In light of this, we have explored the case where the WR mass loss is reduced by a factor of $100$ in post-processing. We found that this leads to more successful models with optically-thick winds and hardly modifies the existing ones but removes almost all successful models without optically-thick winds for MARTA 4327. This is in line with our conclusion that the enrichment of CNO (and He) needed to explain high-$z$ observations is mainly driven by optically-thick winds at earlier stages with $T_{\rm eff}\lesssim10^5\ \rm K$.

On the other hand, we do not consider any special mass loss scheme for luminous RSGs with stellar envelopes approaching the Eddington limit. As a result, the models of $m_\star\gtrsim 40\ \rm M_\odot$ stars that do not trigger optically-thick winds end their evolution as extremely luminous RSGs ($L\gtrsim 10^{5.8}\ \rm L_{\odot}$, $T_{\rm eff}\sim 10^{3.9}$~K) violating the empirical Humphreys-Davidson (HD) limit \citep{Humphreys1979}, which happens to the majority of initially slowly-rotating ($\omega\lesssim 0.3$) stars with $m_\star\lesssim 60\ \rm M_\odot$. This can be avoided if mass loss is enhanced before/after the star crosses the HD limit. In fact, the widely used models of \citet{Limongi2018} for wind metal yields from massive rotating stars introduce special mass loss episodes to remove unbound layers of quasi-super-Eddington RSGs at $\log(T_{\rm eff}\ [\rm K])\sim 3.7-3.9$. This leads to larger/comparable N yields compared with our case even without explicitly considering optically-thick winds (see Appendix~\ref{apdx:comp_model}). Recently, \citet{Pauli2026} developed a new prescription for Eddington-limit-induced mass ejections from inflating stars. With this prescription, single stellar evolution models with moderate/low initial rotation ($\omega\sim 0.1$) can reproduce key features of the observed stellar populations in the SMC and Large Magellanic Cloud, including (faint) single WR stars that motivate optically-thick winds. This mechanism is expected to have similar metal enrichment signatures and can supplement optically-thick winds in producing N/O enhancements. Taking it into account should not modify the general picture outlined in Sec.~\ref{sec:obs}, which we defer to future work.

In addition to the mass loss prescription, metal yields are also strongly affected by the treatments of convection, overshooting, rotational mixing, and supernovae, as shown in Appendix~\ref{apdx:comp_model} through the comparison between the metal yields from the stellar evolution tracks computed by three codes under different assumptions. For instance, the weaker core overshooting and stronger rotational mixing in the \textsc{franec} models of \citet{Limongi2018} with respect to our \textsc{mesa} models lead to larger N yields. Our models set the parameters for overshooting and rotational mixing based on calibrations against asteroseismology and surface abundance observations mainly targeting solar-metallicity stars of $m_\star\sim 10-40\ \rm M_\odot$ \citep{Heger2000,Lechien2026}. It is unclear if these choices are still valid for more massive and metal-poorer stars that are primarily responsible for NOE metal enrichment. We plan to explore a broader range of prescriptions for convection, overshooting, and rotational mixing \citep[e.g.,][]{Agrawal2026} and new calibrations \citep[e.g.,][]{Sciarini2026} in future work. 
The treatment of PPISNe is also highly uncertain in the lack of direct observational constraints. As mentioned in Sec.~\ref{sec:sn} trigger a PPISN when the final He core mass is above $32\ \rm M_\odot$ following \citet{Woosley2017}. However, the core mass threshold for PPISNe is sensitive to the uncertain $\rm^{12}C(\alpha, \gamma)^{16}O$ reaction rate, and can reach as high as $\sim 100\ \rm M_\odot$ with a reaction rate $3\sigma$ lower than the median \citep{Farmer2020,Mehta2022}. Therefore, we have explored the case with no PPISNe, finding that it has negligible effects on the results for the models with optically-thick winds while removing many successful models in the thin-wind-only case. This further strengthens the preference for including optically-thick winds.

Finally, we only consider single stars while most massive stars form in binary/multiple systems in reality. Binary interactions not only provide additional ways of stripping stars \citep[e.g.,][]{Farmer2020,Farmer2021,Yates2024,Ma2025,Briel2026} but also modify their surface chemical abundances \citep[e.g.,][]{jin2026chemical} and pre-SN states \citep[e.g.,][]{Gabrielli2026,Souropanis2026}, thus can strongly regulate the metal yields. For instance, it is found by \citet{Nguyen2024} that non-conservative case B mass transfer significantly boosts the ejecta from metal-poor ($\rm [Fe/H]\sim -1$) stars of $m_\star\sim 10-40\ \rm M_\odot$ by a factor of $\sim 6$, which show clear signatures of hot H burning, namely, strong enrichment of N (by up to a factor of $\sim 20$) accompanied by moderate depletion of C and O (by a factor of $\sim 2-3$), similar to the N-rich winds in our case. Similar effects have been seen in the metal yields of lower-mass ($m_\star\sim0.7- 8.5\ \rm M_\odot$) AGB stars, where binary interactions produce systems with stronger N/O enhancements ($\rm [N/O]\sim 1-2.5$) and lower C/O ratios ($\rm-0.9\lesssim [C/O]\lesssim -0.38$) than those available from single stars \citep{Osborn2025,Osborn2025mp}. This can explain the extremely low [C/O] observed in some NOEGs \citep{Ji2026,Rusakov2026}. Therefore, binary interactions could be an alternative or supplementary channel of NOE metal enrichment with respect to rotation-enhanced optically-thick winds. However, their exact roles are largely unknown, as theoretical predictions suffer from large uncertainties in binary interaction physics in addition to the aforementioned uncertainties in single stellar evolution. For instance, binary effects will be weaker with higher mass transfer efficiency, and the relevant uncertainties in metal yields can be comparable to the binary effects themselves \citep{Kemp2025}. Moreover, it is recently shown by \citet{Bao2025,Lechien2025,LopezOller2026} that the observed properties of Be X-ray binaries and Be+sdOB star binaries favour an intermediate mass transfer efficiency $\sim 0.6$, in contrast to the highly non-conservative mass transfer (limited by critical rotation of the accretor) commonly assumed in previous studies. As ongoing research rapidly updates our understanding of binary interaction physics with accumulating observational data
\citep[e.g.,][]{Moe2025,Schurmann2025,Mapelli2026,vanSon2026,Xu2026}, it would be more practical to investigate the metal enrichment from binary stars in future work.



\subsection{Alternative N/O enhancement mechanisms}\label{sec:alt}
While our models demonstrate that the winds of rotating massive stars with $m_\star\sim 20-100\ \rm M_\odot$ entering the WR phase provide a compelling explanation for the NOE abundance patterns of high-$z$ galaxies, in particular those showing direct spectroscopic signatures of WR stars, it is important to consider alternative physical mechanisms of N/O enhancements that may be required to explain systems without WR signatures or with chemical abundance patterns deviating from the predictions of our models. Below, we discuss two well-known alternative scenarios, focusing on their different predictions for the timing, duration, scale, strength, and broader chemical signatures of the NOE phase.

The first alternative mechanism invokes the enrichment of the interstellar medium by very/extremely massive or supermassive stars, with $m_\star\sim 100-10^5\ \rm M_\odot$ \citep{Vink2018,vink2023,Charbonnel2023}. 
These objects are hypothesised to form in (metal-poor) dense stellar environments through runaway collisions and/or rapid gas accretion \citep[e.g.,][]{Gieles2018,Chon2020,Reinoso2023,Fujii2024,Liu2024sf,Rantala2024}. 
The prevalence of $m_\star>100\ \rm M_\odot$ stars in high-$z$ galaxies is also supported by their observed emission-line features and metal-scaling relations \citep[e.g.,][]{Liu2025mzsfr,Yang2026}. 
The nucleosynthesis from such stars is characterised by extreme CNO-cycle processing, even stronger than that in the less massive stars considered in our case, producing very high N/O ratios up to $\log(\rm N/O)\sim 1-2$ \citep[e.g.,][]{Nagele2023,Nandal2025,Ebihara2026} that are difficult to achieve with WR winds (which produce a maximum of $\log(\rm N/O)\sim 0.1-0.8$). Furthermore, models of supermassive stars ($m_\star\sim 10^{3-5}\ \rm M_\odot$) predict chemical signatures distinct from the WR wind scenario, such as enhanced Ne/O ratios \citep{Nandal2025},  enhanced scandium or vanadium from general relativistic-instability explosions \citep{Nagele2023}, and an anticorrelation between sodium and oxygen or magnesium and aluminium, reminiscent of the patterns observed in globular/nuclear star clusters \citep{Charbonnel2023,Senchyna2024,Ji2026,Kokorev2026,Tang2026lrd}. The NOE phase will be short-lived ($t_{\rm NOE}\sim 1-2\ \rm Myr$) if the stars do not collapse entirely into black holes before post-MS nucleosynthesis removes the hot H-burning signatures. 
However, the conveyor belt mechanism, 
where gas accretion and wind/collisional mass loss reach a balance, can continuously process H and release a large amount of N-rich material over several million years, thus sustaining extended NOE phases that are resilient to ISM mixing/dilution \citep{Gieles2025,Ramirez-Galeano2025,Shi2026,Xu2026agn}. 

The second alternative mechanism attributes the N/O enhancements to AGB stars operating in conjunction with bursty star formation and temporally differential winds \citep{Bhattacharya2026,McClymont2026,Watanabe2026}. In this scenario, a starburst episode produces intermediate-mass ($m_\star\sim 1-8\ \rm M_\odot$) stars. Thereafter, stellar feedback expels the ISM enriched with O by CCSNe, leaving behind a relatively low-mass gas reservoir that will be enriched with N by the AGB winds of the intermediate-mass stars on timescales of $\sim 40-100\ \rm Myr$. This processes happen on galactic ($\sim\rm kpc$) scales, while the WR and SMS scenarios produce localized enrichment on scales of compact star clusters $\lesssim 10\ \rm pc$ \citep{Shi2026vms,Saitoh2026}. The resulting N/O enhancement can reach $\log(\rm N/O)\sim 0.5$, comparable to the WR case, while [C/O] is typically not enhanced \citep{Osborn2025mp,McClymont2026} unlike our case where C-rich optically-thick winds of WR stars boost [C/O] and reduced [N/O] at $t\gtrsim 3-4$~Myr after the starburst \citep[see also][]{Kobayashi2024,Tapia2024}. A key feature of the AGB scenario is the association of N/O enhancements with galactic outflows, which are tentatively detected in $\sim 40\%$ of observed NOEGs \citep{Rusakov2026}. Moreover, the N/O enhancement shows an anti-correlation with star formation rate, with NOEGs preferentially found in the post-starburst lull phase with small H$\beta$ equivalent widths ($\rm EW [H\beta]\lesssim 40\ \AA$) in contrast to the WR-driven metal enrichment during active starbursts. Evidence for such anti-correlation has recently been found by \citep{SinghRai2026} through a stacking analysis of 135 galaxies with JWST/NIRSpec $R\sim 1000$ spectroscopy at $z\sim 6-10$. However, this also means that AGB enrichment alone struggles to explain high-EW [H$\beta$] NOEGs, which may require additional sources of metal yields and/or H$\beta$ emission, such as WR stars and AGN \citep{Isobe2025,Shi2026,Rusakov2026,Zhu2026} or inhomogeneous metal distribution \citep{McClymont2026}. 

It is worth noting that these mechanisms are not mutually exclusive and may operate simultaneously or sequentially on different scales at different evolutionary stages in different galaxies \citep{Kobayashi2024,Schaerer2026,Watanabe2026}. 
Future observations with improved sensitivity and spectral resolution will be essential for distinguishing between these scenarios and establishing the relative contributions of different enrichment mechanisms to the observed population of NOEGs \citep[e.g.,][]{Marques-Chaves2024,Cameron2026,Ji2026,Rusakov2026,SinghRai2026,Umeda2026,Watanabe2026}. 

\section{Conclusions}\label{sec:conclusion}
We compute the metal yields of massive ($m_\star\in [20,100]\ \rm M_\odot$), metal-poor stars entering the Wolf-Rayet (WR) phase, using a series of \textsc{mesa} stellar evolution models equipped with a self-consistent switch between optically-thin and optically-thick wind regimes \citep{Sabhahit2023,Boco2025} on a dense grid of metallicities ($Z\in [0.001,0.0045]$) and initial spins ($\omega\in [0,0.7]$). We trace the chemical evolution of the medium enriched by young ($t_{\rm age}\lesssim 10$~Myr) stellar populations and compare our predictions with JWST observations of high-$z$ galaxies with WR signatures and enhanced N/O ratios. Here are our main findings:
\begin{enumerate}
    \item Our models reveal a general two-phase picture of metal enrichment. The early phase ($t\lesssim 3-4$~Myr after the onset of star formation) is dominated by N-rich winds from MS stars with efficient rotational mixing ($\omega\gtrsim 0.3$), which can drive the N/O ratio to supersolar values as high as $\log(\rm N/O)\sim 0.1-0.8$ (depending on the initial spin distribution). This phase is followed by a rapid transition driven by C/O-rich ejecta from optically-thick winds and/or pulsational pair-instability supernoave (PPISNe). Crucially, the PPISN-dominated case typically releases C and O simultaneously and thus, does not significantly enhance [C/O] as the optically-thick wind scenario. The models including optically-thick winds naturally produce a C-enrichment phase that boosts [C/O] up to $\sim 0.4-1.2$. \vspace{2pt}
    \item The duration of the N/O-enhanced (NOE, $\log\rm (N/O)>-1.1$) and highly N/O-enhanced (HNOE, $\log(\rm N/O)>-0.6$) phases of the enriched medium are sensitive to the efficiency of mixing with the ISM $f_{\rm mix}\equiv M_{\rm ISM}/M_\star$. Considering the observationally motivated spin distribution from O-type stars in the Small Magellanic Cloud (SMC), the (H)NOE phase lasts for $\sim 2-3\ (1-2)$~Myr, provided that $f_{\rm mix}\lesssim 0.01$. These timescales imply that a few percent of high-$z$ galaxies should be caught in the (H)NOE phase, which is consistent with the observed occurrence rate of $\sim 3-18$\% \citep{Rusakov2026}.
    \item The models incorporating optically-thick winds can successfully reproduce the full abundance patterns ([N/O], [C/O], [O/H], and He abundance) of three representative high-$z$ galaxies with direct WR signatures: RXCJ2248-ID, Sunburst Arc, and MARTA 4327 (Sec.~\ref{sec:comp_obs}).  
    The required stellar ages and metallicities are consistent with those independently inferred from SED-fitting. These three galaxies can be interpreted as a coherent evolution sequence captured before ($t\lesssim 3$~Myr), during ($t\sim 3-4$~Myr), and after ($t\gtrsim 5$~Myr) the transition from the NOE phase to the normal low-N/O state. In contrast, models without optically-thick winds struggle to reproduce the observed C/O ratio and the timing of the transition needed to explain Sunburst Arc and MARTA 4327, both of which show spectroscopic signatures of WC stars \citep{Rivera-Thorsen2024,Curti2026}. This underscores the critical role of optically-thick winds in driving the transition and providing the necessary C enrichment.
    \item The total N yield per unit stellar mass from our models reaches $M_{\rm N}/M_\star \sim 10^{-6}$ at $t\sim 2$~Myr after the starburst and starts to saturate around $3-9\times 10^{-5}$ at $t\sim 4-10$~Myr. This is sufficient to explain the N mass budgets $M_{\rm N}/M_\star\sim 10^{-8}-10^{-5}$ inferred from observations of systems like RXCJ2248-ID \citep{Berg2026}, Sunburst Arc \citep{Pascale2023}, GN-z11 \citep{Senchyna2024}, and CEERS-1019 \citep{Marques-Chaves2024}. For RXCJ2248-ID, the predicted N yield at the SED-derived age of $\sim 2-3$~Myr is well within the observed ionised N mass range, implying that the strong N emission in the high-density, high-pressure phase is powered primarily by the N-rich stellar winds with highly inefficient mixing ($f_{\rm mix}\lesssim 0.01$), which is consistent with the results of the hydrodynamic simulations by \citet{Shi2026vms}.
\end{enumerate}

Our models show that the winds of rotating massive stars provide a promising explanation for the NOE abundance patterns of high-$z$ galaxies. However, several caveats warrant consideration. The mass loss from massive stars, especially in the metal-poor regime, remains poorly understood, and our wind prescription, while successful, is not a perfect solution for all evolution stages \citep[e.g.,][]{Pauli2026,Sander2026,Sander2026wno}. The treatments of convection, overshooting, rotational mixing, and supernovae also introduces significant uncertainties, as demonstrated by comparisons with previous predictions of metal yields \citep{Limongi2018,Liu2025mzsfr} from other stellar evolution codes (see Appendix~\ref{apdx:comp_model}). Furthermore, we have only considered single stars, whereas most massive stars form in binary or multiple systems, where binary interactions can regulate metal yields \citep[e.g.,][]{Nguyen2024}. Alternative mechanisms for N/O enhancement, such as enrichment by very/extremely massive or supermassive stars ($m_\star\sim 100-10^5\ \rm M_\odot$) or by asymptotic giant branch (AGB) stars, may also contribute \citep[e.g.,][]{Charbonnel2023, Nandal2025,McClymont2026}. These scenarios are not mutually exclusive and may complement each other and dominate metal enrichment on different scales and at different evolutionary stages.


\begin{acknowledgements}
    The cumulative metal yields as functions of time are publicly available at \url{https://zenodo.org/records/22897057}. 
      The authors acknowledge financial support from the German Excellence Strategy via the Heidelberg Cluster of Excellence (EXC 2181 - 390900948) STRUCTURES. LB acknowledges support by the Deutsche Forschungsgemeinschaft (DFG, German Research Foundation) in the form of a Walter Benjamin position -- Projektnummer 555003977. 
      The authors thank Gustavo Bruzual, Stephane Charlot, and Alessandro Bressan for helpful discussion. 
      We made use of the \textsc{mesa} software (\url{https://docs.mesastar.org/en/latest/}) version r12115; \citep{Paxton2011, Paxton2013, Paxton2015, Paxton2018, Paxton2019}.
      We made use of the \textsc{mesa} inlists \url{https://github.com/Apophis-1/VMS_Paper1}, and \url{https://github.com/Apophis-1/VMS_Paper2} from \citet{Sabhahit2023}. 
      This research made use of \textsc{NumPy} \citep{Harris20}, \textsc{SciPy} \citep{SciPy2020}, and \textsc{Matplotlib} \citep{Hunter2007}.
\end{acknowledgements}

%
\bibliographystyle{bibtex/aa} 
\bibliography{references} 


\begin{appendix}
\nolinenumbers
    
\section{Correction for non-solar initial abundance patterns}\label{sec:non_solar}

Our \textsc{mesa} models use the solar abundance pattern as initial conditions, despite the low metallicities. However, metal-poor stars/nebulae/galaxies show distinct abundance patterns in local observations, with deficits of C, N, and Fe with respect to O \citep[e.g.,][]{Nicholls2017,Mendez-Delgado2024,Cataldi2026}. Similar trends have been seen in high-$z$ metal-poor galaxies \citep[e.g.,][]{Cataldi2026,Ji2026}, although the exact scaling relation and scatter can be different for different stellar/galaxy populations and evolve with redshift. To capture such features, we re-scale the CNO yields (from both winds and SNe) in post-processing to reflect a non-solar pre-starburst ISM abundance pattern using $\rm [Fe/O]_{\rm pre}=-0.2$ and the $\rm[N/O]_{\rm pre}$ and $\rm [C/O]_{\rm pre}$ values predicted by the local scaling relations of \citet{Cataldi2026} and \citet{Nicholls2017}, respectively. These relations predict metal-poor plateaus of $\rm[N/O]_{\rm pre}\sim[C/O]_{\rm pre}\sim -0.5$ at $\log(\rm O/H)+12\lesssim 8$. 
We choose $\rm [Fe/O]_{\rm pre}=-0.2$ based on the semi-analytical galaxy evolution model of \citep{Liu2025mzsfr} calibrated to reproduce the observed metallicity-mass-star-formation-rate relation at $z\sim 4-10$. This model predicts a mean Fe/O ratio $\rm [Fe/O]\sim0.2$ for galaxies at $z\sim 4-12$ with $\rm [Fe/H]\in [-1.15,-0.5]$ (matching our stellar evolution models) in the mass range $M_\star\sim 10^{7.5}-10^{9.5}\ \rm M_\odot$, matching that of the observed NOEGs \citep[see their fig.~2]{Rusakov2026}. This value is also typical among 2nd-generation GC stars and observed NOEGs \citep[see their fig.~4]{Ji2026}. Given the yield of X$=$C, N, O predicted by the stellar evolution model, $m_{\rm X,sim}$, the corrected yield is 
\begin{align}
\begin{split}
    &m_{\rm X}=\begin{cases}
        m_{\rm X,sim}\times 10^{\rm [X/Fe]_{\rm pre}}\ ,& m_{\rm X,sim}\le m_{\rm X,0}\ ,\\
        m_{\rm X,0}\times 10^{\rm [X/Fe]_{\rm pre}}+\Delta m_{\rm X,sim} f_{\rm pro}\ ,& m_{\rm X,sim}>m_{\rm X,0}\ ,
    \end{cases}\\
    &\Delta m_{\rm X,sim} = m_{\rm X,sim}-m_{\rm X,0}\ ,\\
    &m_{\rm X,0}=m_{\rm H,sim}A_{\rm X}\times 10^{\rm [Fe/H]+\log(X/H)_{\odot}}\ ,
\end{split}
\end{align}
where $m_{\rm X,0}$ is the mass of X initially in the ejecta of H mass $m_{\rm H,sim}$, $\Delta m_{\rm X,sim}$ is the predicted mass of newly-synthesized X, and $f_{\rm pro}$ is a scaling factor that takes into account the modification of secondary CNO production by the non-solar initial abundance pattern:
\begin{align}
    &f_{\rm pro}=f_{\rm 2nd,ej}f_{\rm CNO} + 1 - f_{\rm 2nd,ej}\ ,\\
    &f_{\rm 2nd,ej} = n_{\rm CNO,0}/n_{\rm CNO,sim}\, \quad f_{\rm CNO}= x_{\rm CNO,pre}/ x_{\rm CNO,\odot}\ .
\end{align}
Here, $f_{\rm 2nd,ej}$ is the fraction of secondary CNO production (from CNO initially in the star) \textit{in the ejecta}, which will be corrected by $f_{\rm CNO}$, the ratio of total CNO nuclei numbers of the non-solar and solar abundance patterns, while the primary production part $1-f_{\rm 2nd,ej}$ is unchanged. We calculate $f_{\rm 2nd,ej}$ using the effective total number of CNO nuclei in the yield $n_{\rm CNO,sim}$ and that initially in the ejecta $n_{\rm CNO,0}$:
\begin{align}
    &n_{\rm CNO,sim} = m_{\rm C,sim}/A_{\rm C} + m_{\rm N,sim}/A_{\rm N} + m_{\rm O,sim}/A_{\rm O}\ ,\\
    &n_{\rm CNO,0}=m_{\rm H,sim}\times10^{\rm [Fe/H]}\times x_{\rm CNO,\odot}\ ,
\end{align}
where $x_{\rm CNO,\odot}$ is the total CNO abundance of the sun, and $x_{\rm CNO,pre}$ is its non-solar counterpart at the same iron abundance:
\begin{align}
    &x_{\rm CNO,\odot} = \sum_{\rm X'= C,N,O} 10^{\rm \log(X'/H)_{\odot}}\ ,\\
    &x_{\rm CNO,pre}=\sum_{\rm X'= C,N,O} 10^{\rm [X'/Fe]_{\rm pre}+ \log(X'/Fe)_{\odot}+\log(Fe/H)_{\odot}}\ . 
\end{align}
In our case, $f_{\rm CNO}$ remains close to unity ($|f_{\rm CNO}-1|\lesssim 0.3$), as the reduction of C, N relative to O is compensated by the enhancement of O relative to Fe. Therefore, the secondary CNO production in our stellar evolution models with a solar initial abundance pattern should operate similarly as that from the non-solar initial abundance pattern when approaching the CNO equilibrium. When primary CNO production dominates in late evolution stages with for $f_{\rm 2nd,ej}\rightarrow 0$, the correction naturally vanishes as $f_{\rm pro}\rightarrow 1$. 

Note that, our \textsc{mesa} models adopt the solar abundance pattern from \citet{Grevesse1998}, and the \textsc{parsec} models underlying the yields of CCSNe and failed SNe use the solar abundance pattern from \citet{Caffau2011} as initial conditions. The relevant CNO abundances are similar to those of the \citet{Asplund2009} (proto)solar abundance pattern used to express CNO abundance ratios. For simplicity, we always use the $\rm \log(X/H)_{\odot}$ values from \citet{Asplund2009} to compute the CNO yield initially in the ejecta (i.e., $m_{\rm X,SN,0}$, $\tilde{m}_{\rm X,SN,0}$, and $m_{\rm X,0}$). 

\section{Results for continuous star formation}\label{apdx:csfr}

\begin{figure*}
    \centering
    \includegraphics[width=1\textwidth]{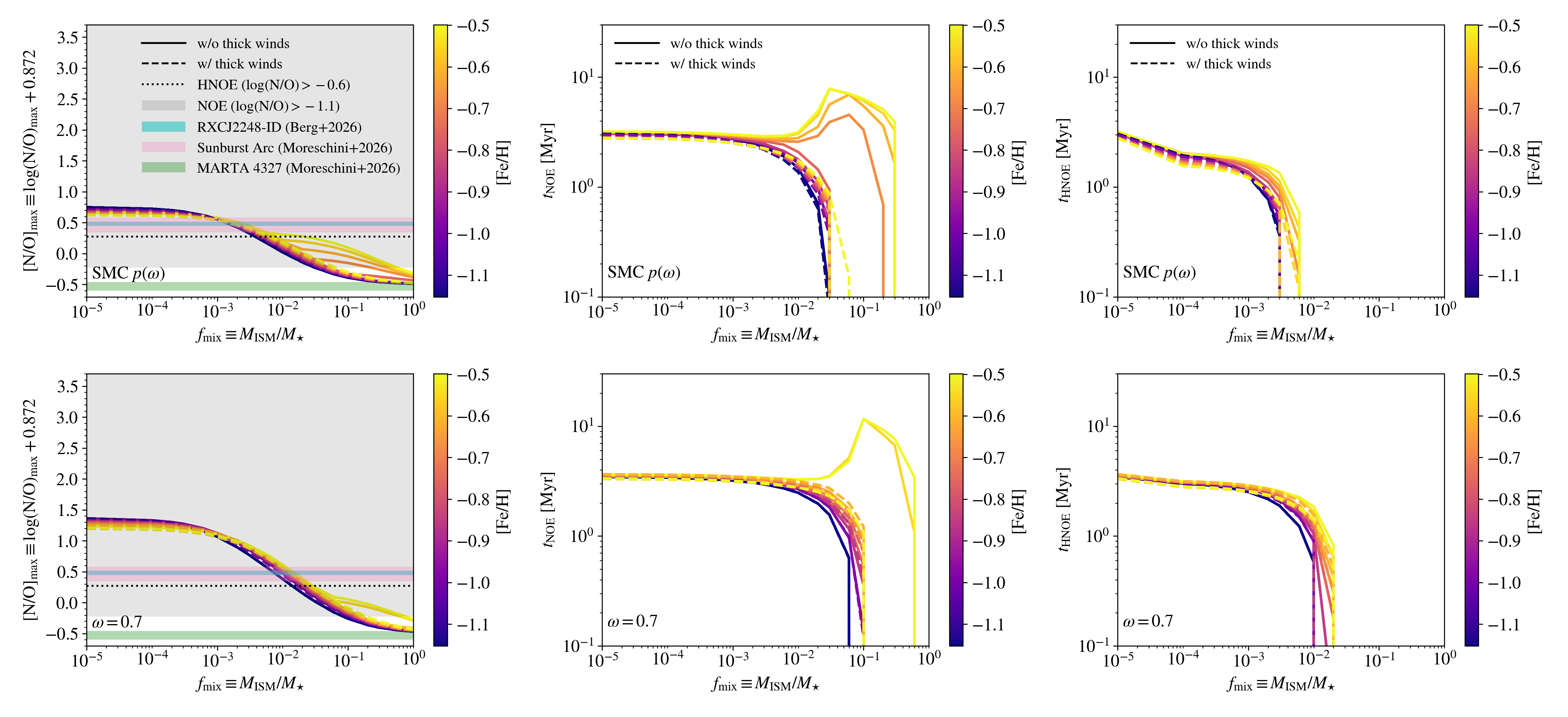}
    \vspace{-20pt}
    \caption{Same as Fig.~\ref{fig:NOE_avg} but for continuous star formation at a constant rate.}
    \label{fig:NOE_avg_csfr}
\end{figure*}

\begin{figure*}
    \centering
    \includegraphics[width=0.78\textwidth]{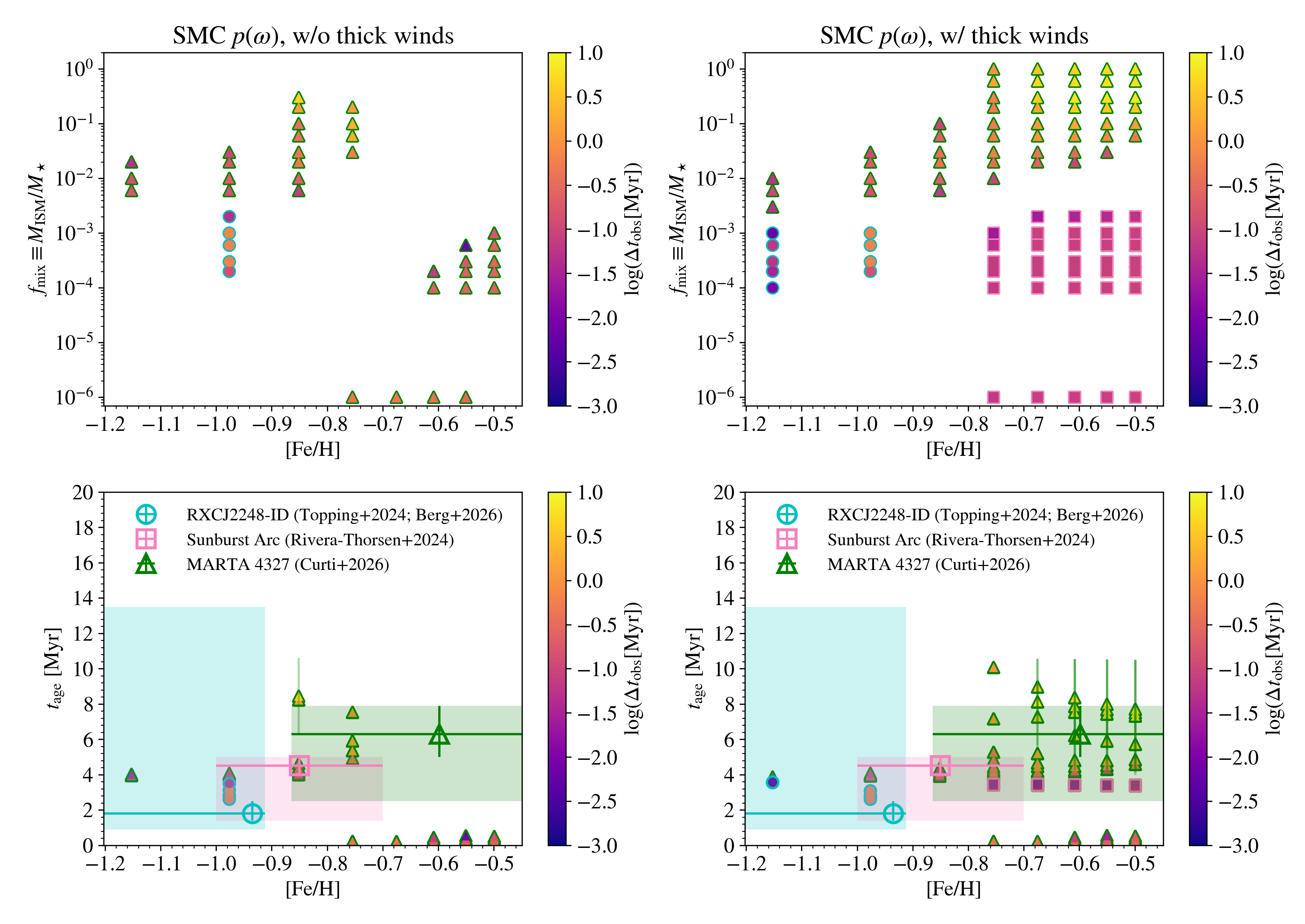}
    \vspace{-10pt}
    \caption{Same as Fig.~\ref{fig:fit_avg} but for continuous star formation at a constant rate.}
    \label{fig:fit_avg_csfr}
\end{figure*}

\begin{figure*}
    \centering
    \includegraphics[width=0.78\textwidth]{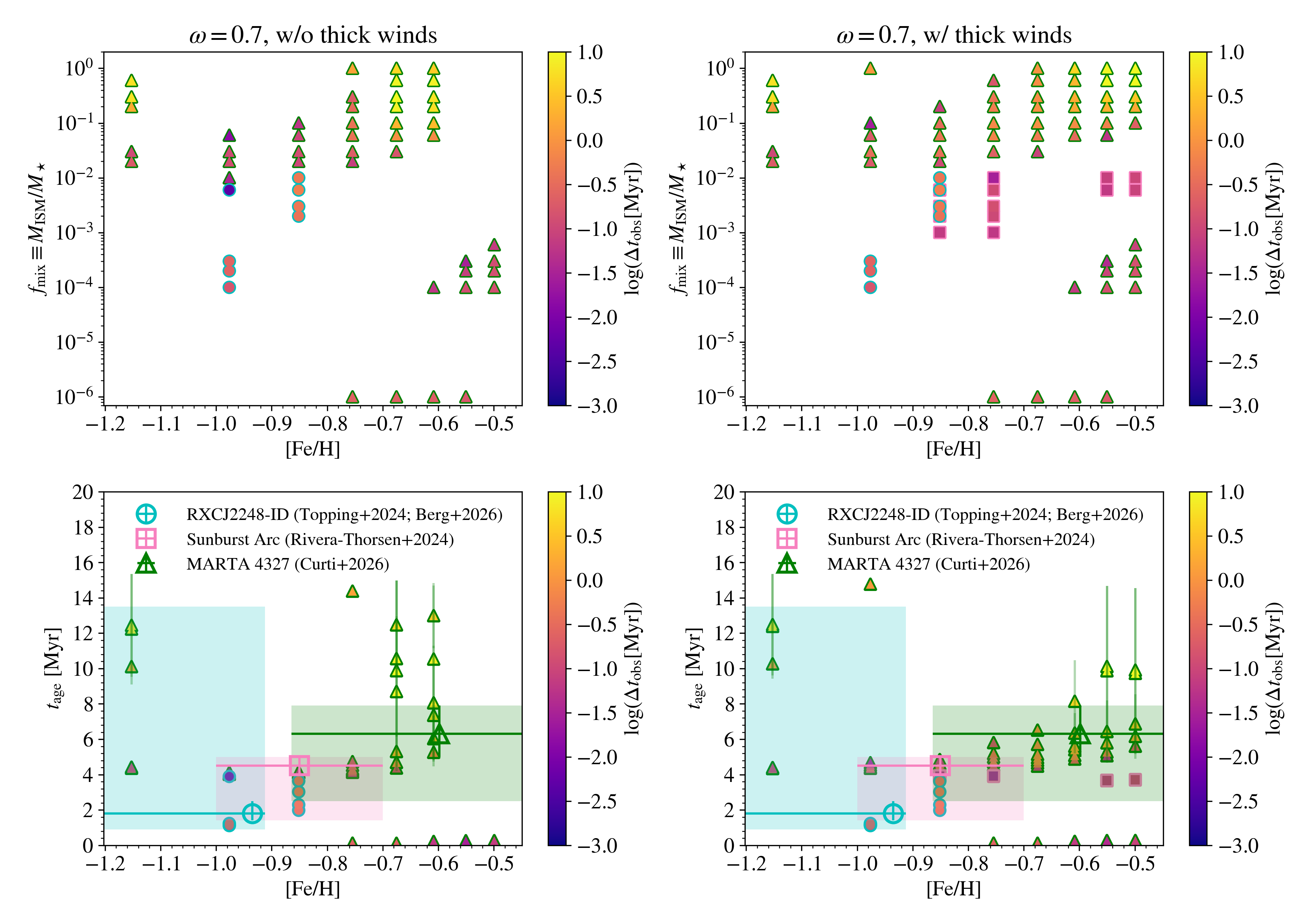}
    \vspace{-10pt}
    \caption{Same as Fig.~\ref{fig:fit_avg} but for continuous star formation at a constant rate and purely fast-rotating stars with $\omega=0.7$.}
    \label{fig:fit_opt_csfr}
\end{figure*}

In addition to the default assumption of an instantaneous starburst adopted for the results presented in the main text, we consider continuous star formation at a constant rate here to check whether our conclusions are robust against variations in the star formation history. 

Fig.~\ref{fig:NOE_avg_csfr} shows general properties of the (H)NOE phase in terms of $\rm [N/O]_{\max}$ (left), $t_{\rm NOE}$ (middle), and $t_{\rm HNOE}$ (right) as functions of $f_{\rm mix}$. Compared with the results for an instantaneous starburst shown in Fig.~\ref{fig:NOE_avg}, $\rm [N/O]_{\max}$ is moderately smaller by $\sim 0.3$~dex. This is caused by the contributions of newly-formed stars whose winds are less enriched with N by rotational mixing (see Fig.~\ref{fig:no_ind}) than those of older stars. As a result, the N/O enhancement is more vulnerable to mixing with the ISM, such that the rapid drop of $t_{\rm NOE}$ and $t_{\rm HNOE}$ below $\sim 1$~Myr generally occurs at smaller $f_{\rm mix}$ by a factor of $\sim 3$. 

Figs.~\ref{fig:fit_avg_csfr} and \ref{fig:fit_opt_csfr} show the successful models that can fit the abundance patterns of RXCJ2248-ID, Sunburst Arc, and MARTA 4327, considering the SMC spin distribution and the extreme case of purely fast-rotating ($\omega=0.7$) stars, respectively. The key general difference from the results for an instantaneous starburst (Figs.~\ref{fig:fit_avg} and \ref{fig:fit_opt}) is that the successful models tend to have slightly larger $t_{\rm age}$ and smaller $f_{\rm mix}$, as if the chemical enrichment process is weakened and slowed down. This is reasonable because continuous star formation smooths the variation of abundance ratios with time, including the early rise of [N/O] and its subsequent decrease caused by C/O enrichment. 

There are subtle differences for individual target systems: For RXCJ2248-ID, the $\rm [Fe/H]=-0.85$ models can no longer reproduce observations with the SMC spin distribution, because the decrease of $\rm [O/H]$ in the early phase ($t\lesssim 3$~Myr) is too weak to reach the observed [O/H] range. For Sunburst Arc, when optically-thick winds are included, there are many more successful models with larger $\Delta t_{\rm obs}$, including ones that have stellar ages and metallicities fully consistent with the SED-fitting results. As explained in Sec.~\ref{sec:obs}, Sunburst Arc is reproduced as a delicate transition phase with moderate enrichment of C/O that is just enough to reproduce the observed [C/O] and [O/H] but too weak to significantly suppress [N/O]. With continuous star formation, the transition is slowed down due to the mixture of the C/O-rich winds from old massive ($m_\star\sim 80-100\ \rm M_\odot$) stars with the N-rich winds of their young counterparts. Still, no models can reproduce Sunburst Arc without optically-thick winds. Similarly, the optically-thin-wind-only scenario is also disfavoured more by MARTA 4327 with continuous star formation unless all stars are initially fast-rotating at $\omega=0.7$. 

In conclusion, considering continuous star formation at a constant rate does not change our main conclusion, namely, the preference for models including optically-thick winds, which can explain the stellar and nebular properties of RXCJ2248-ID, Sunburst Arc, and MARTA 4327 with a coherent evolution sequence featuring an optically-thick-wind-driven transition from NOE to N/O-normal at $t\sim 3-4$~Myr. 




\section{Best-fit models of RXCJ2248-ID, Sunburst Arc, and MARTA 4327 }\label{apdx:best_fit}


\begin{figure*}
    \centering
    \includegraphics[width=0.78\textwidth]{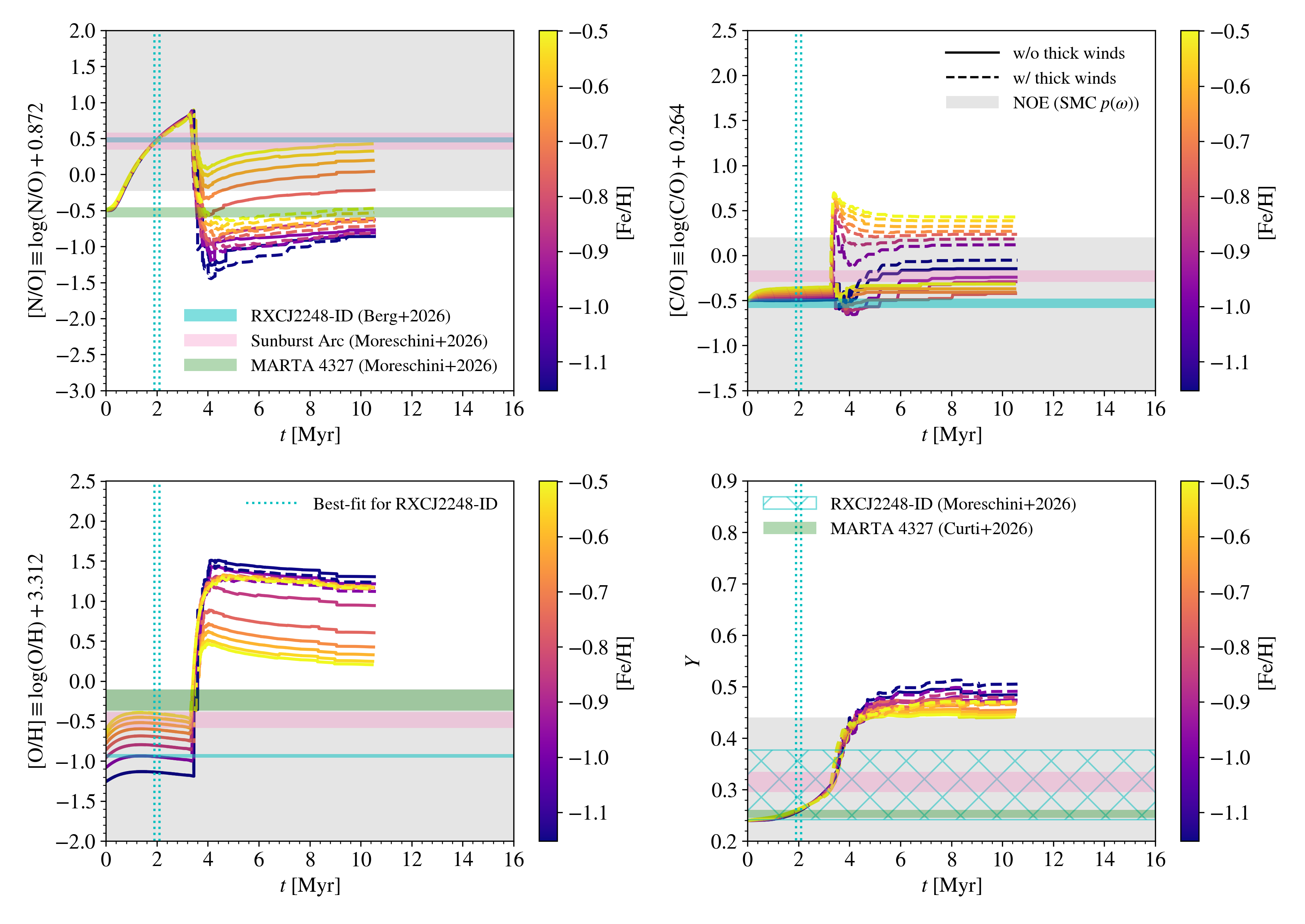}
    \vspace{-10pt}
    \caption{Evolution of [N/O] (top-left), [C/O] (top-right), [O/H] (bottom-left), and $Y$ (bottom-right) in the enriched medium of mixed stellar ejecta and ISM with $f_{\rm mix}=0.001$ for single-age stellar populations without (solid) and with (dashed) optically-thick winds. Each population, colour-coded by the initial stellar metallicity [Fe/H], includes the contributions of stars with different initial spins using the observed spin distribution of O stars in the SMC \citep{Boco2025}. The shaded regions denote the abundance-ratio space generally occupied by observed NOE galaxies (see Fig.~\ref{fig:abundance_avg}). The horizontal cyan, pink, and green shaded/hatched bands show the chemical abundances (within $1\sigma$ uncertainties) of three representative high-$z$ galaxies with Wolf-Rayet signatures, RXCJ2248-ID, Sunburst Arc, and MARTA 4327, respectively \citep{Berg2026,Curti2026,Moreschini2026}. The vertical dotted lines mark the period ($\Delta t_{\rm obs}\sim 0.2$~Myr) in which the $\rm [Fe/H]\simeq-0.98$ models with and without optically-thick winds can reproduce RXCJ2248-ID within $1\sigma$.}
    \label{fig:abundance_t_avg}
\end{figure*}

\begin{figure*}
    \centering
    \includegraphics[width=0.78\textwidth]{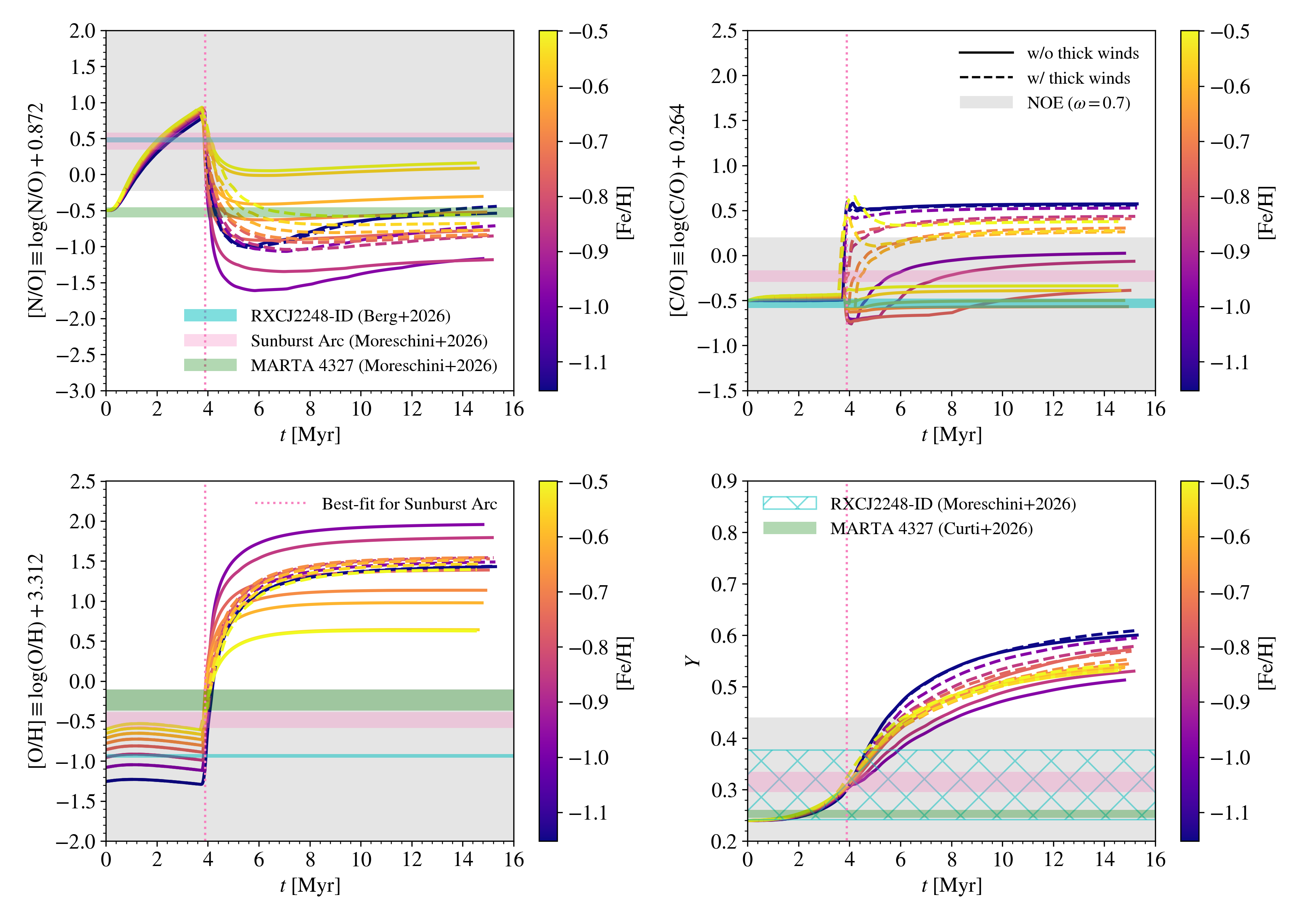}
    \vspace{-10pt}
    \caption{Same as Fig.~\ref{fig:abundance_t_avg} but for $f_{\rm mix}=0.003$, continuous star formation at a constant rate, and purely fast-rotating stars with $\omega=0.7$. The vertical dotted line marks the period ($\Delta t_{\rm obs}\sim 0.06$~Myr) in which the $\rm [Fe/H]\simeq-0.76$ model with optically-thick winds can reproduce Sunburst Arc within $1\sigma$.}
    \label{fig:abundance_t_opt_fmix0.003_csfr}
\end{figure*}

\begin{figure*}
    \centering
    \includegraphics[width=0.78\textwidth]{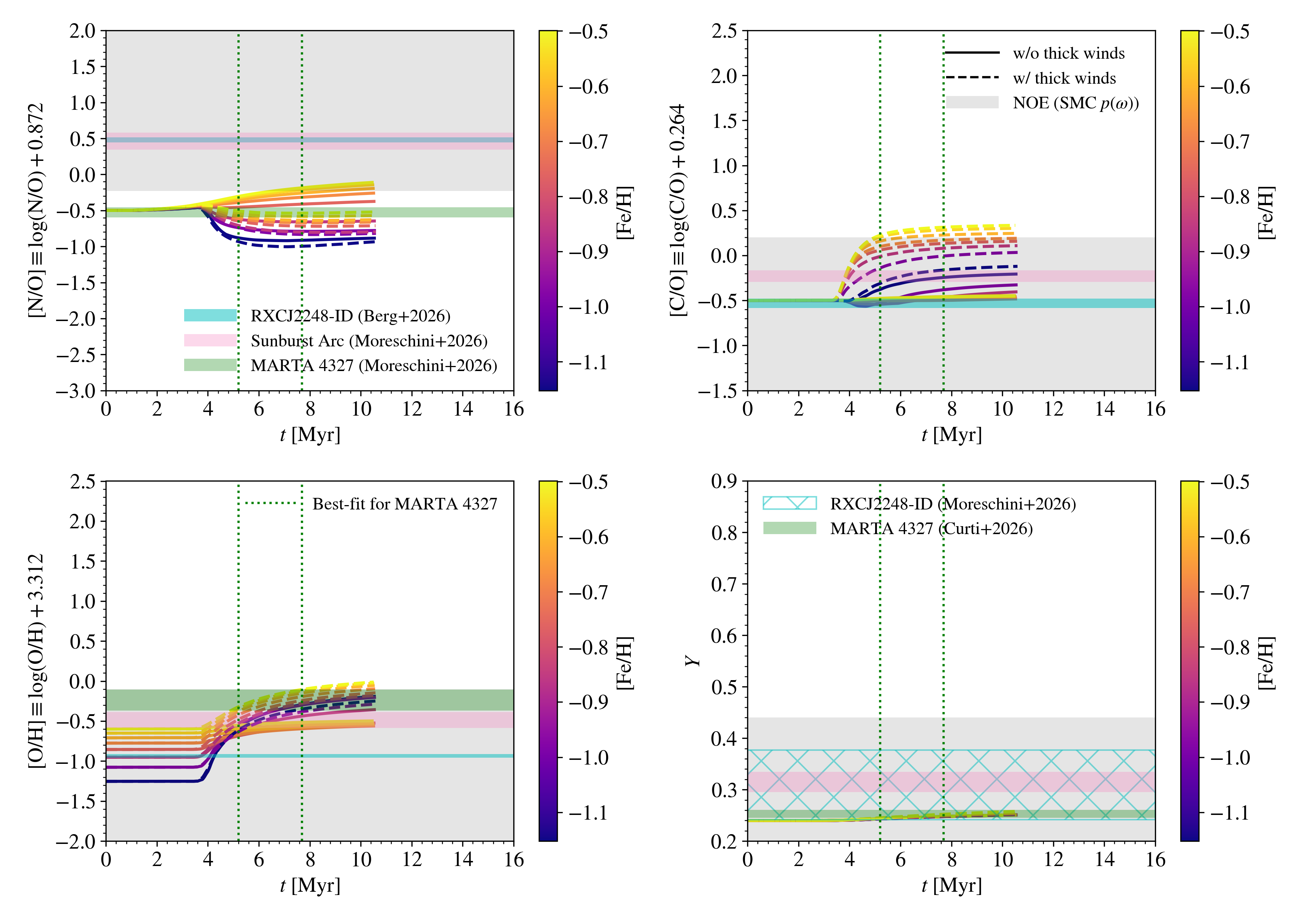}
    \vspace{-10pt}
    \caption{Same as Fig.~\ref{fig:abundance_t_avg} but for $f_{\rm mix}=0.3$ and continuous star formation at a constant rate. The vertical dotted lines mark the period ($\Delta t_{\rm obs}\sim 2.5$~Myr) in which the $\rm [Fe/H]\simeq-0.5$ model with optically-thick winds can reproduce MARTA 4327 within $1\sigma$.}
    \label{fig:abundance_t_avg_fmix0.3_csfr}
\end{figure*}

Several models have been identified in Sec.~\ref{sec:obs} and Appendix~\ref{apdx:csfr} to reproduce the abundance patterns of three representative high-$z$ galaxies with WR signatures within $2\sigma$: RXCJ2248-ID, Sunburst Arc, and MARTA 4327 given proper choices of ISM mixing efficiency $f_{\rm mix}$, stellar metallicity [Fe/H], initial spin distribution, wind prescription, and star formation history. Here, we present selected best-fit models that simultaneously fit the observed  [N/O], [C/O], [O/H], and $Y$ within $1\sigma$ with stellar metallicities and ages consistent with SED-fitting results. 

For RXCJ2248-ID, Fig.~\ref{fig:abundance_t_avg} shows the time evolution of abundance ratios in the enriched medium with $f_{\rm mix}=0.001$ for all single-age stellar populations without (solid) and with (dashed) optically-thick winds, assuming the SMC spin distribution. Among them, both models with and without optically-thick winds at $\rm [Fe/H]\simeq-0.98$ can reproduce the abundance pattern of RXCJ2248-ID within $1\ (2)\sigma$ in a WN phase lasting $\Delta t_{\rm obs}\sim 0.2\ (0.4)$~Myr at $t\sim 2$~Myr after an instantaneous starburst, in good agreement with the stellar metallicity and age inferred from SED-fitting \citep{Topping2024,Berg2026,Claeyssens2026} (see Fig.~\ref{fig:fit_avg}). 

For Sunburst Arc, we instead require $f_{\rm mix}=0.003$, purely fast-rotating ($\omega=0.7$) stars, and continuous star formation to achieve the best fitting quality. As shown in Fig.~\ref{fig:abundance_t_opt_fmix0.003_csfr}, the model at $\rm [Fe/H]\simeq-0.76$ including optically-thick winds can reproduce the abundance pattern of Sunburst Arc within $1\ (2)\sigma$ at $t\sim 3.9$~Myr with $\Delta t_{\rm obs}\sim 0.06\ (0.09)$~Myr. The stellar age and metallicity are consistent with those needed to fit the observed spectra \citep{Chisholm2019,Pascale2023,Rivera-Thorsen2024} (see Fig.~\ref{fig:fit_opt_csfr}). Under the same conditions, the models without optically-thick winds cannot simultaneously reproduce the moderate [C/O] and high [N/O] of Sunburst Arc because PPISN-driven C enrichment is always accompanied by O enrichment that strongly suppresses [N/O]. 

For MARTA 4327, stronger mixing is needed to reduce [N/O], and we identify the best-fit model with $f_{\rm mix}=0.3$, $\rm [Fe/H]=-0.5$, SMC spin distribution, optically-thick winds included, and continuous star formation. As shown in Fig.~\ref{fig:abundance_t_avg_fmix0.3_csfr}, the abundance pattern of MARTA 4327 is fitted within $1\sigma$ during $t\sim 5.2-7.7$~Myr after the onset of star formation, lasting for $\Delta t_{\rm obs}\sim 2.5$~Myr, in perfect agreement with the stellar metallicity and age inferred from the WR bump features \citep{Curti2026} (see Fig.~\ref{fig:fit_avg_csfr}). The best-fit model predicts a moderate C/O enhancement with $[\rm C/O]\sim 0.3$ in MARTA 4327, which could be used to test our model in future observations. The optically-thin-wind-only scenario is again disfavoured because it cannot provide enough O yields to reproduce the low (high) [N/O] ([O/H]) of MARTA 4327 with respect to high-$z$ NOEGs.


\section{Comparison with previous metal yields predictions}\label{apdx:comp_model}

\begin{figure*}
    \centering
    \includegraphics[width=1\textwidth]{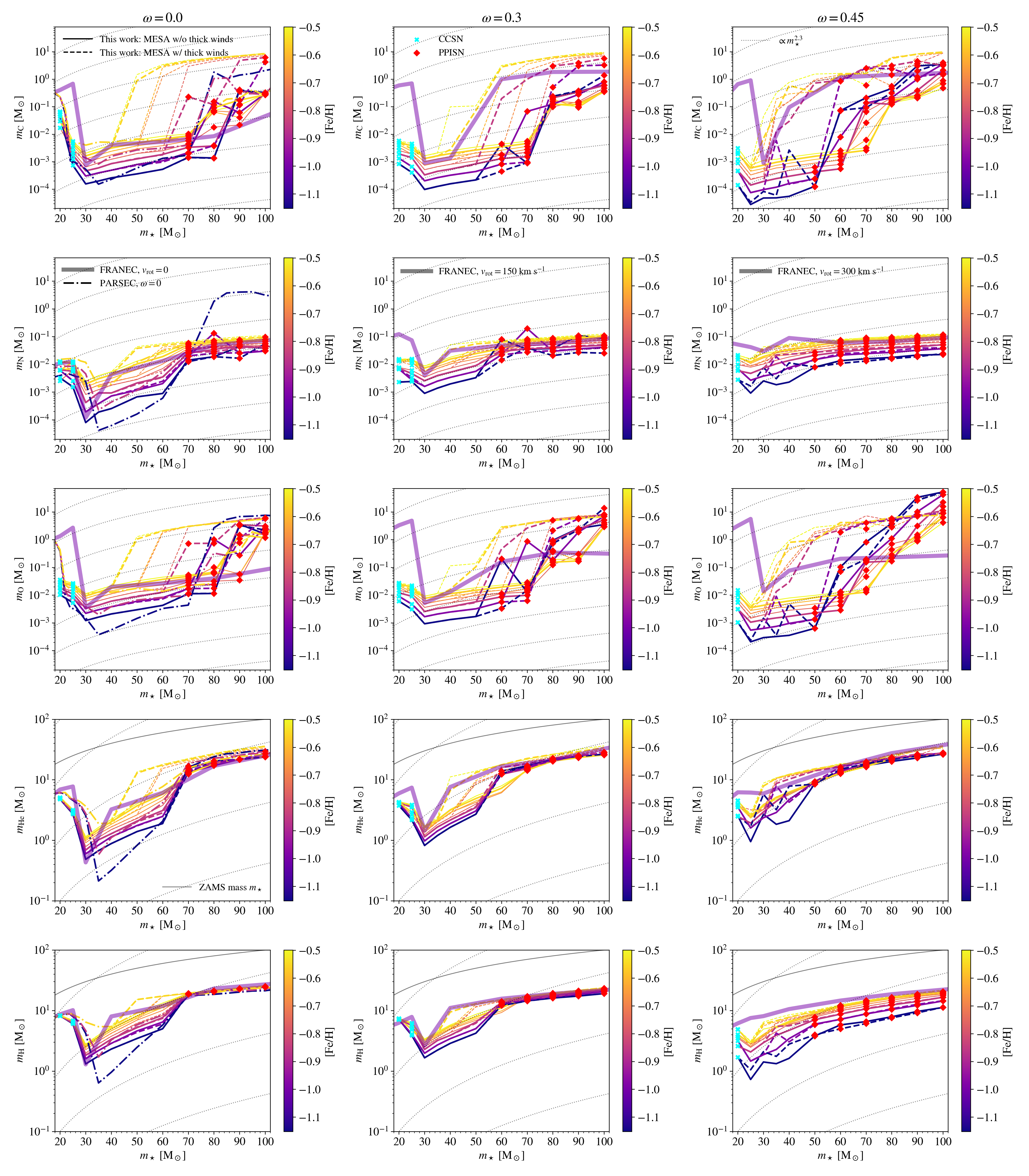}
    \vspace{-20pt}
    \caption{Total C , N, O, He, and H yields as functions of ZAMS mass for initially non-rotating ($\omega=0$, left column), moderately rotating ($\omega=0.3$, middle column), and fast-rotating ($\omega=0.45$, right column) massive stars. The solid and dashed curves show the results of our \textsc{mesa} models without and with optically-thick winds, colour-coded by metallicity. The results for $\rm [Fe/H]=-1.15$, -0.98, -0.85, and -0.55 are highlighted with thicker curves to be compared directly with literature results. The stars that become CCSNe and PPISNe are labeled by the crosses and diamonds. The thickest solid curves show the results of \textsc{franec} tracks \citep[set R]{Limongi2018} at $\rm [Fe/H]=-1$ for initial rotation velocities of $v_{\rm rot}=0$, 150, and $300\ \rm km\ s^{-1}$, roughly corresponding to $\omega=0$, 0.3 and 0.45. The dashed-dotted curves in the left panels for $\omega=0$ show the yields used by \citet{Liu2025mzsfr} computed following the method of \citet{Goswami2021} from non-rotating \textsc{parsec} tracks \citep{Costa2025} at $\rm [Fe/H]=-1.15$, -0.85, and -0.55. }
    \label{fig:mz_m_comp}
\end{figure*}

We compare our yields based on \textsc{mesa} tracks \citep{Boco2025} with those predicted by \textsc{parsec (v2.0)} \citep{Costa2025} and \textsc{franec} \citep{Limongi2018} tracks, focusing on C, N, O, He, and H. In particular, we consider the \textsc{parsec} yields from non-rotating stars at three metallicity bins $\rm [Fe/H]=-1.15$, -0.85, and -0.55 within the metallicity range of our tracks. These yields are computed with the method of \citet{Goswami2021} and used by \citet{Liu2025mzsfr} to model chemical evolution of high-$z$ galaxies. For \textsc{franec} tracks, we take the set R yields of \citet{Limongi2018}\footnote{We take the data in the table \texttt{tab\_yieldstot\_iso\_exp.nod} from \url{https://orfeo.iaps.inaf.it/}.} for $\rm [Fe/H]=-1$ and three initial rotation velocities $v_{\rm tot}=0$, 150, and $300\ \rm km\ s^{-1}$, which roughly correspond to $\omega=0$, 0.3, and 0.45 (for $m_\star\sim 20\ \rm M_\odot$) in our case. The goal of this comparison is to estimate and understand the uncertainties in current predictions of metal yields from massive stars in the ZAMS range $m_\star\sim 20-100\ \rm M_\odot$. A systematic comparison of the underlying stellar evolution models is beyond the scope of this work \citep[see, e.g., sec. 4.1 of][]{Costa2025}. Below, we summarise the key differences in the assumptions made by the three sets of models, and the reader is referred to the original papers for details.  
\begin{itemize}
    \item \textbf{Winds:} The treatments of baseline optically-thin winds are identical, but optically-thick winds, WR winds, and rotational enhancement are implemented differently: Unlike our \textsc{mesa} models computed with a self-consistent switch to optically-thick winds developed by \citet{Sabhahit2023}, the \textsc{parsec} models include the dependence of mass loss rate on $\Gamma_{\rm e}$ from \citet{Grafener2008,Vink2011}. 
    The \textsc{franec} models trigger special mass loss episodes to remove unbound/super-Eddington layers during the RSG phase ($\log(T_{\rm eff}\ [\rm K])\sim 3.7-3.9$). 
    The \textsc{parsec} and \textsc{franec} models use the WR ($T_{\rm eff}>10^5\ \rm K$) wind prescriptions from \citet{Sander2019} and \citet{Nugis2000}, respectively, while we combine the models of \citet{Vink2017} and \citet{Sander2020}. The \textsc{franec} models use the prescription of \citet{Heger2000} for rotational enhancement of mass loss, while we use that of \citet{Maeder2000}.
    \item \textbf{Convection:} Both our \textsc{mesa} models and the \textsc{franec} models use the Ledoux criterion for convective regions but with different mixing length parameters, $\alpha_{\rm MLT}=1.5$ and 2.3, respectively. In contrast, the \textsc{parsec} models use the Schwarzschild Criterion with $\alpha_{\rm MLT}=1.74$, which tend to produce larger convective cores and more efficient mixing.
    \item \textbf{Overshooting:} Our \textsc{mesa} models use the exponential formalism for core overshooting with $f_{\rm ov}=0.03$ and 0.01 for MS and post-MS evolution. The \textsc{parsec} models instead use penetrative overshooting \citep{Bressan1981} with an equivalent core overshooting length $l_{\rm ov}\sim 0.25 H_P$, roughly corresponding to $f_{\rm ov}=0.025$ in the exponential formalism. They also include step overshooting from convective envelopes with $l_{\rm ov}= 0.7 H_P$. The \textsc{franec} models only consider step overshooting from the H-buring core with $l_{\rm ov}= 0.2 H_P$, corresponding to $f_{\rm ov}\sim 0.02$. In general, we have \textsc{mesa} $\gtrsim$ \textsc{parsec} $\gtrsim$ \textsc{franec} for the strength of overshooting during MS, and \textsc{parsec} $\gtrsim$ \textsc{mesa} $\gtrsim$ \textsc{franec} in post-MS evolution.
    \item \textbf{Rotational mixing\footnote{Note that the two rotational mixing parameters, $f_{c}$ and $f_\mu$, are degenerate when calibrated to limited data for surface N abundance. Besides, the calibration is model-dependent as other processes also affect surface abundances. For instance, when calibrated to the same ($\rm [Fe/H]=0$) target considered by \citet{Heger2000}, \citet{Chieffi2013} obtained $f_c=1$ and $f_\mu=0.03$ rather than $f_{c}=1/30$ and $f_\mu=0.05$ adopted in our models from \citet{Heger2000}. To reproduce the same observations in the Large Magellanic Cloud, \citet{Limongi2018} obtained $f_c=1.5$ and $f_\mu=0.01$ using \textsc{franec}, while \citet{Costa2019} got $f_c=1.7$ and $f_\mu=0.45$ using \textsc{parsec}.}:} Our \textsc{mesa} models use $f_{c}=1/30$ and $f_\mu=0.05$ following the calibration of \citet[]{Heger2000} that produces surface enhancements of N by a factor of $\sim 2-3$ in solar-metallicity models with $m_\star\sim 10-20\ \rm M_\odot$. The \textsc{franec} models instead use $f_c=1.5$ and $f_\mu=0.01$, which are calibrated, using $m_\star=13\ \rm M_\odot$ stars, to reproduce the surface N enhancements observed in the Large Magellanic Cloud \citep{Hunter2009} following \citet{Brott2011}. It is shown by \citet[see their fig.~3]{Limongi2018} that the latter calibration leads to stronger rotational mixing that typically enhances the surface N abundance by $\sim 0.3$~dex. 
    \item \textbf{Supernovae:} Both our \textsc{mesa} models and the \textsc{parsec} models use the the method of \citet{Goswami2021} to compute SN yields, which interpolate the yield tables of \citet{Limongi2003} and \citet{Chieffi2004} for CCSNe, and model PPISNe as super winds based on the remnant masses of \citet{Woosley2017}. The difference is that we use the delayed formalism of \citet{Fryer2012} for CCSN explodability and remnant mass, while the \textsc{parsec} models follow the approach of \citet{Limongi2003} and \citet{Chieffi2004} to eject $\sim 0.07\ \rm M_\odot$ of $\rm ^{56}Ni$ per SN. The \textsc{franec} models instead adopt the mixing and fallback scheme \citep{Umeda2002} to compute their own CCSN yields for $m_\star\sim 13-25\ \rm M_\odot$ stars (again ensuring a fixed $\rm ^{56}Ni$ ejecta of $\sim 0.07\ \rm M_\odot$) and do not model the SN yields from more massive stars. 
\end{itemize}

Fig.~\ref{fig:mz_m_comp} shows the total C , N, O, He, and H yields \footnote{Previous studies (e.g., \citealp[]{Kobayashi2006,Nomoto2013,Costa2025}) typically publish tables of total/net yields released throughout the stars' lifetimes without information for time evolution, except for \citet{Limongi2018}. Therefore, most chemical evolution models inject all metal yields from a star at the end of its life (or MS). This simplification can be problematic if the metal yields from early stellar evolution stages are important. This is the case of rapidly rotating stars under quasi-chemically homogeneous evolution, which already launch N-rich winds early on (see Fig.~\ref{fig:no_ind}). In light of this, we publish the cumulative yields as functions of time (\url{https://zenodo.org/records/22897057}) and call for future work to treat the time evolution seriously.} 
per star as functions of ZAMS mass for three initial spins $\omega=0$ (left), 0.3 (middle), and 0.45 (right). 

In the non-rotating case, the He (and C/O) yields from \textsc{parsec} tracks are larger than those of our \textsc{mesa} tracks at $m_\star\lesssim 30\ (20)\ \rm M_\odot$. The reason is that the weaker overshooting during MS in the \textsc{parsec} models leads to smaller He cores and weaker fallback in CCSNe. 
In the wind-dominated regime of $m_\star\sim 30-70\ \rm M_\odot$, the \textsc{parsec} models typically produce less H, He, N, and O yields than our \textsc{mesa} results at $\rm [Fe/H]=-1.15$. Considering our models without optically-thick winds, this trend is weaker and occasionally reverses at larger $m_\star$ and $\rm [Fe/H]$, likely caused by a complex interplay between the prescriptions for overshooting and winds. Interestingly, the C yields are generally larger in the \textsc{parsec} models by $\sim 0.3$~dex, which could be caused by their stronger post-MS overshooting. 
Once optically-thick winds are active in our \textsc{mesa} models, the \textsc{parsec} yields are far behind. This is generally true for more massive stars with $m_\star\sim 70-100\ \rm M_\odot$ that can undergo PPISNe as well. However, the $\rm [Fe/H]=-1.15$ case is an exception, where the \textsc{parsec} models produce (significantly) larger C/O (N) yields. The highly enhanced N production is likely caused by overshooting from convective envelopes that mix hot H-burning shells with the He core. Meanwhile, the stronger post-MS core overshooting produces larger CO cores and, therefore, stronger mass loss from PPISNe. The \textsc{parsec} models at $m_\star\sim 70-100\ \rm M_\odot$ tend to produce larger C, O, and He yields than our \textsc{mesa} models without optically-thick winds, likely due to their stronger post-MS overshooting and $\Gamma_{\rm e}$-dependent wind mass loss rates. 

The \textsc{franec} models produce more overall ejecta (dominated by H and He) by up to a factor of $\sim 3$ at $m_\star\lesssim 25\ \rm M_\odot$ and $m_\star\sim 40-60$ than our \textsc{mesa} results at $\rm [Fe/H]\sim -1$, where optically-thick winds hardly make any difference. The larger yields at $m_\star\lesssim 25\ \rm M_\odot$ are caused by stronger CCSN mass ejection, partially due to smaller He cores with weaker overshooting in the \textsc{franec} models. 
The mixing and fallback scheme adopted by \textsc{franec} further boost the C/O yields from CCSNe with respect the \textsc{mesa} and \textsc{parsec} results. The difference at $m_\star\sim 40-60\ \rm M_\odot$ indicates that the \textsc{franec} models undergo stronger mass loss, which can be attributed to the special prescription for removing unbound layers of RSGs. The CNO yields are correspondingly larger, and the same trend exists in the rotating-models without optically-thick winds. For more massive stars with $m_\star\sim 70-100\ \rm M_\odot$, our \textsc{mesa} models produce similar overall ejecta as the \textsc{franec} models. Besides, we have much larger C/O yields (by up to 2 orders of magnitude) and moderately larger (smaller) He (N) yields by $\sim 0.3$~dex. The reason could be that our stars lose more mass from larger He cores via optically-thick winds and/or PISNe. The larger N yields in the \textsc{franec} models are likely caused by hotter H-burning shells next to smaller He cores. 

For rotating stars, the enhanced mass loss from RSGs approaching the Eddington limit in the \textsc{franec} models is even stronger, leading to larger yields of H, He, N, and C than our \textsc{mesa} results (even with optically-thick winds) in most cases at $m_\star\gtrsim 30\ \rm M_\odot$. Hotter H-burning shells and higher rotational mixing efficiency in the \textsc{franec} models can also contribute to this trend. Interestingly, the differences are generally smaller for heavier elements and larger $m_\star$, and the trend (strongly) reverses in the C (O) yields at $m_\star\gtrsim 60-90~\rm M_\odot$ for our models with optically-thick winds. In our case, massive stars do not become RSGs whenever optically-thick are triggered, as winds efficiently strip these stars during core He burning and keep them compact/hot ($T_{\rm eff}\gtrsim 10^{4.4}\ \rm K$). This normally happens to metal-poor ($\rm [Fe/H]\lesssim -0.85$) massive stars with $\omega\gtrsim 0.3$ and $m_\star\gtrsim 50\ \rm M_\odot$. The compactness (which enhances rotational mixing) and core overshooting could explain why our models can produce larger C/O yields compared with those of the \textsc{franec} models without post-MS overshooting in this regime. In the \textsc{franec} models, $m_\star\gtrsim 30\ \rm M_\odot$ stars expand significantly to become RSGs and only contract after the major N-rich (super-Eddington) mass loss episodes, among which the models with $m_\star\gtrsim 40\ \rm M_\odot$ always go deeply beyond the HD limit before getting evacuated by the enhanced mass loss. Interestingly, the yields of all elements considered here almost increase monotonically with initial rotation velocity in the \textsc{franec} models, while a similar trend is only seen in the C, N, O, He yields in our \textsc{mesa} models when optically-thick models are active. At $m_\star\lesssim 50\ \rm M_\odot$ where the yields are dominated by optically-thin winds, CCSNe and failed SNe, the C/O yields are smaller for larger $\omega$, because stronger rotational mixing (aided by overshooting) leads to larger He cores and stronger fallback. These mechanisms can also explain the decrease of H yields with $\omega$ at $\omega\gtrsim 0.3$.


Finally, we derive the average N yield per O-type star (as WR progenitors) $\bar{m}_{\rm N}\equiv M_{\rm N}/N(m_\star \ge20\ {\rm M_\odot})\simeq M_{\rm N}/N_{\rm O-type}$ in the ZAMS range $m_\star\sim 20-100\ \rm M_\odot$. Assuming the \citet{Kroupa2001} IMF and the SMC spin distribution, our \textsc{mesa} models with (without) optically-thick winds have $\bar{m}_{\rm N}\sim 0.01-0.035\ (0.027)\ \rm M_\odot$ (larger at higher $\rm [Fe/H]$) for $\rm [Fe/H]\in [-1.15,-0.5]$. This corresponds to a N yield per total stellar mass formed $M_{\rm N}/M_\star\simeq \bar{m}_{\rm N}/(341\ \rm M_\odot)\sim 3-8\times 10^{-5}$. Under the same condition, the \textsc{franec} models at $\rm [Fe/H]=-1$ predict $\bar{m}_{\rm N}\simeq 0.045\ \rm M_\odot$, larger than our value at the same metallicity by a factor of $\sim 4$. If we only consider non-rotating stars, our \textsc{mesa} models predict $\bar{m}_{\rm N}\sim 0.003-0.021\ \rm M_\odot$, while the \textsc{franec} $\rm [Fe/H]=-1$ models have $\bar{m}_{\rm N}\simeq 0.01\ \rm M_\odot$. The \textsc{parsec} models at $\rm [Fe/H]=-0.85$ and $-0.55$ have $\bar{m}_{\rm N}\simeq 0.007$ and $0.016\ \rm M_\odot$, respectively. The $\rm [Fe/H]=-1.15$ case of \textsc{parsec} shows a much larger N yield $\bar{m}_{\rm N}\simeq 0.17\ \rm M_\odot$ driven by PPISNe of $m_\star\sim 70-100\ \rm M_\odot$ stars in which envelope overshooting can enhance N production. As discussed in Sec.~\ref{sec:MN}, $M_{\rm N}/N_{\rm O-type}\sim 0.01\ \rm M_\odot$ is sufficient to explain the N mass budget in GN-z11 \citep{Senchyna2024}, and other NOEGs require comparable or smaller N yields. This condition is well fulfilled by the massive stellar evolution models considered here when rotation and optically-thick winds (i.e., enhanced mass loss for high $\rm \Gamma_{\rm e}$) are taken into account. The interplay of winds and rotation is particularly important for metal-poor ($\rm [Fe/H]\lesssim -1$) stars, without which O-type stars with $\bar{m}_{\rm N}\lesssim 0.01\ \rm M_\odot$ may not be able to provide enough N yields for the most metal-poor NOEGs in observations \citep{Ji2026,Rusakov2026}. 

\end{appendix}
\end{document}